\documentclass[numberedappendix]{emulateapj}

\usepackage{amssymb,amsmath,natbib,threeparttable,rotating,url}
\usepackage[graphicx]{realboxes}
\usepackage[normalem]{ulem}
\usepackage{float}
\usepackage{comment}
\usepackage{amsmath}
\usepackage{mathtools}
\usepackage{paralist}
\usepackage[caption = false]{subfig}
\usepackage{morefloats}
\usepackage{tabularx}

\def\rf{rest-frame}
\def\um{$\mu$m}
\def\uJy{$\mu$Jy}

\def\lir{L$_{\mathrm{IR}}$}
\def\luv{L$_{\mathrm{UV}}$}
\def\msol{$M_{\odot}$}
\def\msun{$M_{\odot}$}
\def\lsol{L$_{\odot}$}

\def\lirtot{L$_{8-1000\mu m}$}

\def\fs{{\tt FourStar}}

\newcommand{\ks}{$K_s$}

\def\b4{Balmer/4000$\mathrm{\AA}$ break}

\begin{document}

\title{The FourStar Galaxy Evolution Survey\altaffilmark{12} (ZFOURGE): ultraviolet to far-infrared catalogs, medium-bandwidth photometric redshifts with improved accuracy, stellar masses, and confirmation of quiescent galaxies to \MakeLowercase{z}$\sim3.5$}

\author{Caroline M. S. Straatman\altaffilmark{1,2}, Lee R. Spitler\altaffilmark{3,4}, Ryan F. Quadri\altaffilmark{5}, Ivo Labb\'e\altaffilmark{2}, Karl Glazebrook\altaffilmark{6}, S. Eric Persson\altaffilmark{7}, Casey Papovich\altaffilmark{5}, Kim-Vy H. Tran\altaffilmark{5}, Gabriel B. Brammer\altaffilmark{8}, Michael Cowley\altaffilmark{3,4}, Adam Tomczak\altaffilmark{5}, Themiya Nanayakkara\altaffilmark{6}, Leo Alcorn\altaffilmark{5}, Rebecca Allen\altaffilmark{6}, Adam Broussard\altaffilmark{5}, Pieter van Dokkum\altaffilmark{9}, Ben Forrest\altaffilmark{5}, Josha van Houdt\altaffilmark{2}, Glenn G. Kacprzak\altaffilmark{6}, Lalitwadee Kawinwanichakij\altaffilmark{5}, Daniel D. Kelson\altaffilmark{7}, Janice Lee\altaffilmark{8}, Patrick J. McCarthy\altaffilmark{7}, Nicola Mehrtens\altaffilmark{5}, Andrew Monson\altaffilmark{7}, David Murphy\altaffilmark{10}, Glen Rees\altaffilmark{4}, Vithal Tilvi\altaffilmark{5}, Katherine E. Whitaker\altaffilmark{11}\footnotemark[$\dagger$]}

\altaffiltext{1}{straatman@mpia.de}
\altaffiltext{2}{Leiden Observatory, Leiden University, PO Box 9513, 2300 RA Leiden, The Netherlands}
\altaffiltext{3}{Australian Astronomical Observatory, PO Box 915, North Ryde, NSW 1670, Australia}      
\altaffiltext{4}{Department of Physics \& Astronomy, Macquarie University, Sydney, NSW 2109, Australia}
\altaffiltext{5}{George P. and Cynthia W. Mitchell Institute for Fundamental Physics and Astronomy, Department of Physics and Astronomy, Texas A\&M University, College Station, TX 77843}
\altaffiltext{6}{Centre for Astrophysics and Supercomputing, Swinburne University, Hawthorn, VIC 3122, Australia}
\altaffiltext{7}{Carnegie Observatories, Pasadena, CA 91101, USA}
\altaffiltext{8}{Space Telescope Science Institute, 3700 San Martin Drive, Baltimore, MD 21218, USA}
\altaffiltext{9}{Department of Astronomy, Yale University, New Haven, CT 06520, USA}
\altaffiltext{10}{School of Mathematics and Science, Chaffey College, 5885 Haven Avenue, Rancho Cucamonga, CA 91737}
\altaffiltext{11}{Department of Astronomy, University of Massachusetts, Amherst, MA 01003, USA}
\altaffiltext{12}{This paper contains data gathered with the 6.5 meter Magellan Telescopes located at Las Campanas observatory, Chile.}
\footnotetext[$\dagger$]{Hubble Fellow}

\begin{abstract}
The \textit{\fs\ galaxy evolution survey (ZFOURGE)} is a 45 night legacy program with the \fs\ {near-infrared} camera on Magellan {and one of the most sensitive surveys to date. ZFOURGE covers a total of } $400\ \mathrm{arcmin}^2$ in {cosmic} fields CDFS, COSMOS and UDS, overlapping CANDELS. {We present photometric catalogs comprising $>70,000$ galaxies, selected from ultradeep $K_s$-band detection images ($25.5-26.5$ AB mag, $5\sigma$, total), and $>80\%$ complete to $K_s<25.3-25.9$ AB}. We use 5 near-IR medium-bandwidth filters ($J_1,J_2,J_3,H_s,H_l$) as well as broad-band $K_s$ {at} $1.05\ - 2.16\ \micron$ to $25-26$ AB at a seeing of $\sim0\farcs5$. {Each field has} ancillary imaging in {$26-40$} filters {at} $0.3-8\ \micron$. We derive photometric redshifts and stellar population properties. Comparing with spectroscopic redshifts indicates {a photometric redshift uncertainty} $\sigma_z={0.010,0.009}$, and {0.011} in {CDFS, COSMOS, and UDS}. As spectroscopic samples are often biased towards bright and blue sources, {we also inspect the photometric redshift differences between close pairs of galaxies, finding} $\sigma_{z,pairs}= 0.01-0.02$ at $1<z<2.5$. We quantify how $\sigma_{z,pairs}$ depends on redshift, magnitude, SED type, and the inclusion of \fs\ medium bands. $\sigma_{z,pairs}$ {is} smalle{st} for bright, blue star-forming samples, while red star-forming galaxies have the worst $\sigma_{z,pairs}$. Including \fs\ medium bands reduces $\sigma_{z,pairs}$ by 50\% at $1.5<z<2.5$.
We calculate SFRs based on ultraviolet {and} ultradeep {far-IR} $Spitzer$/MIPS and Herschel/PACS data. We derive rest-frame $U-V$ and $V-J$ colors, and illustrate how these correlate with specific SFR and dust emission to $z=3.5$. We confirm the existence of quiescent galaxies at $z\sim3$, {demonstrating} their SFRs are suppressed by $>\times15$.
\end{abstract}


\section{Introduction}
Over the last few decades it has been possible to obtain new insights into the formation and evolution of galaxies in a statistically significant way by using large samples of sources from multiwavelength photometric surveys, for example with SDSS \citep{York00}. Improved near-IR facilities on the ground, as well as advanced space-based instruments have enabled galaxy surveys probing the universe at higher resolution, fainter magnitudes and towards higher redshifts ($z>1.5$) \citep[e.g.,][]{Lawrence07,Wuyts08,Grogin11,Koekemoer11,Whitaker11,Muzzin13b,Skelton14}. These in turn have led to great progress in tracing the structural evolution of galaxies \citep[e.g.,][]{Daddi05,vanDokkum08,Franx08,Bell12, Wuyts12,vanderWel12,vanderWel14}, luminosity and stellar mass functions \citep[e.g.,][]{Faber07,Perez-Gonzalez08,Marchesini09,Muzzin13a,Tomczak14}, the environmental effects on galaxy evolution \citep[e.g.,][]{Postman05,Peng10b,Cooper12,Papovich10,Quadri12,Kawinwanichakij14,Allen15} and the correlation between stellar mass and star-formation rate \citep[e.g.,][]{Noeske07,Wuyts11,Whitaker12} over cosmic time. 

The redshift range $1<z<3$, when the universe was between 2.1 and 5.6 Gyr old, is an important epoch for studies of galaxy evolution. During this period 60\% of all star-formation took place \citep[e.g.,][]{Madau98,Sobral13}, an early population of quiescent galaxies started to appear \citep[e.g.,][]{Daddi05,Kriek06,Marchesini10} and galaxies evolved into the familiar elliptical and spiral morphologies that we see in the universe today \citep[e.g.,][]{Bell12}. A fundamental observational limitation to understanding galaxy evolution is the availability of accurate distance estimates for mass{-limited galaxy} samples. These can be obtained with spectroscopy, but observations are limited to a biased population of galaxies: bright and most often star-forming, with strong emission lines.

Instead many galaxy surveys rely exclusively on the photometric sampling of the spectral energy distributions (SEDs) of galaxies to derive redshifts. Even when {deep imaging spanning the optical and near-infrared} is used to derive photometric redshifts, these surveys {are generally hampered by} systematic effects from the use of broadband filters. These can lead to large random errors, of the order of $\sigma_z/(1+z)\sim0.1$. {Moreover the photometric redshift accuracy is generally estimated by comparison to a small and unrepresentative spectroscopic sample, which does not allow for an analysis of the errors as a function of magnitude, redshift, or galaxy type. Redshift errors may introduce biases in derived luminosities and stellar masses} \citep{Chen03,Kriek08}.

A better sampling of the SED improves the accuracy of the photometric redshifts greatly and can be obtained by the use of medium-bandwidth filters. These were first applied in the optical for the COMBO17 survey \citep{Wolf04}. A notable feature in the SED of a galaxy is the Balmer/4000$\mathrm{\AA}$ break at rest-frame $4000\mathrm{\AA}$, which shifts into the near-IR at $z\gtrsim1.5$. For high redshift surveys, it is therefore advantageous to split up the canonical broadband J and H filters into multiple near-IR medium-bandwith filters \citep{vanDokkum09}, which stradle the Balmer/4000$\mathrm{\AA}$ break at $1.5\lesssim z\lesssim3.5$. A set of near-IR medium-bandwidth filters was used for the {NEWFIRM Medium-Band Survey} NMBS, a survey using NEWFIRM on the Kitt Peak Mayall 4m Telescope, with a limiting $5\sigma$ depth in K of 23.5 AB mag for point sources and a photometric redshift accuracy of $\sigma_z/(1+z)\sim1-2\%$ up to $z=3$ \citep{Whitaker11}.

The \fs\ Galaxy Evolution Survey (ZFOURGE) aims to advance further the study of intermediate to high redshift galaxies by pushing to much fainter limits (25-26 AB), well beyond the typical limits of groundbased spectroscopy. This provides a unique opportunity to study the higher redshift and lower mass galaxy population in unprecedented detail, at cutting edge mass completeness limits. The power of this deep survey is demonstrated by \cite{Tomczak14}, who showed the stellar mass functions of star forming and quiescent galaxies can be accurately traced down to $10^9$ \msol\ at z=2, well below $M^*$. {\citet{Papovich15} showed that at this depth one can trace the evolution of progenitors of present-day $M^*$ galaxies (like M31 and the Milky Way Galaxy) out to $z\sim 3$.} Furthermore \cite{Straatman14} showed that a {population} of {massive} quiescent galaxies with $M>10^{10.6}$ was already in place at $z\sim4$, while \cite{Tilvi13} used the \fs\ medium-bandwidth filters to pinpoint Lyman Break galaxies at $z\sim7$ and distinguish them from cool dwarf stars.

In this paper we present the ZFOURGE data products\footnote{available for download at zfourge.tamu.edu}, comprising 45 nights of observations with the \fs\ near-infrared Camera on the 6.5m Magellan Baade Telescope at Las Campanas in Chile \citep{Persson13}. The survey was conducted over three extragalactic fields: CDFS ($\mathrm{RA\ (J2000)}=03$:$32$:$30,\ \mathrm{Dec (J2000)}=-27$:$48$:$30$) \citep{Giacconi02}, COSMOS  ($\mathrm{RA}=10$:$00$:$30,\ \mathrm{Dec}=+02$:$17$:$30$) \citep{Scoville07} and UDS ($\mathrm{RA}=02$:$17$:$00,\ \mathrm{Dec}=-05$:$13$:$00$) \citep{Lawrence07}, to reduce the effect of cosmic variance, and benefit from the large amount of public UV, optical and IR data already available. We present $K_s$-band selected near-IR catalogs, supplemented with public UV to IR data at $0.3-8\micron$, far-IR data from $Spitzer$/MIPS at $24\micron$ for all fields and Herschel/PACS at $100\micron$ and $160\micron$ for CDFS. 

In Sections \ref{sec:data} and \ref{sec:conv}, we discuss the survey and image processing and optimization. In Section \ref{sec:phot} we discuss source detection and photometry and include a description of the ZFOURGE data products. In Section \ref{sec:completeness} we test the completeness limits of the survey. We derive photometric redshifts and rest-frame colors in Section \ref{sec:eazy} and stellar masses, stellar ages and star formation rates in Section \ref{sec:fast}. In Section \ref{sec:uvj} we show how to effectively distinguish quiescent from star forming galaxies using a UVJ diagram, validating this classification with far-IR $Spitzer$/MIPS and Herschel/PACS data. A summary is provided in Section \ref{sec:sum}. Throughout, we assume a standard $\mathrm{\Lambda CDM}$ cosmology with $\mathrm{\Omega_M=0.3,\ \Omega_{\Lambda}=0.7}$ and $H_0=70\mathrm{km\ s^{-1} Mpc^{-1}}$. The adopted photometric system is AB \citep{Oke95}.

\section{Data}\label{sec:data}

\subsection{ZFOURGE}
The \fs\ Galaxy Evolution Survey (ZFOURGE, PI: I. Labb\'e) is a 45 night program with the \fs\ instrument  \citep{Persson13} on the 6.5~m Magellan Baade Telescope at Las Campanas, Chile. \fs\ has 5 near-IR medium bands: $J_1,J_2,J_3,H_s$ and $H_l$, covering the same range as the more classical J and H broadband filters, and a $K_s$-band. The central wavelengths of these filters range from $1.05~\micron$ ($J_1$) to $2.16~\micron$ ($K_s$). 

\begin{figure*}
\includegraphics[width=\textwidth]{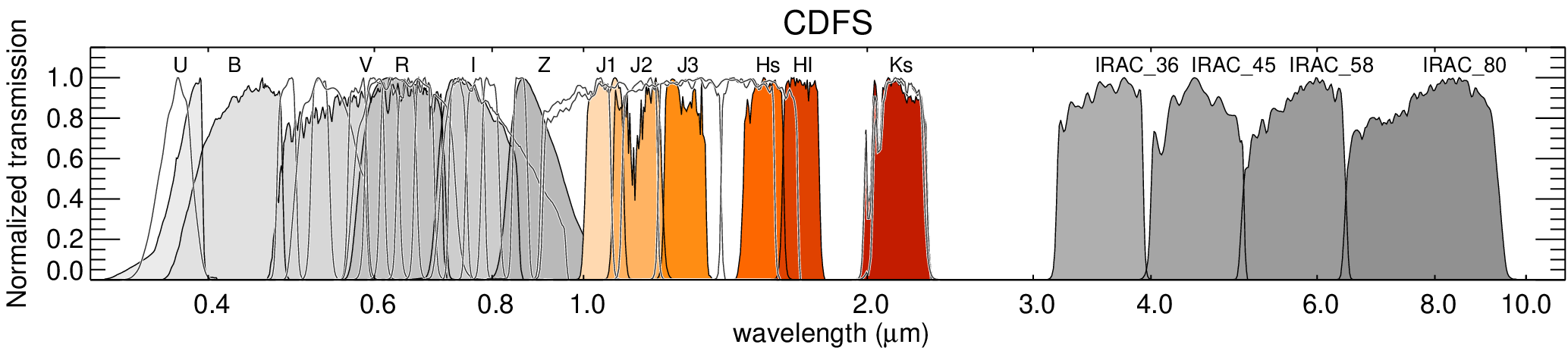}
\includegraphics[width=\textwidth]{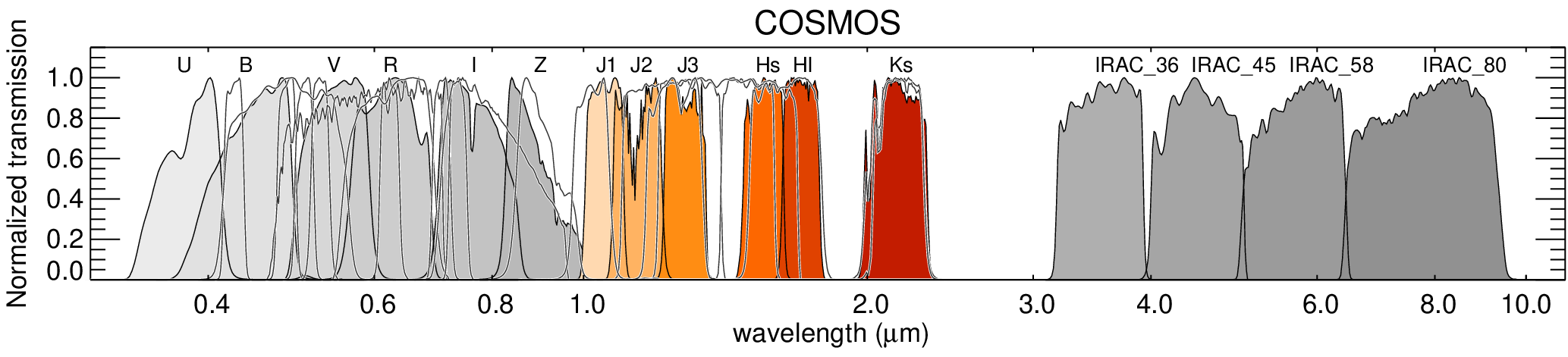}
\includegraphics[width=\textwidth]{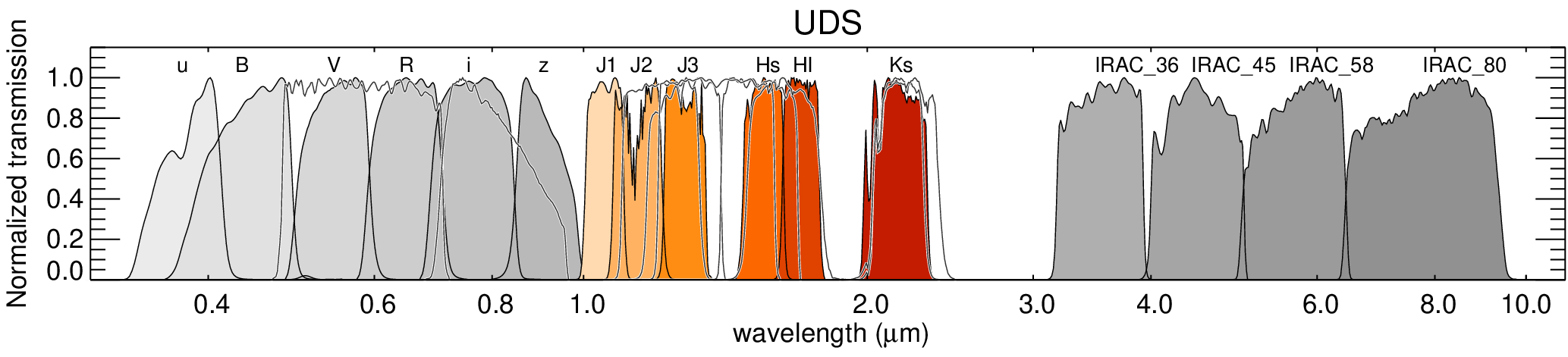}
\caption{Normalized transmission corresponding to the \fs\ medium-bandwidth and ancillary filters, each panel representing a different field. From top to bottom: CDFS, COSMOS and UDS. We show the \fs\ $J_1,J_2,J_3,H_s,H_l$ and $K_s$ medium-bandwidth filters with different shades of red. The UV to optical $U,B,V,R,I$ and $Z$ filters and the $Spitzer$/IRAC filters are {shown with} gray shaded curves. These correspond to different instruments in each field. The \fs\ filters overlap with other broadband near-IR filters, e.g., $HST$/WFC3/F125W$-$F160W, while providing a higher resolution sampling. Atmospheric transmission was included in all \fs\ filter curves. All filters are mentioned separately in Tables \ref{tab:anci0} (CDFS), \ref{tab:anci1} (COSMOS) and \ref{tab:anci2} (UDS).}
\label{fig:filters}
\end{figure*}

\begin{table}
\caption{\fs\ observations}
\begin{threeparttable}
\begin{tabular}{l c c c}
\hline
\hline
Cosmic field & Filter & Total integration time & $5\sigma$ depth \\
& & (hrs) & (AB mag) \\
\hline
CDFS & $J_1$ & 6.3 & 25.6 \\ 
CDFS & $J_2$ & 6.5 & 25.5 \\
CDFS & $J_3$ & 8.8 & 25.5 \\
CDFS & $H_s$ & 12.2 & 24.9 \\
CDFS & $H_l$ & 5.9 & 25.0 \\
CDFS & $K_s$ & 5.0 & 24.8 \\
\hline
COSMOS & $J_1$ & 13.9 & 26.0 \\
COSMOS & $J_2$ & 16.0 & 26.0 \\
COSMOS & $J_3$ & 13.8 & 25.7 \\
COSMOS & $H_s$ & 12.1 & 25.1 \\
COSMOS & $H_l$ & 12.1 & 24.9 \\
COSMOS & $K_s$ & 13.4 & 25.3 \\
\hline
UDS & $J_1$ & 7.9 & 25.6 \\		
UDS & $J_2$ & 8.7 & 25.9 \\
UDS & $J_3$ & 9.3 & 25.6 \\
UDS & $H_s$ & 11.0 & 25.1 \\
UDS & $H_l$ & 10.4 & 25.2 \\
UDS & $K_s$ & 3.9 & 24.7 \\
\hline
\end{tabular}
\end{threeparttable} 
\label{tab:survey}
\end{table}

The filter curves are shown in Figure \ref{fig:filters}; we have also added the filter curves of the ancillary dataset (see Section \ref{sec:ancil}), showing that we cover the full UV to near-IR wavelength range. The \fs\ filters overlap with broadband filters such as $HST$/WFC3/F125W, F140W and F160W in wavelength space, except they are narrower and sample the near-IR in more detail. The effective filter curves we use are modified to include the \cite{Lord92} atmospheric transmission functions with a water column of 2.3mm. The total integration time in each filter is shown in Table \ref{tab:survey}. 

\begin{figure}
\begin{center}
\includegraphics[width=0.49\textwidth]{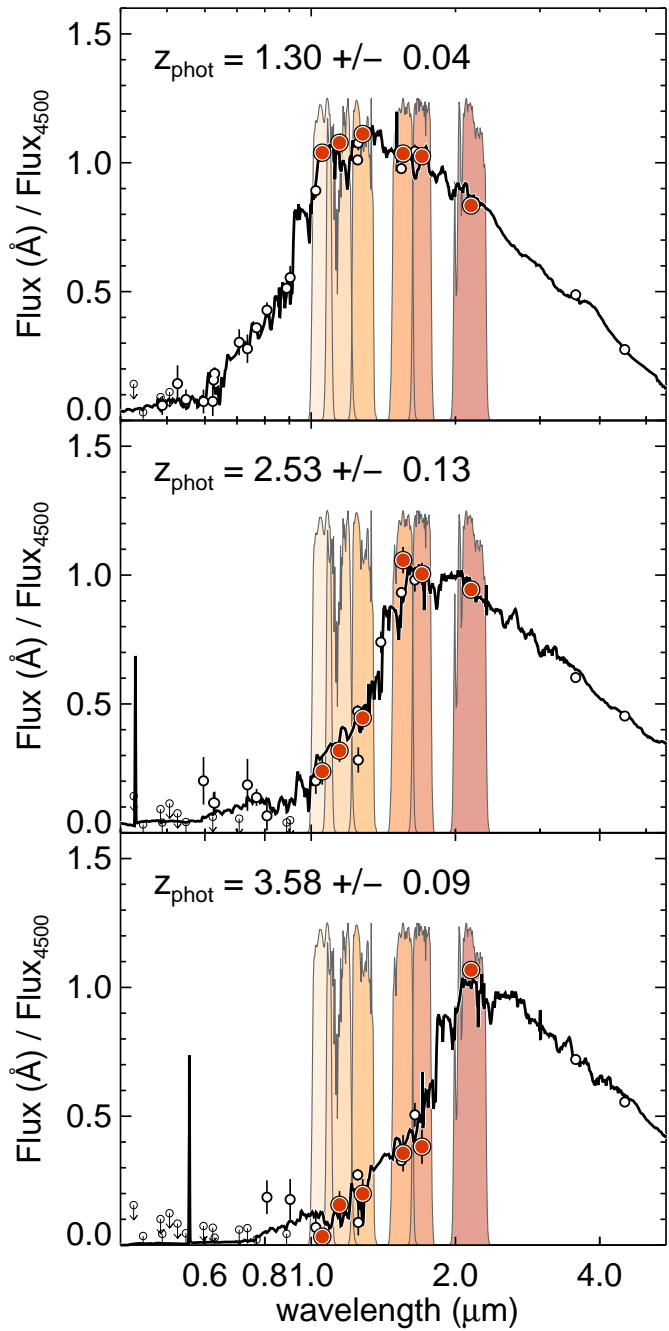}
\end{center}
\caption{The \fs\ filters provide detailed sampling of the \b4 of galaxies at $z\gtrsim1.5$. Here we show the SEDs of three observed galaxies in COSMOS with large \b4s, at $z=1.30$, $z=2.53$ and $z=3.58$. With increasing redshift, the \b4\ moves through the range defined by the \fs\ bands. Observed datapoints are shown as white or red dots with errorbars for ancillary and \fs\ filters, respectively. Upper limits (mostly in the UV) are indicated with downward arrows. The solid curves are the EAZY best-fit SEDs (see Section \ref{sec:eazy}). Observed and fitted SEDs are normalized at rest-frame $4500\mathrm{\AA}$.}
\label{fig:sf}
\end{figure}

The sampling of the \fs\ medium-bandwidth filters is illustrated in Figure \ref{fig:sf}, where we show the SEDs of observed galaxies in COSMOS with large \b4s at $z\gtrsim1.5$. The \fs\ near-IR {photometry is} highlighted in red. {The medium-band filters are} shown in the background. They are particularly well suited to trace the \b4\ at higher redshifts, which is crucial to derive photometric redshifts.

\subsection{FourStar Image reduction}
\subsubsection{Pipeline}

The \fs{} data were reduced using a custom {\tt IDL} pipeline written by one of the authors (I. Labb\'e) and also used in the NMBS \citep{Whitaker11}. It employs a two-pass sky subtraction scheme based on the {\tt IRAF} package {\tt xdimsum}. 

The pipeline processes the {data,} which consist of dithered frames {for each of the} 4 \fs{} detectors, separately for each $\sim1-1.5$ hour observing block. Observed frames taken with each of the detectors {were} reduced and subsequently combined into a single mosaic. 

Linearity corrections from the \fs{} website\footnote{\tiny \url{http://instrumentation.obs.carnegiescience.edu/FourStar/calibration.html}} {were} applied to the raw data. Dark current was determined to be variable so we did not remove any dark pattern. We also found constant bias levels along columns and rows in the raw data. We therefore subtracted the median of a column/row from itself. 

Master flat field data were produced using twilight observations. For the \ks{}-band, where thermal contributions play a role, we attempted to mitigate the impact of illumination from the warm telescope. By combining multiple dithered observations of a blank field at the end of a night when the telescope had cooled, we were able to characterize the telescope illumination pattern. Shortly afterwards we took twilight flats and subtracted the telescope illumination pattern from each exposure. The flats with the telescope contribution removed were normalized and combined into the master \ks{}-band flats.

Sky background models were {then} subtracted from individual science exposures. The sky background was computed by averaging up to 8 images taken before and after that exposure. Masking routines were run to remove: (1) bad pixels via a static mask from the \fs{} website (2) satellite trails (3) guider cameras entering the field of view and (4) persistence from saturated objects in previous exposures. Bad pixels make up between 0.3 and 1.7 \% of the detectors \citep{Persson13}. In addition, the individual exposures were visually screened for any remaining tracking issues, asteroids, airplanes and satellites.

Corrections for geometric distortion and absolute astrometric solutions {were} computed by crossmatching sources using astrometric reference images. In COSMOS we used the CFHT/$i$-band as reference \citep{Erben09,Hildebrandt09}, in CDFS we used ESO/MPG/WFI/I from the ESI survey \citep{Erben09,Hildebrandt06} and in UDS the UKIDDS data release 8 $K_s$-band image (Almaini, in prep). The observations {were} interpolated onto a pixel grid with a resolution of $0\farcs15\ \mathrm{pix}^{-1}$, which is close to the native scale of \fs\ of $0\farcs159\ \mathrm{pix}^{-1}$. The new grid shares the WCS tangent point ({\tt CRVAL}) with the CANDELS $HST$ images \citep{Koekemoer11,Grogin11} and places {\tt CRVAL} at a half-integer pixel position ({\tt CRPIX}). 

\begin{figure}
\includegraphics[width=0.49\textwidth]{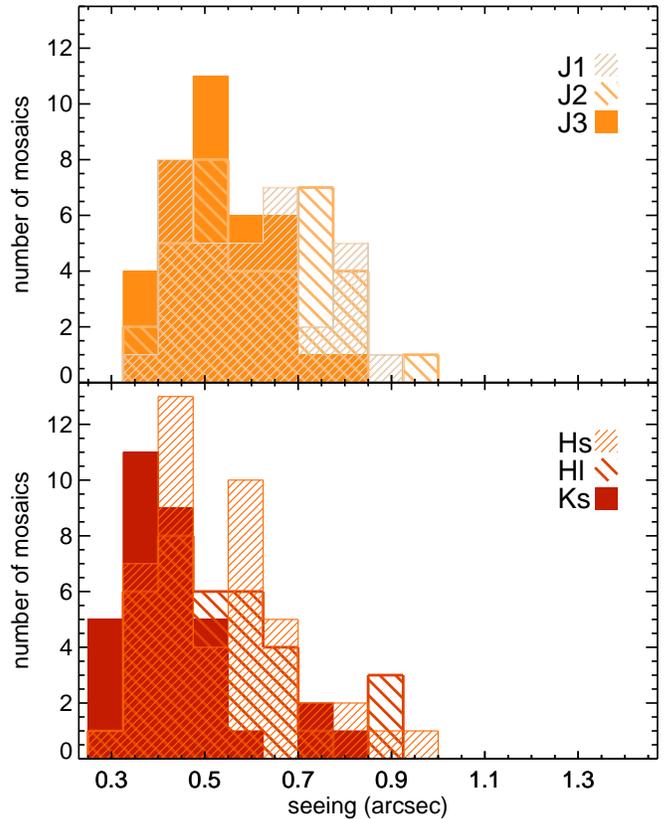}
\caption{Seeing histograms of the \fs\ single images, corresponding to $\sim1-1.5$ hour observing blocks. Many of the images have a seeing of $\sim0\farcs4-0\farcs5$.}
\label{fig:seeing}
\end{figure}

To optimize the signal-to-noise (SNR) of the images for each observing block (and for the final mosaics), they {were} weighted by their seeing, sky background levels and ellipticity of the PSF before they are combined. The seeing conditions at Las Campanas were extraordinarily good, with a median seeing FWHM {for the entire set of observations} of $0\farcs5$ as shown in the histogram in Figure \ref{fig:seeing}. Since the \ks{}-band cannot be observed with the $HST$, we {paid special attention to this filter and} only observed the $K_s$-band when the seeing was excellent. This resulted in a {a very low} median seeing {for the} \fs/$K_s-${filter} of $0\farcs4$.

Finally, we subtracted a background in the final mosaics using Source Extractor \citep[SE;][]{Bertin96} to ensure any remaining structure in the background did not impact the aperture photometry. In short, SE iteratively estimated the median of the distribution of pixel values in areas of $48\times48$ pixels in CDFS and COSMOS and $96\times96$ pixels in UDS. {The dimensions of these areas were chosen to avoid overestimating the background near bright sources.} These estimates {were} smoothed on a scale of $3\times$ the background area, after which the background for the full images {was} calculated using a bicubic spline interpolation.

\subsubsection{Photometric calibration}
Here we describe how we derived the near-IR photometric zeropoints of the final mosaics. Since these vary significantly with changes in local precipitable water vapor and airmass, we employed a {differential} photometric calibration scheme, {using secondary standard stars}. First, we selected a nearby standard star. We selected relatively faint ($K_s=14.5-17$ mag) spectrophotometric standard stars from the CALSPEC Calibration Database\footnote{ \url{http://www.stsci.edu/hst/observatory/crds/calspec.html } }. We then observed this primary standard star under photometric conditions immediately before or after a science observation in a particular filter. The science dataset was reduced and photometrically calibrated using the primary standard star observations and using an atmospheric watercolumn of 2.3mm. 
Secondly, we then selected bright, unsaturated stars in each of the chips of the science field for use as secondary standard stars. All other science observations of an observing block were then calibrated to the primary standard star via the secondary standard stars within each of the science fields.

In Section \ref{sec:eazy} we derive additional corrections to the zeropoints, that are typically of the order of 0.05 magnitude. We added these to the photometric zeropoints calculated here.

\subsubsection{Image depths}\label{sec:depths}
We measured the depths of the \fs\ images by determining the {root-mean-square (RMS)} of the background pixels. Since pixels may be correlated on small scales, e.g., due to confusion or systematics introduced during the reduction process, we used a method in which we randomly placed 5000 apertures of $0\farcs6$ diameter in each background subtracted image. Due to the dither pattern the images have less coverage from individual frames at the edges. We therefore considered only regions with coverage within 80\% of the maximum exposure. Sources were also masked, based on the SE segmentation maps after object detection (see Section \ref{sec:sd}).

The resulting aperture flux distributions, representing the variation in the noise, were fit with a Gaussian, from which we derived the standard deviation ($\sigma$). We then applied the point-spread-functions (PSFs) derived from bright stars (further explained in Section \ref{sec:conv}), to determine a flux correction for missing light outside of the aperture. $\sigma$ was then multiplied by 5 and converted to magnitude using the effective zeropoint (the photometrically derived zeropoints as desribed above, with a correction applied) of each \fs\ mosaic, to obtain {an estimate of} the $5\sigma$ limiting depth. The resulting depth in AB magnitude can thus be summarized as 
\begin{equation}
\mathrm{depth (5\sigma)} = zp- 2.5\mathrm{log_{10}}[5\sigma apcorr]
\label{eq:depth}
\end{equation}
with $zp$ the zeropoint of the image and $apcorr$ the aperture flux correction (typically factors of $1.7-2.6$, depending on the seeing). The $5\sigma$ depths are summarized in Table \ref{tab:survey} and have typical values of $25.5-26.0$ AB mag in $J_1,J_2,J_3$ and $24.9-25.2$ AB mag in $H_s,H_l$ and $24.7-25.3$ AB mag $K_s$.

\begin{figure*}
\includegraphics[width=\textwidth]{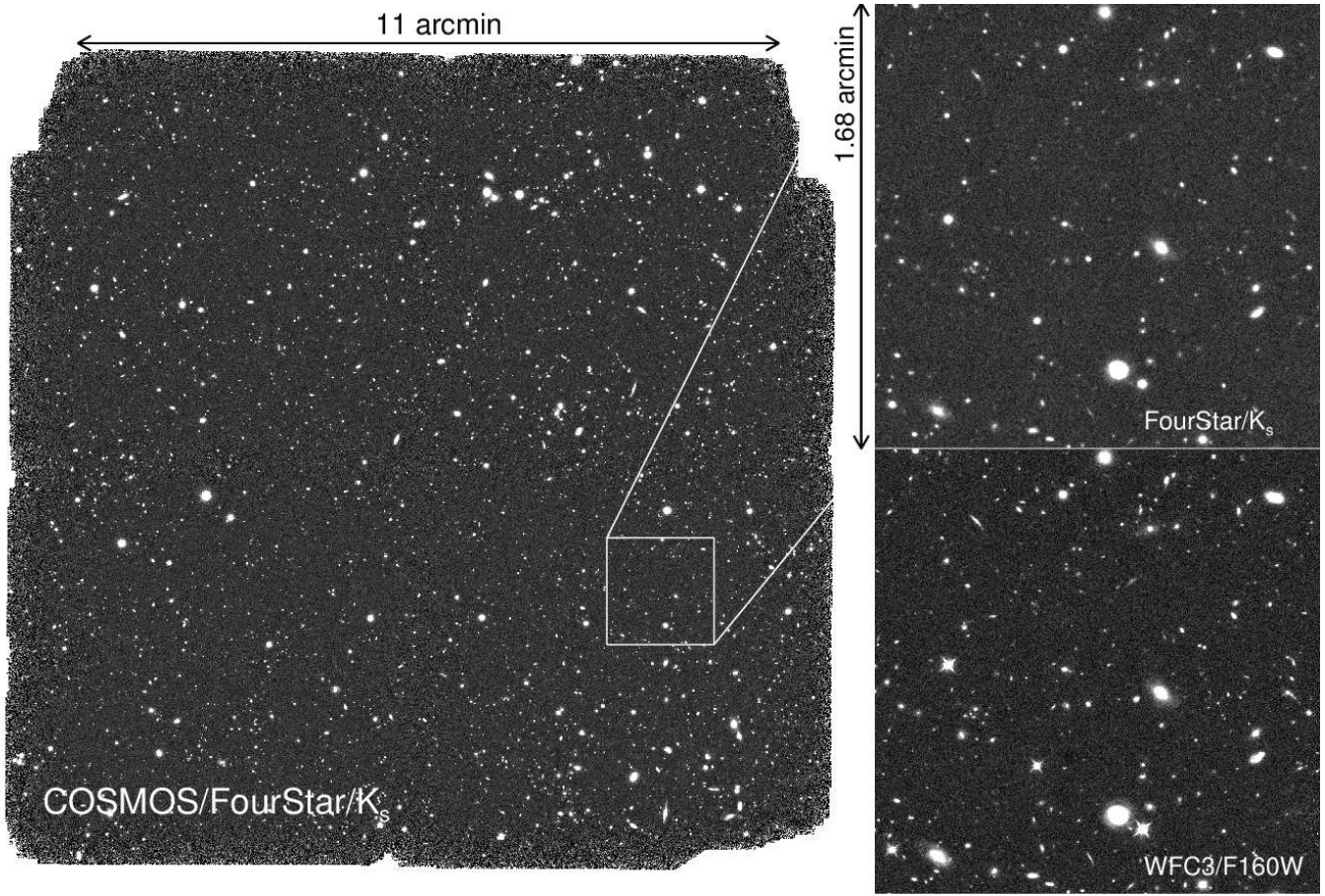}
\caption{Left: The \fs/$K_s$-band reduced image in COSMOS. The FourStar footprint is $13\arcmin\times13\arcmin$. Top right: zooming in on a $1.68\arcmin\times1.68\arcmin$ region in the COSMOS field. Bottom right: the same region with $HST$/WFC3/F160W.}
\label{fig:cosmos}
\end{figure*}

\begin{figure*}
\includegraphics[width=\textwidth]{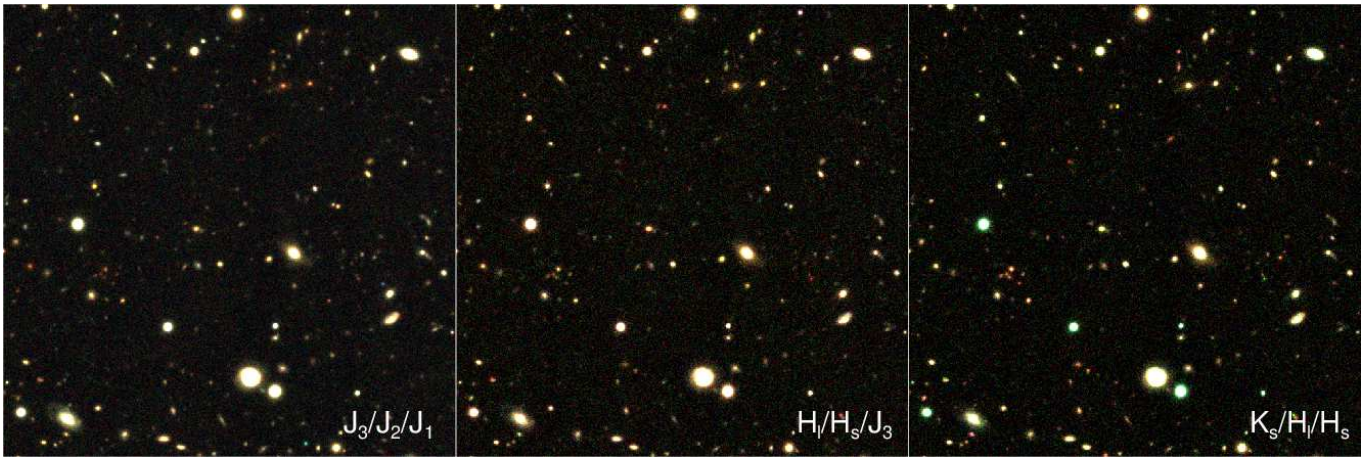}
\caption{False color images of the same cutout region as shown in Figure \ref{fig:cosmos}, demonstrating the high quality obtained with the \fs\ filters, as well as the {usefulness} of using medium-bandwidth filters to {characterize} the colors of galaxies within a classical J or H broadband. The filter combinations that were used in each panel are indicated at the bottom (red/green/blue).}
\label{fig:imcc}
\end{figure*}

In Figure \ref{fig:cosmos} we show as an example the \fs/$K_s$-band image in COSMOS. We also compare with the near-IR CANDELS/$HST$/WFC3/F160W observations, with FWHM=$0\farcs19$ and a limiting $5\sigma$ depth of 26.4 AB mag. The deeper space-based F160W image has a higher resolution, but as a result of the very deep magnitude limits combined with excellent seeing conditions we can achieve almost a similar quality for our near-IR ground-based observations. The $K_s$-band images in CDFS and UDS have similar depth. To highlight the wealth of information provided by the fine spectral sampling of the \fs\ medium-bandwidth filters we show again in Figure \ref{fig:imcc} the same cut-out region of Figure \ref{fig:cosmos}, using different filter combinations.

\subsection{$K_s$-band detection images}\label{sec:kstacks}
We combine our \fs/$K_s$-band observations with deep pre-existing K-band imaging to create super-deep detection images. In CDFS we use VLT/HAWK-I/$K$ from HUGS (with natural seeing between $0\farcs3$ and $0\farcs5$) \citep{Fontana14}, VLT/ISAAC/$K$ (v2.0) from GOODS, including ultra deep data in the HUDF region  (seeing$=0.5\ \arcsec$ \citep{Retzlaff10}, CFHST/WIRCAM/K from TENIS (seeing$=0\farcs9$) \citep{Hsieh12}, and Magellan/PANIC/K in HUDF (seeing$=0\farcs4$) (PI: I. Labb\'e). In COSMOS we add VISTA/K from UltraVISTA (DR2) (seeing$=0\farcs7$) \citep{McCracken12} and in UDS we add imaging with UKIRT/WFCAM/K from UKIDSS (DR10) (seeing$=0\farcs7$) (Almaini et al, in prep) and also natural seeing VLT/HAWK-I/$K$ imaging from HUGS.

\begin{figure*}
\begin{center}
\includegraphics[width=0.51\textwidth]{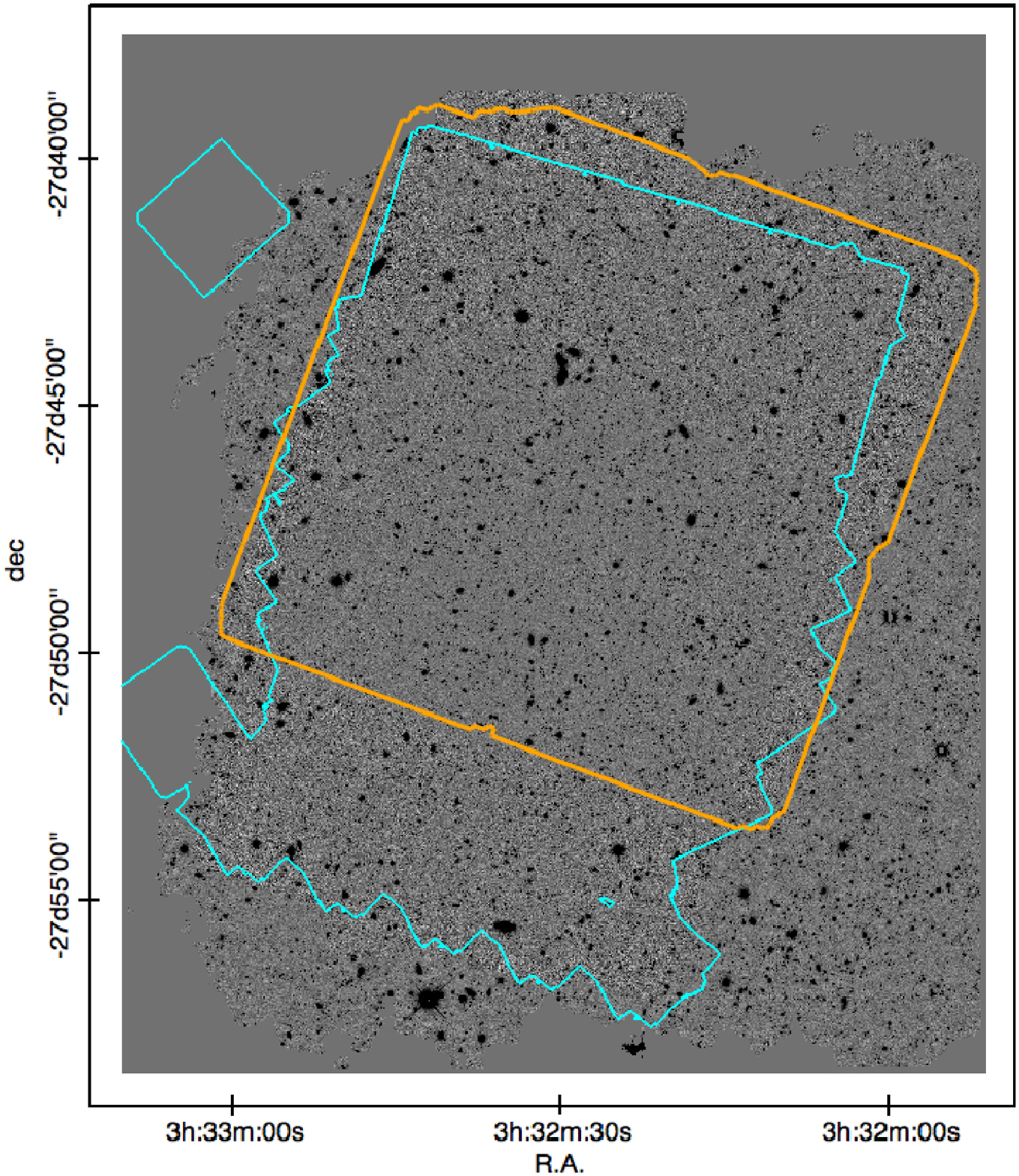}
\caption{{Deep $K_s$-band detection image in CDFS. The orange outline shows the ZFOURGE footprint. With cyan outlines we show the $HST/WFC3/F160W$ footprint from CANDELS. North is up and East is to left.}}
\label{fig:ftp0}
\end{center}
\end{figure*}

\begin{figure*}
\begin{center}
\Rotatebox{90}{\includegraphics[width=0.71\textwidth]{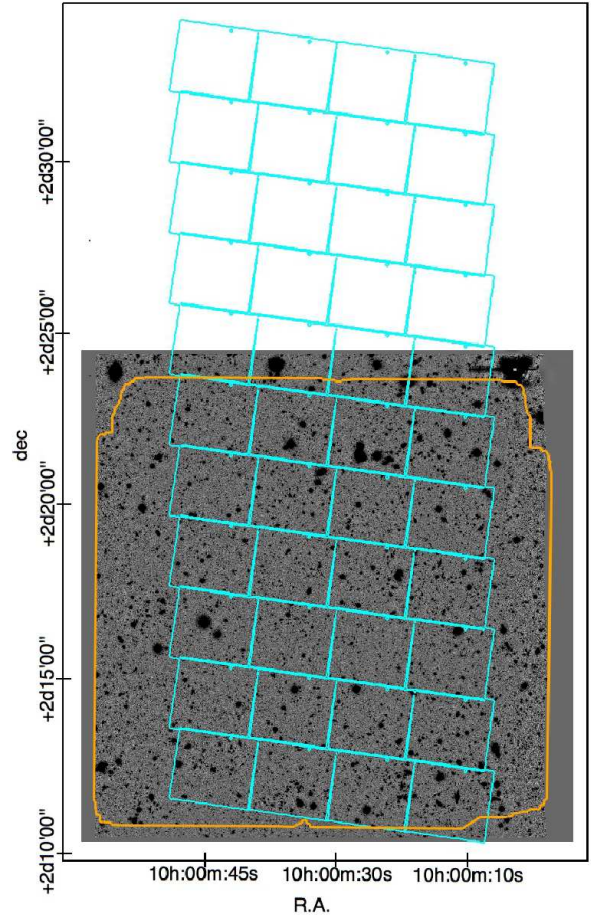}}	
\caption{{Deep $K_s$-band detection image in COSMOS and {outlines} as in Figure \ref{fig:ftp0}.}}
\label{fig:ftp1}
\end{center}
\end{figure*}

\begin{figure*}
\begin{center}
\includegraphics[width=0.74\textwidth]{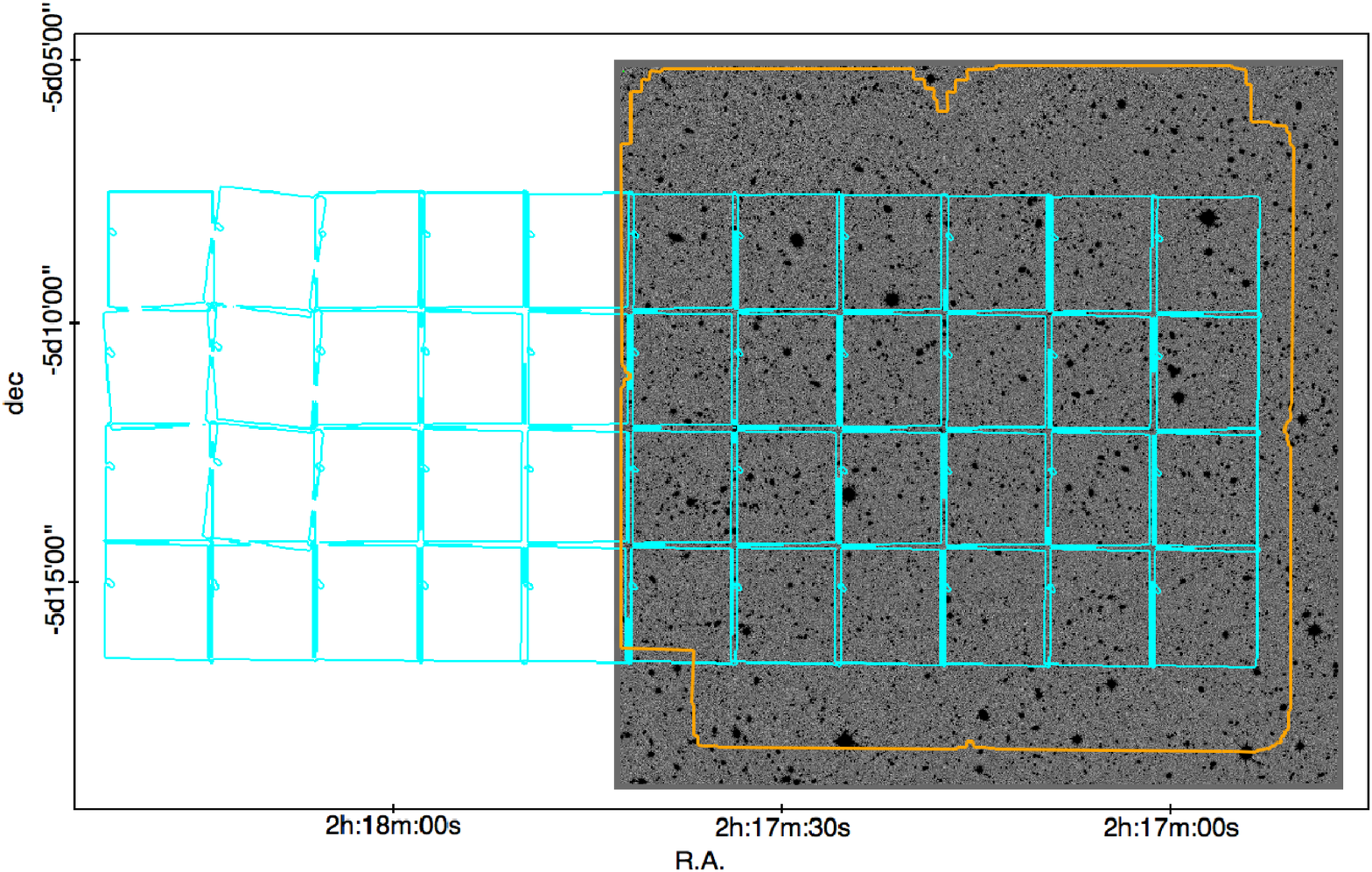}
\caption{{Deep $K_s$-band detection image in UDS and {outlines} as in Figure \ref{fig:ftp0}.}}
\label{fig:ftp2}
\end{center}
\end{figure*}

Using sources common to the images a distortion map was determined. Subsequent bicubic spline interpolation was used to register the images to better than $0\farcs03$ across the field.  We then determined the background RMS flux variation ($\sigma_{RMS}$) and the seeing in each image, and we used these to assign a weight using $weight=1/(\sigma_{RMS}\times seeing^2)$. {Note that the images were not PSF-matched prior to combining.} The final combined image stacks were obtained by a weighted average of the individual K- and $K_s$-band science images. Weight maps were obtained by averaging the individual exposure maps in the same way as the science images. The final $K_s$-band stacks have maximum limiting depths at $5\sigma$ significance of 25.5 and 25.7 AB mag in COSMOS and UDS, respectively, which are 0.2 and 1.0 magnitudes deeper than the individual \fs/$K_s$-band observations. The depth in CDFS varies between 26.2 and 26.5, 1.4 to 1.7 magnitudes deeper than the \fs/$K_s$-band image only. The average seeing in the three fields is $FWHM=0\farcs45,\ 0\farcs58$ and $0\farcs60$. We use these images for source detection (Section \ref{sec:phot}), after calculating and subtracting the background. {They are shown in Figures \ref{fig:ftp0} to \ref{fig:ftp2}, with the ZFOURGE footprint indicated as well as the $HST/WFC3/F160W$ footprint from CANDELS.

\subsection{Ancillary data{: UV, optical, NIR, and IR imaging}}\label{sec:ancil}

\begin{table}
\caption{CDFS passband parameters}
\label{tab:anci0}
\begin{tabular}{r c c c c c}
\hline
\hline
Filter & $\lambda_c$ & FWHM & zeropoint & offset & galactic\\
& ($\micron$) & ($\arcsec$)& (AB mag) & & extinction \\
\hline
$B$ &   0.4318 &  0.73 & 22.097 & -0.029 & -0.032 \\
$I$ &   0.7693 &  0.73 & 22.151 &  0.019 & -0.014 \\
$R$ &   0.6443 &  0.65 & 27.321 & -0.148 & -0.020 \\
$U$ &   0.3749 &  0.81 & 25.932 & -0.181 & -0.037 \\
$V$ &   0.5919 &  0.73 & 22.968 & -0.010 & -0.022 \\
$Z$ &   0.9036 &  0.73 & 21.378 &  0.041 & -0.011 \\
$H_s$ &   1.5544 &  0.60 & 26.618 & -0.031 & -0.004 \\
$H_l$ &   1.7020 &  0.50 & 26.588 & -0.051 & -0.004 \\
$J_1$ &   1.0540 &  0.59 & 26.270 & -0.041 & -0.009 \\
$J_2$ &   1.1448 &  0.62 & 26.558 & -0.043 & -0.006 \\
$J_3$ &   1.2802 &  0.56 & 26.521 & -0.067 & -0.006 \\
$K_s$ &   2.1538 &  0.46 & 26.851 & -0.083 & -0.003 \\
$NB118$ &   1.1909 &  0.47 & 24.668 &  0.000 & -0.006 \\
$NB209$ &   2.0990 &  0.45 & 24.786 &  0.000 & -0.003 \\
$F098M$ &   0.9867 &  0.26 & 25.670 &  0.011 & -0.008 \\
$F105W$ &   1.0545 &  0.24 & 26.259 & -0.002 & -0.007 \\
$F125W$ &   1.2471 &  0.26 & 26.229 &  0.004 & -0.005 \\
$F140W$ &   1.3924 &  0.27 & 26.421 & -0.027 & -0.004 \\
$F160W$ &   1.5396 &  0.27 & 25.942 & -0.000 & -0.004 \\
$F814W$ &   0.8057 &  0.22 & 25.931 & -0.004 & -0.011 \\
$IA484$ &   0.4847 &  0.81 & 25.463 & -0.013 & -0.024 \\
$IA527$ &   0.5259 &  0.87 & 25.639 & -0.059 & -0.022 \\
$IA574$ &   0.5763 &  1.01 & 25.543 & -0.148 & -0.019 \\
$IA598$ &   0.6007 &  0.69 & 25.962 & -0.040 & -0.018 \\
$IA624$ &   0.6231 &  0.67 & 25.887 &  0.014 & -0.017 \\
$IA651$ &   0.6498 &  0.67 & 26.072 & -0.062 & -0.016 \\
$IA679$ &   0.6782 &  0.86 & 26.105 & -0.080 & -0.015 \\
$IA738$ &   0.7359 &  0.83 & 26.003 & -0.003 & -0.013 \\
$IA767$ &   0.7680 &  0.77 & 26.000 & -0.028 & -0.012 \\
$IA797$ &   0.7966 &  0.74 & 25.986 & -0.022 & -0.012 \\
$IA856$ &   0.8565 &  0.74 & 25.713 & -0.007 & -0.010 \\
$WFI\_V$ &   0.5376 &  0.96 & 23.999 & -0.076 & -0.021 \\
$WFI\_Rc$ &   0.6494 &  0.84 & 24.597 & -0.038 & -0.016 \\
$WFI\_U38$ &   0.3686 &  0.98 & 21.587 & -0.291 & -0.032 \\
$tenisK$ &   2.1574 &  0.86 & 24.130 &  0.233 & -0.002 \\
$KsHI$ & 2.1748 & 0.45 & 31.419 & 0.022 & -0.003 \\
$IRAC\_36$ &   3.5569 &  1.50 & 20.054 & -0.016 &  0.000 \\
$IRAC\_45$ &   4.5020 &  1.50 & 20.075 &  0.005 &  0.000 \\
$IRAC\_58$ &   5.7450 &  1.90 & 20.626 &  0.023 &  0.000 \\
$IRAC\_80$ &   7.9158 &  2.00 & 21.803 &  0.022 &  0.000 \\
\hline
\end{tabular} 
\begin{tablenotes}
\item Zeropoints are the effective zeropoints. These have galactic extinction and zeropoint corrections derived in Section \ref{sec:eazy} incorporated, i.e., $zp=zp_I+offset+GE$, with $zp_I$ representing the photometrically derived zeropoint of image $I$, $offset$ the zeropoint correction and $GE$ the galactic extinction value. The corrections (in units of AB magnitude) are indicated in separate columns.
\end{tablenotes}
\end{table}

\begin{table}
\caption{COSMOS passband parameters}
\label{tab:anci1}
\begin{tabular}{r c c c c c}
\hline
\hline
Filter & $\lambda_c$ & FWHM & zeropoint  & offset & galactic\\
& ($\micron$) & ($\arcsec$)& (AB mag) & & extinction \\
\hline
$B$ &   0.4448 &  0.61 & 31.129 & -0.195 & -0.076 \\
$G$ &   0.4870 &  0.90 & 26.290 & -0.015 & -0.069 \\
$I$ &   0.7676 &  0.77 & 25.759 &  0.091 & -0.034 \\
$IA427$ &   0.4260 &  0.79 & 31.119 & -0.202 & -0.079 \\
$IA484$ &   0.4847 &  0.75 & 31.214 & -0.116 & -0.069 \\
$IA505$ &   0.5061 &  0.82 & 31.252 & -0.083 & -0.065 \\
$IA527$ &   0.5259 &  0.74 & 31.281 & -0.058 & -0.061 \\
$IA624$ &   0.6231 &  0.72 & 31.348 & -0.002 & -0.050 \\
$IA709$ &   0.7074 &  0.81 & 31.343 & -0.015 & -0.042 \\
$IA738$ &   0.7359 &  0.80 & 31.347 & -0.014 & -0.039 \\
$R$ &   0.6245 &  0.79 & 25.903 &  0.023 & -0.047 \\
$U$ &   0.3828 &  0.82 & 24.913 & -0.235 & -0.086 \\
$V$ &   0.5470 &  0.80 & 31.418 &  0.077 & -0.059 \\
$Rp$ &   0.6276 &  0.83 & 31.453 &  0.100 & -0.047 \\
$Z$ &   0.8872 &  0.74 & 24.859 &  0.121 & -0.030 \\
$Zp$ &   0.9028 &  0.90 & 31.557 &  0.187 & -0.030 \\
$H_l$ &   1.7020 &  0.60 & 26.624 &  0.033 & -0.010 \\
$H_s$ &   1.5544 &  0.54 & 26.673 &  0.062 & -0.012 \\
$J_1$ &   1.0540 &  0.57 & 26.358 &  0.026 & -0.020 \\
$J_2$ &   1.1448 &  0.55 & 26.590 &  0.038 & -0.018 \\
$J_3$ &   1.2802 &  0.53 & 26.573 &  0.011 & -0.016 \\
$K_s$ &   2.1538 &  0.47 & 26.918 & -0.011 & -0.006 \\
$NB118$ &   1.1909 &  0.58 & 24.637 &  0.000 & -0.018 \\
$NB209$ &   2.0990 &  0.52 & 24.849 &  0.000 & -0.006 \\
$F125W$ &   1.2471 &  0.26 & 26.236 & -0.000 & -0.011 \\
$F140W$ &   1.3924 &  0.26 & 26.455 & -0.000 & -0.010 \\
$F160W$ &   1.5396 &  0.26 & 25.948 & -0.000 & -0.008 \\
$F606W$ &   0.5921 &  0.20 & 26.437 & -0.016 & -0.038 \\
$F814W$ &   0.8057 &  0.21 & 25.951 &  0.032 & -0.024 \\
$UVISTA\_J$ &   1.2527 &  0.82 & 30.052 &  0.062 & -0.011 \\
$UVISTA\_H$ &   1.6433 &  0.81 & 29.995 &  0.003 & -0.008 \\
$UVISTA\_Ks$ &   2.1503 &  0.79 & 30.028 &  0.035 & -0.006 \\
$UVISTA\_Y$ &   1.0217 &  0.85 & 30.045 &  0.061 & -0.016 \\
$IRAC\_36$ &   3.5569 &  1.70 & 21.530 & -0.051 &  0.000 \\
$IRAC\_45$ &   4.5020 &  1.70 & 21.537 & -0.044 &  0.000 \\
$IRAC\_58$ &   5.7450 &  1.90 & 21.577 & -0.004 &  0.000 \\
$IRAC\_80$ &   7.9158 &  2.00 & 21.520 & -0.061 &  0.000 \\
\hline
\end{tabular}    
\end{table}

\begin{table}
\caption{UDS passband parameters}
\label{tab:anci2}
\begin{tabular}{r c c c c c}
\hline
\hline
Filter & $\lambda_c$ & FWHM & zeropoint  & offset & galactic\\
& ($\micron$) & ($\arcsec$)& (AB mag) & & extinction \\
\hline
$u$ &   0.3828 &  1.06 & 24.905 & -0.268 & -0.089 \\
$B$ &   0.4408 &  0.91 & 24.803 & -0.123 & -0.074 \\
$V$ &   0.5470 &  0.93 & 24.870 & -0.072 & -0.058 \\
$R$ &   0.6508 &  0.96 & 24.914 & -0.038 & -0.049 \\
$i$ &   0.7656 &  0.98 & 24.986 &  0.021 & -0.035 \\
$z$ &   0.9060 &  0.99 & 24.974 &  0.001 & -0.027 \\
$J_1$ &   1.0540 &  0.55 & 26.121 & -0.036 & -0.022 \\
$J_2$ &   1.1448 &  0.53 & 26.408 & -0.029 & -0.019 \\
$J_3$ &   1.2802 &  0.51 & 26.481 & -0.019 & -0.015 \\
$H_s$ &   1.5544 &  0.49 & 26.591 & -0.000 & -0.011 \\
$H_l$ &   1.7020 &  0.51 & 26.448 & -0.036 & -0.010 \\
$K_s$ &   2.1538 &  0.44 & 26.804 & -0.067 & -0.006 \\
$J$ &   1.2502 &  0.91 & 30.863 & -0.052 & -0.015 \\
$H$ &   1.6360 &  0.89 & 31.262 & -0.108 & -0.010 \\
$K$ &   2.2060 &  0.86 & 31.825 & -0.059 & -0.006 \\
$F125W$ &   1.2471 &  0.26 & 26.214 & -0.000 & -0.016 \\
$F140W$ &   1.3924 &  0.26 & 26.439 & -0.000 & -0.013 \\
$F160W$ &   1.5396 &  0.26 & 25.935 & -0.000 & -0.011 \\
$F606W$ &   0.5893 &  0.20 & 26.383 & -0.054 & -0.054 \\
$F814W$ &   0.8057 &  0.23 & 25.926 &  0.015 & -0.033 \\
$Y$ &   1.0207 &  0.58 & 27.004 &  0.026 & -0.022 \\
$KsHI$ & 2.1748 & 0.46 & 27.520 & 0.026 & -0.006 \\
$IRAC\_36$ &   3.5569 &  1.70 & 21.539 & -0.042 &  0.000 \\
$IRAC\_45$ &   4.5020 &  1.70 & 21.556 & -0.025 &  0.000 \\
$IRAC\_58$ &   5.7450 &  1.90 & 21.458 & -0.123 &  0.000 \\
$IRAC\_80$ &   7.9158 &  2.00 & 21.522 & -0.059 &  0.000 \\
\hline
\end{tabular}
\end{table}

In addition to the 6 \fs\ filters, we incorporate imaging in 20-34 filters into each catalog, from publicly available surveys at $0.3-8\micron$. In CDFS we have a total of 40 bands, in COSMOS a total of 37 and in UDS a total of 26. These are summarized in Tables \ref{tab:anci0}, \ref{tab:anci1} and \ref{tab:anci2}, where we additionally show, for every image, the central wavelength, PSF FWHM (see Section \ref{sec:conv}), effective zeropoint, galactic extinction value and zeropoint offset derived in Section \ref{sec:eazy}. The galactic extinction values were calculated using the $E(B-V)$ values from \citet{Schlafly11}, interpolated between the given bandpasses and the central wavelengths of our filterset.

The CDFS UV-to-optical filters include VLT/VIMOS/$U,R$-imaging \citep{Nonino09}, $HST$/ACS/$B,V,I,Z$-imaging \citep{Giavalisco04,Wuyts08}, ESO/MPG/WFI/$U_{38},V,R_c$-imaging \citep{Erben05,Hildebrandt06}, $HST$/WFC3/$F098M,F105W$,$F125W,F140W,F160W$ and $HST$/ACS$F606W,F814W$-imaging \citep{Grogin11,Koekemoer11,Windhorst11,Brammer12}, 11 Subaru/{Suprime-Cam} optical medium bands \citep{Cardamone10} with seeing $<1\farcs1$ {(from a set of 18, including seeing $>1.1\arcsec$ images)} and CFHT/WIRCAM/$K$-band imaging \citep{Hsieh12}.

In COSMOS we added CFHT/$u,g,r,i,z$-imaging \citep{Erben09,Hildebrandt09}, Subaru/{Suprime-Cam}/$B,V,r+,z+$-imaging and 7 {Subaru/Suprime-Cam} optical medium-bandwidth filters \citep{Taniguchi07} with seeing $<1\farcs1$ {(from a set of 12, including seeing $>1\farcs1$ images)}, $HST$/WFC3/$F125W,F140W,F160W$ and $HST$/ACS$F606W,F814W$-imaging \citep{Grogin11,Koekemoer11,Brammer12} and UltraVISTA/$Y,J,H,K_s$-imaging \citep{McCracken12}.

In UDS the additional filters are CFHT/MegaCam/$U$ (Almaini/Foucaud, in prep), Subaru/Surpime-Cam/$B,V,R,i,z$ \citep{Furusawa08}, UKIRT/WFCAM/$J,\ H,\ K_s$ (Almaini, in prep), $HST$/WFC3/$F125W$, $F140W$, $F160W$ and $HST$/ACS$F606W$, $F814W$ \citep{Grogin11,Koekemoer11,Brammer12} and VLT/HAWK-I/Y \citep{Fontana14}.

In CDFS and UDS we have additionally available \fs\ narrow-bandwidth data at $1.18\micron$ (\fs/NB118) and $2.09\micron$ (\fs/NB209) \citep{Lee12}. The narrowbands are sensitive to emission line flux. Small bandwidths in combination with high SNR for some galaxies may lead to biased photometric redshift and stellar mass estimates, because the models we use for determining redshifts and stellar population parameters do not contain well-calibrated strong emission lines. As such, they are incorporated into the catalogs, but are not used to derive photometric redshifts or stellar masses. The images have $5\sigma$ image depths of 25.2 and 24.8 AB mag in NB118 and CDFS and COSMOS, respectively and 24.4 and 24.0 AB mag in NB209.

The {\it Spitzer}/IRAC/$3.6$ and $4.5\micron$ images used in CDFS are the ultradeep mosaics from the IUDF (PI: Labb\'e), using data from the cycle 7 IUDF program, IGOODS (PI: Oesch), GOODS (PI: Dickinson), ERS (PI: Fazio), S-CANDELS (PI: Fazio), SEDS (PI: Fazio) and UDF2 (PI: Bouwens). In CDFS we further use {\it Spitzer}/IRAC/$5.8$ and $8.0\micron$ images from GOODS \citep{Dickinson03}. In COSMOS and UDS we use the $3.6$ and $4.5\micron$ images from SEDS \citep{Ashby13}. The $5.8$ and $8.0\micron$ data in COSMOS are from S-COSMOS \citep{Sanders07} and in UDS from spUDS (Dunlop et al, in prep).

The ancillary images are registered and interpolated to the same grid as the \fs\ mosaics, using the program {\tt wregister} in IRAF. Backgrounds {for the UV, optical and near-IR images} were estimated with SE and manually subtracted.

We further supplement the optical/near-IR catalogs with deep far-IR imaging from {\it Spitzer}/MIPS at $24\micron$ (GOODS-S: PI Dickinson, COSMOS: PI Scoville, UDS: PI Dunlop). Median 1$\sigma$ flux uncertainties in $24\micron$ for the COSMOS and UDS pointings are roughly 10\uJy. The CDFS pointing is deeper with a median 1$\sigma$ flux uncertainty of 3.9\uJy. In CDFS we additionally make use of public {\it Herschel}/PACS observations from PEP \citep{Magnelli13} at $100\micron$ and $160\micron$, with 1$\sigma$ flux uncertainties of 205 and 354\uJy\, respectively. {In COSMOS and UDS deep {\it Herschel}/PACS data {are} not yet publicly released.}

\section{Photometry}\label{sec:phot}

\subsection{PSF matching}\label{sec:conv}
The full UV/optical to near-IR dataset contains images of varying seeing quality. The FWHMs of the PSF corresponding to each image varies between $0\farcs2$ for the $HST$ bands to $1\farcs05$ for some of the UV/optical images. To measure aperture fluxes consistently over the full wavelength range, i.e., measuring the same fraction of light per object in each filter, the images have to be convolved so that the PSFs match. To achieve a consistent PSF we first characterize the PSF in all individual images, we then define a theoretical model PSF as a reference, and finally convolve all bands to match the reference PSF.

The average PSF for each image was produced by selecting unsaturated stars with high SNR ($>150$) (see Section \ref{sec:stars}, in which we describe how stars were identified {in the images}), in postage stamps of $10\farcs65\times10\farcs65$. For each star we measured a curve of growth, i.e the total integrated light as a function of radius, with nearby objects masked using the SE segmentation map. Outliers, such as saturated stars, were then determined based on the shape of their light profile compared with the median curve of growth, and rejected from the sample. We median averaged the remaining stars, and, after {normalizing the flux}, used this to fill in masked regions. After renormalizing each tile by the total integrated flux at sufficiently large radius (25 pixels or $3\farcs75)$ we again stacked the postage stamps to obtain a median star. Finally, to obtain {a} clean sample, we again compared the light profiles of individual stars against the median light profile, and iteratively rejected stars if the average deviation from the median curve of growth squared exceeded 5\%. The result was a tightly homogeneous sample of stars, from which we obtained the final median 2-dimensional PSF. 

We generated as a reference PSF a model Moffat profile {\citep{Moffat69}} with full-width-at-half-maximum (FWHM) $=0\farcs9$ and $\beta=2.5$.  The advantage of using a model PSF rather than the average PSF from an image, is that a theoretical model is noiseless. To convolve the images to match the target model PSF, we first derive a kernel for each image individually. For this we use a deconvolution code developed by I. Labb\'e, which fits a series of Gaussian-weighted Hermite polynomials to the Fourier transform of the PSF. The original images were then convolved with this kernel {to match the target PSF}. This method results in very low residuals and is optimal for images with {either} a smaller PSF, or a PSF {that is at most} slightly larger ($\lesssim$ 15\%); {further details are shown} in Appendix \ref{apx:psf}. {We find that} 12\% of the {images have a PSF that is broader than our target PSF}. {Our method} improves the accuracy of the final convolved PSFs, compared with e.g., maximum likelihood algorithms. For example, \cite{Skelton14} find $<1\%$ accuracy when convolving $HST$/WFC3 images, using the same technique as employed here, compared to e.g., \cite{Williams09} {and	} \cite{Whitaker11}, who use maximum likelihood methods to match point sources to within $2-5\%$ accuracy.

\begin{figure*}
\includegraphics[width=\textwidth]{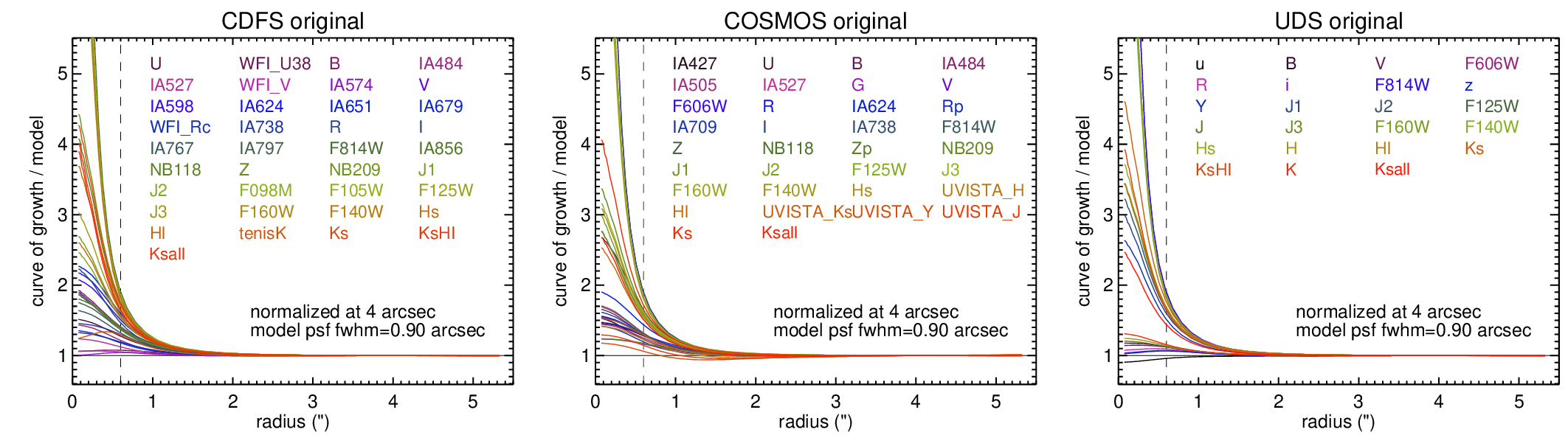}
\includegraphics[width=\textwidth]{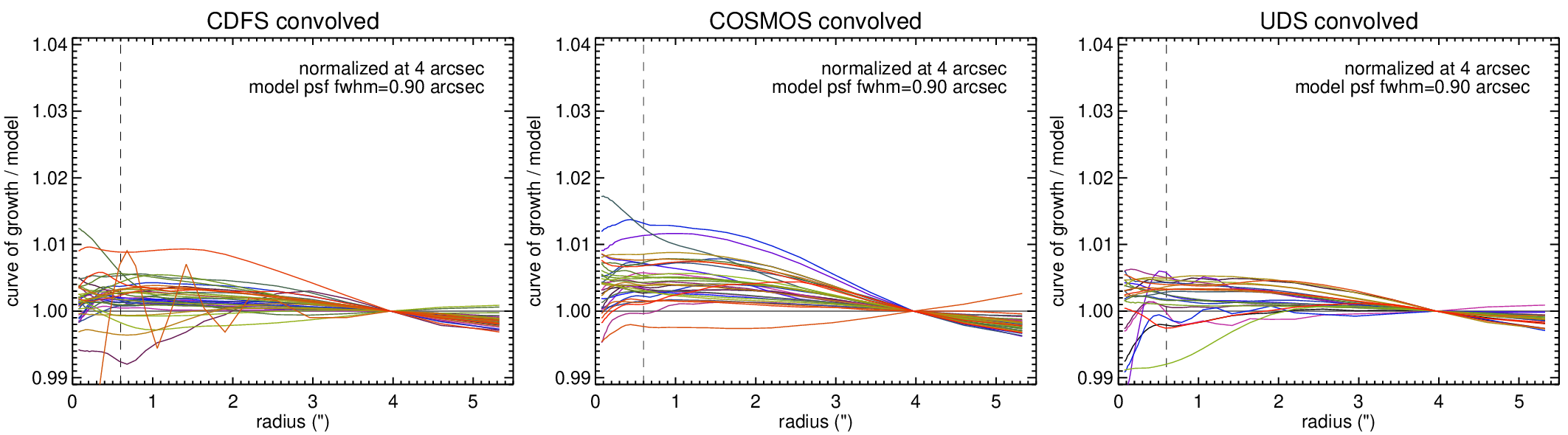}
\caption{Top: curves of growth of the median stacked PSF of stars in the unconvolved images, normalized at $4\arcsec$ radius and divided by the curve of growth of the target moffat PSF. The vertical dashed lines represent the radius at which we measure flux (Section \ref{sec:apflux}). The spread in integrated flux is very large between different images, which would lead to biased color measurements. Bottom: Here we show the curves of growth of the convolved images, where each PSF is convolved to match a Moffat profile. The correspondence with the target PSF is almost one-on-one, with at most a 1.5\% deviation at $r=0\farcs6$.}
\label{fig:gc}
\end{figure*}

PSF curves of growth before and after convolution  are shown in Figure \ref{fig:gc}, normalized by the model PSF. For each convolved image we obtain excellent agreement, within 1.5\% at $r<0\farcs6$. IRAC photometry, with FWHM$>1\farcs5$ is treated separately in Section \ref{sec:farir}.

\subsection{Source detection}\label{sec:sd}

We created detection images from the superdeep background subtracted $K_s$-band {images}, as described in Section \ref{sec:kstacks}, by noise equalizing the images, i.e., multiplying the images with the square root of the corresponding weight images. We then ran SE to create a list of sources and their locations. We optimized source detection by setting the deblending parameters of SE to {\tt DEBLEND\_THRESH$=64$} and {\tt DEBLEND\_MINCONT$=0.0000001$} and the clean parameter ({\tt CLEAN}) to {\tt N}. We also generated a segmentation map {with SE} representing the location and area of each source. The total number of sources in the catalogs is 30,911 in CDFS, 20,786 in COSMOS and 22,093 in UDS. {Our} SE parameter {files} are included in the ZFOURGE data release.

\subsection{$K_s$-band total flux determination}\label{sec:kstot}
To measure the $K_s$-band total flux, SE was run in dual image mode {on the superdeep $K_s$-band images}, using the {noise equalized} images {(Section \ref{sec:sd}) for source detection.} We used a flexible elliptical aperture \citep{Kron80}, to obtain SE's {\tt FLUX\_AUTO}.

This estimate is not yet the total $K_s$-band flux and we have to account for missing flux outside the aperture. We derived a correction factor from the stacked $K_s$-band PSF separately for each field. This aperture correction varies between sources and is a function of the size of the auto-aperture that was estimated by SE. 

We determined the {aperture} correction by {using the curve of growth of the PSF.} 
Total $K_s$-band fluxes were then calculated using 
\begin{equation}
F_{Ks,tot}=F_{Ks,auto}\frac{F_{PSF}(<4\ \arcsec)}{F_{PSF}(<r_{Kron})}
\end{equation}
{\citep{Labbe03a,Quadri07}}, with $F_{Ks,tot}$ the total $K_s$-band flux, $F_{Ks,	auto}$ the flux within the auto-aperture, i.e., {\tt FLUX\_AUTO} from SE, $F_{PSF}(<4\ \arcsec)$ the {flux of the PSF within a} $4\ \arcsec$ {radius} and $F_{PSF}(<r_{Kron})$ the flux within the circularized Kron radius.

We additionally measured the total flux using a fixed circular aperture, of $\sim1.5\times$ the PSF FWHM of the deep $K_s$-band images. In CDFS we therefore used a $0\farcs7$ diameter aperture and in COSMOS and UDS a $0.\farcs9$ diameter aperture. {These} aperture fluxes were also corrected for flux outside of the aperture. 

{Therefore} we have two estimates for the total flux, one using the auto aperture flux, and one using a fixed circular aperture. For small, low SNR sources, the autoscaling aperture size may be {very small}, leading to extreme aperture corrections. Therefore, we only considered the circular aperture measurements for sources if their circularized Kron radius was very small, i.e., smaller than the circular aperture radius.

\subsection{Aperture fluxes}\label{sec:apflux}
In addition to the total $K_s$-band flux, we derived flux estimates in all filters in the three ZFOURGE fields. We ran SE in dual image mode, using the combined $K_s$-band images for source detection and the PSF matched images to measure photometry. We use the PSF matched images to make sure the captured light within the apertures is consistent over all the images. We also included the convolved versions of the deep $K_s$-band stacks. We use circular apertures of $1\farcs2$ diameter, which are suffiently large to capture most of the light (the PSFs of the convolved images have a FWHM$=0\farcs9$), but small enough to optimize SNR.

We correct all aperture fluxes to total, using the ratio between the total flux in the original deep $K_s$-band stacked images to the aperture flux in the PSF matched $K_s$-band stack, i.e.,:
\begin{equation}
F_{F,tot}=F_{F,aper}*\frac{F_{Ks,tot}}{F_{Ks,aper}}
\end{equation}
Here, $F_{F,tot}$ is the aperture flux in filter $F$ scaled to total, $F_{F,aper}$ the unscaled aperture flux, $F_{Ks,tot}$ the total $K_s$-band flux described in Section \ref{sec:kstot} and $F_{Ks,aper}$ the aperture flux from the PSF-matched $K_s$-band image stacks.

\subsection{Flux uncertainties}\label{sec:error}
The uncertainty on the flux measured in an aperture has contributions from the background, the Poisson noise of the source, and the instrument read noise. The relative contribution from the latter two effects will be very small for the faint galaxies and medium band filters used in this study \citep{Persson13}. If the adjacent pixels in an image are uncorrelated, the background noise $\sigma_{RMS}$ measured in an aperture containing $N$ pixels will scale in proportion to $\sqrt{N}$. In a more realistic scenario, pixels are expected to be correlated on small scales due to interpolation or PSF smoothing and on large scale due to imperfect background subtraction, flux from extended objects, undetected sources, or systematics introduced in the reduction process, such as flat field errors.  For perfectly correlated pixels, the background noise is expected to scale as $\sigma_{RMS}\propto N$. The actual scaling of the noise in an image lies somewhere in between and can be parameterized by 
\begin{equation}\label{eq:nmad}
\sigma_{NMAD}=\sigma_1\alpha N^{\beta/2}
\end{equation}
with $\sigma_{NMAD}$ the normalized median absolute deviation and $\beta$ taking on a value between $1<\beta<2$. $\alpha$ is a normalization parameter and $\sigma_1$ is the standard deviation of the background pixels. \citep{Labbe03a,Quadri07,Whitaker11}
We estimated the noise as a function of aperture size empirically by {placing }circular apertures of varying diameter {at} 2000 {random locations} in each image that was used for photometry. These are the convolved images for the aperture fluxes and the unconvolved $K_s$-band stacks that were used to measure total flux. We used the SE segmentation map to mask sources. We also excluded regions with low weight, such as the edges of the \fs\ detectors.

For each aperture diameter, we fit a Gaussian to the measured flux distribution and obtained the standard deviation ($\sigma_{RMS}$). We then fit Equation \ref{eq:nmad} to the various estimates of $\sigma_{RMS}$ as a function of $N$ pixels in each aperture, to obtain $\sigma_1,\alpha$ and $\beta$. 

For circular apertures with radius $r$ pixels, the uncertainty ($e_F$) on the flux measurement in filter $F$ is
\begin{equation}
e_F=\sigma_{NMAD}(r)/\sqrt{w_F}=\sigma_1\alpha(\pi r^2)^{\beta/2}/\sqrt{w}
\label{eq:flux_uncertainty}
\end{equation}
with $w_F$ the median normalized weight. We did not include a Poisson error in our flux uncertainties, as faint sources are background-limited, while uncertainties on bright sources are dominated by systematics.

Weights were obtained from the median normalized exposure images and were measured as the median in apertures with sizes corresponding to those used to measure flux. The radius $r$ used in Equation \ref{eq:flux_uncertainty} was chosen to match the aperture size used for the different flux determinations. $1\farcs2$ diameter apertures are used for the aperture fluxes and, for the total flux, we use SE's {\tt KRON\_RADIUS}, which is based on autoscaling kron-like apertures.

The aperture flux uncertainties obtained from Equation \ref{eq:flux_uncertainty} were scaled to total for a consistent relative error.

\subsection{IRAC and MIPS photometry}\label{sec:farir}

The $Spitzer$/IRAC and MIPS images {(in all fields)} and Herschel/PACS images {(available for CDFS only)} have much broader PSFs than the UV, optical and near-IR images and source blending is a {significant} effect. The FWHM in the IRAC images is typically $>1\farcs5$ and in MIPS $>4\arcsec$. To obtain photometry, we use a source fitting routine that models and subtracts profiles of neighbouring objects prior to measuring photometry for a target \citep{Labbe06,Wuyts08,Whitaker11,Skelton14,Tomczak16}.

The position and extent of each source was based on the SE segmentation maps derived from the super deep $K_s$-band detection images. The $K_s$-band images are assumed to provide a good prior for the location and extent of the unresolved far-IR flux, as sources that are bright in K are also typically bright at redder infra-red wavelengths.
Each source in the $K_s$-band image was extracted using the segmentation map and convolved to match the PSF of the lower resolution far-IR image, assuming negligible morphological corrections. All sources were then fit simultaneously to create a model for the lower resolution image. Next, for each source in the lower resolution image, the modelled light of neighbouring sources was subtracted, after which we measured the flux on the cleaned maps within circular apertures with diameter $D$, using $D=1\farcs8$ for IRAC and $D=7\arcsec$ for MIPS. 

To correct the far-IR aperture flux to total, the measurements were {multiplied} by the ratio of the total $K_s$-band flux to the $D=1\farcs8$ aperture flux on the PSF convolved $K_s$-band template image. Because the MIPS PSF has significant power in the wings at large radii, which are not represented in the convolution kernel, we apply an additional fixed correction of $\times 1.2$ to account for missing flux at $r>15\arcsec$ (using values for point-sources from the MIPS instrument handbook).

{Flux uncertainties were estimated from background maps. These were individually generated for each source on scales of three times the $30\ \arcsec$ tile size used for the modelling, using the cleaned tiles. From these we measured RMS variations using apertures at random locations. Variations on larger scales were corrected by spatially adjusting the zeropoint using the iterative procedure described in Section \ref{sec:eazy}.}

\subsection{Stars}\label{sec:stars}

\begin{figure*}
\includegraphics[width=\textwidth]{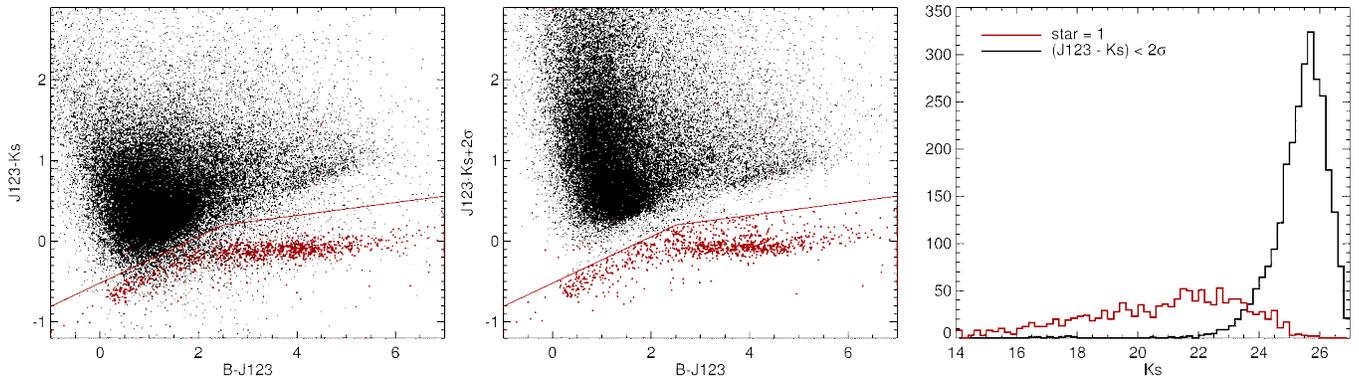}
\caption{Left and Middle panel: $B-J123$ versus $J123-K_s$ for stars (red) and galaxies (black). The red line indicates the star/galaxy separation. In the first panel some sources are scattered below the red line. These are effectively removed from the star classification by selecting in $J123-K_s$ at $>2\sigma$ confidence (middle panel). Right panel: $K_s$-band magnitudes of stars (red) and sources that scatter below the red line (black). By only selecting at $>2\sigma$ confidence in $J123-K_s$, we exclude the faintest sources from the star sample.}
\label{fig:stars}
\end{figure*}

The majority of stars were identified by their observed $B-J123$ and $J123-K_s$ colors. $J123$ here is derived as the median of the flux in the $J1$, $J2$ and $J3$ filters. Stars form a tight sequence in $J123-K_s$ compared to galaxies. In the first two panels of figure \ref{fig:stars} our selection criterion is indicated as a red line, with stars having:
\begin{align*}
(J123-K_s) &< 0.288(B-J123)-0.52 & [(B-J123)<2.5]\\
(J123-K_s) &< 0.08(B-J123) & [(B-J123)>2.5]
\end{align*}

Here we only classified sources as stars if they are below the red line at $>2\sigma$ confidence in $J123-K_s$. By selecting at $>2\sigma$ confidence, we automatically reject faint sources that scatter below the red line from the star sample. This is illustrated by the histograms in the third panel of Figure \ref{fig:stars}, where we show the magnitude counts of stars against sources that are not now classified as stars, but would have been otherwise. These have a distribution of magnitudes that peaks around $K_s=25.5-26$ magnitude.

For sources that were not covered by the $J1$, $J2$ and $J3$ bands, we used broadband $J$ or F125W where available, but only considered sources brighter than 25 mag in $K_s$. For sources without B-band coverage, we used $J123-K_s<0$ to classify stars, considering only sources brighter than 22 magnitude in $K_s$. An finally, if sources did have B-band coverage, but were saturated in B, we also used $J123-K_s<0$.

Red cool stars may not be selected in this way, as they have red $J-K$ colors. To ensure that we cover all types of stars, we fit the observed SEDs of all sources with EAZY using the speX stellar library\footnote{\url{http://pono.ucsd.edu/$\sim$adam/browndwarfs/spexprism/}}. For a few sources that were not flagged already by their $B-J123$ and $J123-K$ colors, the reduced $\chi^2$ indicated a stellar template was a better fit to the data than the best-fit galaxy template (Section \ref{sec:eazy}) and we flagged these sources as stars as well.

A source that is not selected by any of the methods above, is considered a saturated star if it is brighter than 16 magnitude in $J_1$ or $K_s$ and at the same time could not be fit to a galaxy template, having a large reduced $\chi^2$, which we empircally estimated by inspecting many SEDs to be $\chi^2>3000$.

In total, 1.8\% of the sources in the catalogs are classified as stars.

\subsection{Catalog format}\label{sec:header}

\begin{table*}
\caption{Explanation of the photometric catalog header}
\begin{threeparttable}
\begin{tabular}{l l}
\hline
\hline
id&			ID number \\
x,y& 			pixel coordinates (scale: $0\farcs15$ / pixel) \\
ra,dec& 		right ascension, declination (J2000) \\
SEflags& 		Source Extractor flags\\
iso\_area& 		isophotal area above Source Extractor analysis treshold (pix$^2$) \\
fap\_Ksall\tnote{a}& (convolved) $K_s$-band aperture flux within a $1\farcs2$ diameter circular aperture\\
eap\_Ksall& uncertainty on fap\_Ksall\\
apcorr & aperture correction applied to fauto\_Ksall to obtain f\_Ksall (f\_Ksall = fauto\_Ksall * apcorr)\\
Ks\_ratio & ratio between fap\_Ksall and f\_Ksall (Ks\_ratio = f\_Ksall / fap\_Ksall)\\	
fapcirc0D\_Ksall\tnote{b} & aperture flux measured within a $D\arcsec$ diameter {(seeing dependent)\tnote{b}} circular aperture\\
eapcirc0D\_Ksall & uncertainty on fapcirc0D\_Ksall \\
apcorr0D & aperture correction applied to fapcirc0D\_Ksall to obtain fcirc0D\_Ksall (fcirc0D\_Ksall = fapcirc0D\_Ksall * apcorr0D)\\
fcirc0D\_Ksall\tnote{a,b} & total (aperture corrected) $K_s$-band flux within a $D\arcsec$diameter {(seeing dependent)\tnote{b}} circular aperture\\
ecirc0D\_Ksall & uncertainty on fcirc0D\_Ks \\
fauto\_Ksall & $K_s$-band flux within a Kron-like elliptical aperture\\
flux50\_radius & radius (pixels) enclosing 50\% of the $K_s$-band flux \\
a\_vector & major axis of a Kron-like elliptical aperture\\
b\_vector & minor axis of a Kron-like elliptical aperture\\
kron\_radius & radius of a circularized Kron-like elliptical aperture\\
f\_Ksall\tnote{a} & total (aperture corrected) $K_s$-band flux within a Kron-like elliptical aperture\\
e\_Ksall & uncertainty on f\_Ksall\\
w\_Ksall & weight corresponding to f\_Ksall, median normalized\\
f\_[] & (convolved) aperture flux in filter [] within a $1\farcs2$ diameter circular aperture, corrected to total (fap\_[] = f\_[] / Ks\_ratio)\\
e\_[] & uncertainty on f\_[] (also scaled with Ks\_ratio)\\
w\_[] & weight corresponding to f\_[], median normalized\\
wmin\_optical & minimum w\_[] of groundbased optical filters\\
wmin\_hst\_optical & minimum w\_[] of $HST$ optical filters\\
wmin\_fs & minimum w\_[] of \fs\ filters\\
wmin\_jhk & minimum w\_[] of broadband J, H \& K filters \\
wmin\_hst & minimum w\_[] of $HST$ near-IR filters\\
wmin\_irac & minimum w\_[] of $Spitzer$/IRAC filters\\
wmin\_all & minimum w\_[] of all filters\\
star & this flag is set to 1 if the source is likely to be a star, to 0 otherwise, following the criteria described in Section \ref{sec:stars}\\
nearstar & this flag is set to 1 if the source is located within $r(\arcsec)<10-(m-16)$ of a bright star with $m$ the \\ & apparent magnitude of the star and $m<17.5$ in $J_1,J_2,J_3,J$ or $K_s$ \\
use\tnote{c} & sources that pass the following criteria are set to 1: \\
& \begin{minipage}{0.85\textwidth}
\begin{compactitem}[-]
\item star = 0
\item nearstar = 0
\item SNR $\geq 5$
\item wmin\_fs $>$ 0.1 {(A minimum exposure time of at least $0.1\times$ the median exposure in the \fs\ bands)\tnote{d}}
\item wmin\_optical $>$ 0 {(coverage in all optical bands)}
\item not a {catastrophic} EAZY fit: $\chi^2$ (reduced) $\leq 1000$\tnote{e}
\item {not} a {catastrophic} FAST fit, {i.e., a finite and positive stellar mass estimate above $10^6 M_{\sun}$}
\item consistent flux ratios between similar bands of different instruments, namely the $J-$, $H-$ and $K-$bands of FourStar and VISTA, and $F814W-$ and groundbased $I-$bands
\item no $5\sigma$ detection at wavelengths bluer than the restframe 912 $\AA$ Lyman limit
\item not at $z<0.1$
\end{compactitem}
\end{minipage} \\ \vspace{10pt}
snr & signal-to-noise (=fapcirc0D\_Ksall / eapcirc0D\_Ksall) \\ 
\hline
\end{tabular}
\begin{tablenotes}
\item[a] Note that these $K_s$-band fluxes are derived from the superdeep combined Ks-band images. Within the catalogs only f\_Ks corresponds to \fs/$K_s$.
\item[b] In CDFS $D=0\farcs7$, in COSMOS and UDS $D=0\farcs9$ (i.e., $1.5\times$ the seeing FWHM).
\item[c] A standard selection of galaxies can be obtained by selecting sources with use = 1.
\item[d] {Effectively this means that every source has at least 20 minutes exposure in each \fs\ band. Because of the dither pattern, sources with lower weight that are removed by this criterion lie at the edges of the images. For wmin\_fs we used wmin\_ksall instead of the weight of the FourStar $K_s-$band.}
\item[e] {Based on an empirical estimation from inspecting many fits.}
\end{tablenotes}
\end{threeparttable}
\label{tab:header}
\end{table*}

We provide separate photometric catalogs for each cosmic field. These contain the coordinates, total fluxes, flux uncertainties, weight estimates, flags and SNR estimates of each source. Individual sources are indicated by their ID, starting at $\mathrm{ID}=1$. A description of the columns is given in Table \ref{tab:header}.

The CDFS catalog contains 30,911 sources, the COSMOS catalog 20,786 and the UDS catalog contains 22,093 sources. Magnitudes for each source can be obtained by applying a zeropoint of 25 in the AB system (corresponding to a flux density of $3.631\times 10^{-30} erg\ s^{-1}\ Hz^{-1}\ cm^{-2}$ or $0.3631 \mu Jy$). e.g., the stacked $K_s$-band total magnitude is $25-2.5\times\mathrm{log_{10}}{\tt f\_Ksall}$. 

All fluxes in the catalogs are scaled to total. They can be converted back to aperture flux ($1\farcs2$ diameter) by dividing by {\tt Ks\_ratio} for each source. The exceptions are {\tt fap\_Ksall}. {\tt fauto\_Ksall} and {\tt fapcirc0D\_Ksall}. The first is the actual (convolved) $K_s$-band aperture flux, and can only be converted in the other direction, towards total. The second is the auto aperture flux from SE, and we need only to apply the aperture correction, {\tt apcorr}, described in Section \ref{sec:kstot}, to obtain {\tt f\_Ksall}. The last is an alternative to {\tt fauto\_Ksall}, and is measured in a fixed circular aperture with diameter D, instead of the flexible elliptical Kron-like aperture from SE (using apertures of $D=0\farcs7$ in CDFS and $D=0\farcs9$ in COSMOS and UDS). From {\tt fapcirc0D\_Ksall} we can obtain the total {\tt fcirc0D\_Ksall} by multiplying with {\tt apcorr0D}.

Each flux measurement of each source in each filter has been assigned a weight, reflecting the depth in the images at the source locations. The weigths are normalized to the median of the corresponding weight images. In the catalogs we also indicated the minimum weight for sets of filters. For example, the lowest weight of the \fs\ filters is indicated by {\tt wmin\_fs}. If this value is greater than 0, it means a positive weight in all \fs\ images.

In addition to photometric catalogs, we provide the EAZY (Section \ref{sec:eazy}) and FAST (Section \ref{sec:fast}) output files, containing the photometric redshifts and stellar population properties.

\subsection{A standard selection of galaxies}
For convenient use of the catalogs, we have designed a {\tt use} flag. This takes into account SNR, the star/galaxy classifications described above and the depth of the images at the respective source locations. This flag also includes sources that are well within the \fs\ footprint and are observed with each of the near-IR medium-bandwidth filters. The $K_s$-band stacks cover a somewhat larger area, especially in CDFS, which means that not all sources in the catalogs have \fs\ imaging (although the majority do).	 A standard selection of galaxies can be obtained by selecting on {\tt use=1} {(see full definition in Table \ref{tab:header})} from the catalogs.
 
The  {\tt use} flag allows for a straightforward sample selection, representing galaxies with good photometry, i.e., high SNR sources from well exposed regions of the images. For specific science goals a different selection may be optimal. We also warn that the {\tt use=1} sample may still contain problematic sources, with for example uncertain photo-z's and poorly constrained EAZY or FAST fits, and we recommend to always inspect the individual SEDs. However, the {\tt use=1} sample should be a reliable representation of the galaxy population in large statistical studies.

The total area of the Ks-band detection images is $280.9\ \arcmin^2$ for CDFS, $176.5\ \arcmin^2$ for COSMOS and $189.3\ \arcmin^2$ for UDS. Selecting only galaxies with {\tt wmin\_fs}$>0.1$ that are not near bright stars, reduces the area to $132.2\ \arcmin^2$, $139.2\ \arcmin^2$ and $135.6\ \arcmin^2$.

\subsection{Quality verification}

\subsubsection{Flux comparisons}

Here we test whether the total fluxes derived above are reliable, {by (1) comparing our magnitudes in various bands to independent estimates by a different survey directly, and (2) by comparing our total magnitudes in the detection band to a completely different method to measure total flux. For the former we used the 3D-HST data
set, as both surveys use many of the same images and cover similar fields. Many of the same basic imag{e reduction methods} were used to derive photometry for 3DHST, but we performed our own alignment and registration to the $0\farcs15\ \mathrm{pix}^{-1}$ scale of FourStar and an independent estimate of the background. In detail, photometric methods between the two surveys differ significantly, and in addition we derived total flux from the ultra-deep $K_s−$band images, whereas source detection and total flux derivation for 3D-HST was based on \textit{HST/WFC3/F160W} images.} In general we find excellent correspondence between the two surveys. We show diagnostic plots in Appendix \ref{sec:flux_comparisons}.

We tested our method of extracting total flux through SE by comparing to total flux derived with GALFIT \citep{Peng10a}, a program which fits two-dimensional model light profiles to galaxy imaging. The fitting process benefits from high resolution imaging, so we make use of the $HST$/WFC3/F160W size catalogs from \citet{vanderWel14}, based on the source catalog of 3D-HST, which contains parameters derived with GALFIT. {Galaxies may have different morphologies in different bands. However,} as the F160W and $K_s$-band filters lie very closely together in wavelength space, we assume that the correction to total in our catalogs, which is based on the ratio between $K_s$-band aperture and total flux, {also} produces {an accurate approximation} of total F160W-band flux. {As the comparison with 3DHST shows (Appendix \ref{sec:flux_comparisons}), these magnitudes are accurate to within $\sim1\%$, with magnitude offsets of 0.017, 0.005 and $-$0.011 in CDFS, COSMOS and UDS, respectively.}

\begin{figure*}
\includegraphics[width=\textwidth]{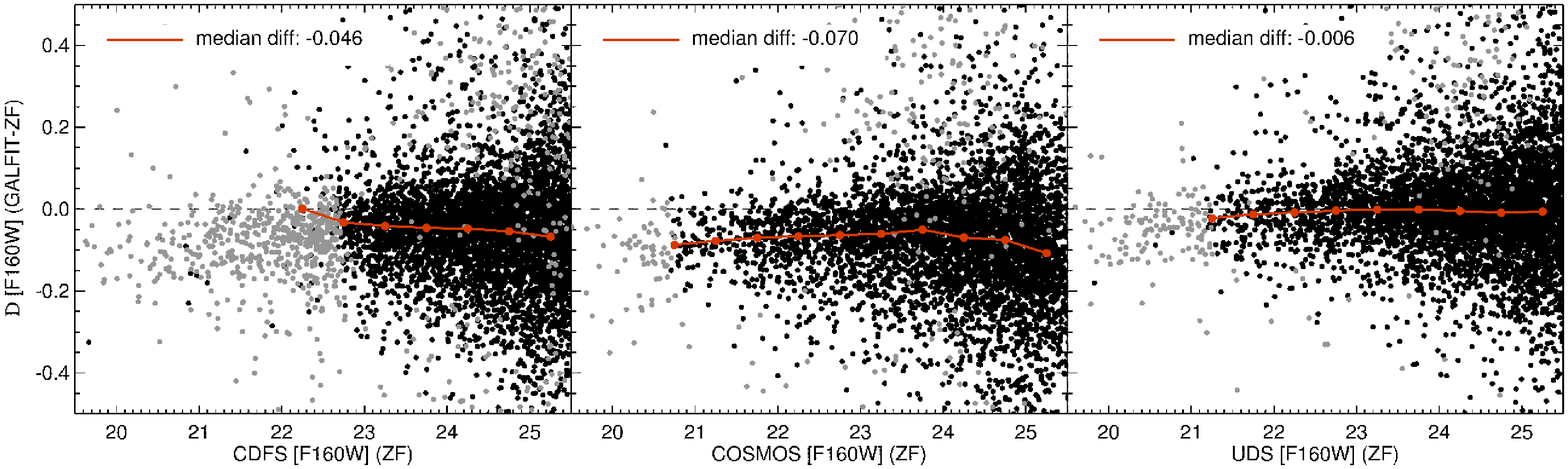}
\caption{The difference between ZFOURGE and GALFIT F160W magnitudes plotted as a function of ZFOURGE magnitude, for galaxies with {\tt use=1} and SEflag=0 (excluding blended or contaminated sources). We show sources with GALFIT flag $=1$ (a suspicious fit) in gray and sources with GALFIT flag $=0$ (a good fit) with black datapoints. Bad fits (GALFIT flag $>1$) were ignored.  The median magnitude difference for galaxies with GALFIT flag$=0$ is shown by the red solid line and filled bulletpoints in bins of 0.5 mag. We also indicate the median offset in the legend. We find slightly brighter magnitudes with GALFIT, of $0.006-0.070$ magnitude {on average, with the difference presumably attributable to different techniques to derive total magnitudes and potential color gradients}.}
\label{fig:gfcheck}
\end{figure*}

The comparison with GALFIT magnitudes is shown in Figure \ref{fig:gfcheck}. We use the goodness of fit flag included in the size catalogs to select sources with a good (GALFIT flag $=0$) or suspicous fit (GALFIT flag $=1$), but not sources with bad fits (GALFIT flag $>1$). We find a median offset between ZFOURGE total F160W magnitude and GALFIT magnitude of $-$0.046,$-$0.070 and $-$0.006 magnitude, for CDFS, COSMOS and UDS, respectively. \citet{Skelton14} show the same comparison, with similar trends with magnitude, and find magnitude offsets for the three fields of $-$0.03,$-$0.04 and 0.00. 
The small offsets that we find between GALFIT magnitude and magnitude derived with SE, {are likely attributable to details of our photometric procedure to determine total fluxes, which are somewhat dependent on galaxy profile and SNR \citep[see][]{Labbe03b,Skelton14} and possible color gradients.}

\subsubsection{Flux uncertainty verification}

\begin{figure}
\includegraphics[width=0.49\textwidth]{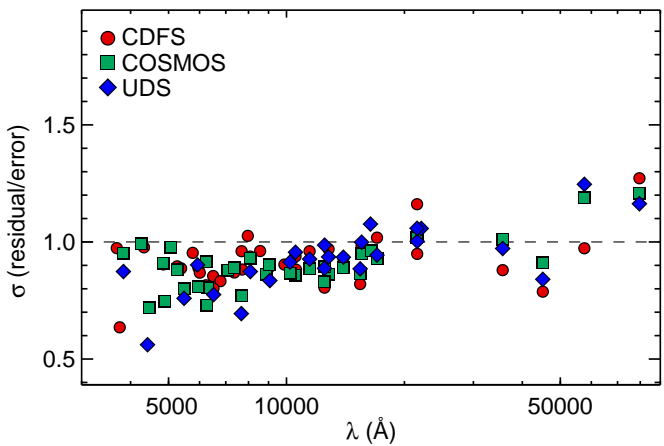}
\caption{NMAD scatter in the error-normalized flux residuals as a function of wavelength, for the three ZFOURGE fields. $\sigma$ is close to unity for most filters, indicating the photometric errors are accurate.}
\label{fig:echeck}
\end{figure}

Here we test the {accuracy} of the flux uncertainties derived in Section \ref{sec:error}. We used the outcome of the SED fitting described below in Section \ref{sec:eazy}. The residual between the best-fit template flux and the observed flux in a filter should reflect the photometric errors in the catalogs. If these are accurate, then normalizing the distribution of the residuals by the photometric error, should result in a Gaussian with a width of unity. We derived the normalized median absolute deviaton (NMAD) of the distribution of the error-normalized residuals, and show the scatter ($\sigma$), for each filter in the catalog, as a function of wavelength in Figure \ref{fig:echeck}. {Overall these look very good, with the average $\sigma_{NMAD}$ very close to unity, and the vast majority ($>90\ \%$) of bands within 20 \% of unity.}

\subsubsection{Close pair contamination}
It is naturally expected that some sources lie in close angular proximity of each other, and may contaminate the aperture flux of their close neighbor. {For UV to near-IR photometry,} this may lead to systematic errors on the aperture photometry of a source, especially if the neighbor is much brighter {(for \textit{IRAC} and \textit{MIPS} photometry we used a source fitting routine that takes into account flux from neighboring sources; see Section \ref{sec:farir})}. The aperture diameter that we used above is $1\farcs2$. We inspected the catalogs for pairs of galaxies that lie closer than $1\farcs2$ distance away from each other. We only looked at sources that are not already classified as stars or as being located in the neighbourhood of a bright star, as we already accounted for these sources that their flux estimate may be affected. The percentage of sources with a neighbour at $<1\farcs2$ distance is 3.8 \% in CDFS, 4.1 \% in COSMOS, and 4.4 \% in UDS. If only the fainter part of a projected galaxy pair is affected, we estimate that $\sim2$ \% of the sources in each field may suffer flux effects from nearby sources.

\section{Completeness}\label{sec:completeness}

\begin{figure}
\includegraphics[width=0.49\textwidth]{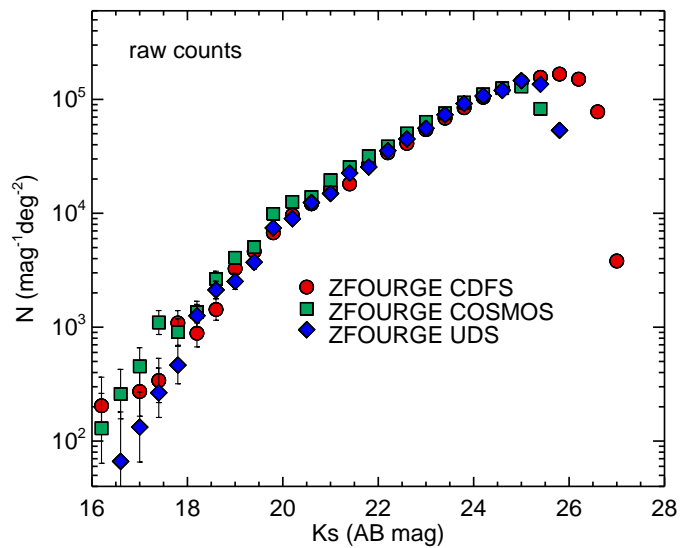}
\caption{$K_s$-band total magnitude number counts of sources with {\tt use=1}. We observe a turn-over in the histograms around $25-26$ magnitude, with less detections for fainter sources.}
\label{fig:kcounts}
\end{figure}

We counted the number of sources with {\tt use=1} per $Ks$-band total magnitude bin in each catalog. This result, taking into account the effective area corresponding to the {\tt use} flag, is shown in Figure \ref{fig:kcounts}. For the different fields, the histograms turn over at $25.5-26$ magnitude, indicating it becomes more difficult to detect fainter sources.

To test how well sources are recovered from the images, we perform completeness tests, using the super-deep $K_s$-band detection images. We drop 10,000 mock sources, {obtained from median stacking low SNR ($9<SNR_{K_s}<11$) sources with {\tt use=1}, in the detection images. The stacks were scaled to a magnitude range of $18<mag\mathrm{(AB)}<27.5$}. We used a powerlaw distribution of magnitudes, matching the slope of the number counts in Figure \ref{fig:kcounts} between $K_s=21$ AB and $K_s=25$ AB. The distribution follows $dlogN/K_s=0.24$, i.e., a factor 1.7 more sources per unit magnitude, with $N$ the number of sources and $K_s$ the total $K_s$-band magnitude, in agreement with previously deteremined values \citep[e.g.,][]{Fontana14}. We ran SE using the same input parameters used to generate the catalogs. We measured the observed magnitude of the input sources that were retrieved with SE. We then compared these with the input source distribution to calculate the correction as a function of observed magnitude, accounting for both completeness and scatter.

We performed the simulation in two ways. First by simply dropping mock sources randomly in the images, only excluding a few small areas around a few very bright stars. To prevent artificial crowding of simulated sources, we only dropped in 500 sources per run, and repeated the simulations a large number of times.

Next we investigated what fraction of incompleteness is due to crowding, where bright sources prevent the detection of fainter sources nearby. We masked all detected sources, using the segmentation map from SE and constrained the location of the simulated sources, such that they do not overlap. In this way, we purely tested if sources can be detected above the noise level in the images. 

\begin{figure*}
\includegraphics[width=0.48\textwidth]{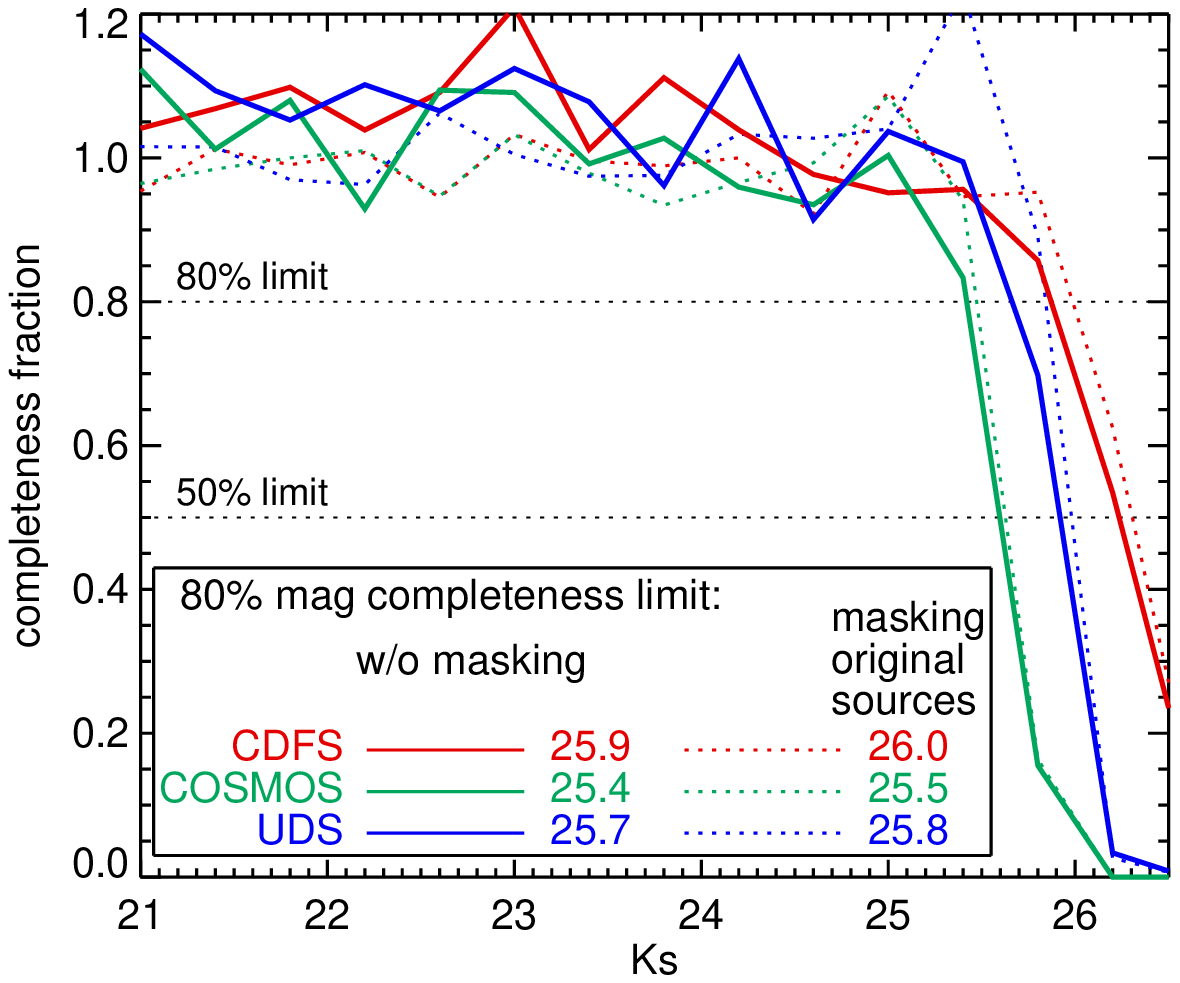}
\includegraphics[width=0.48\textwidth]{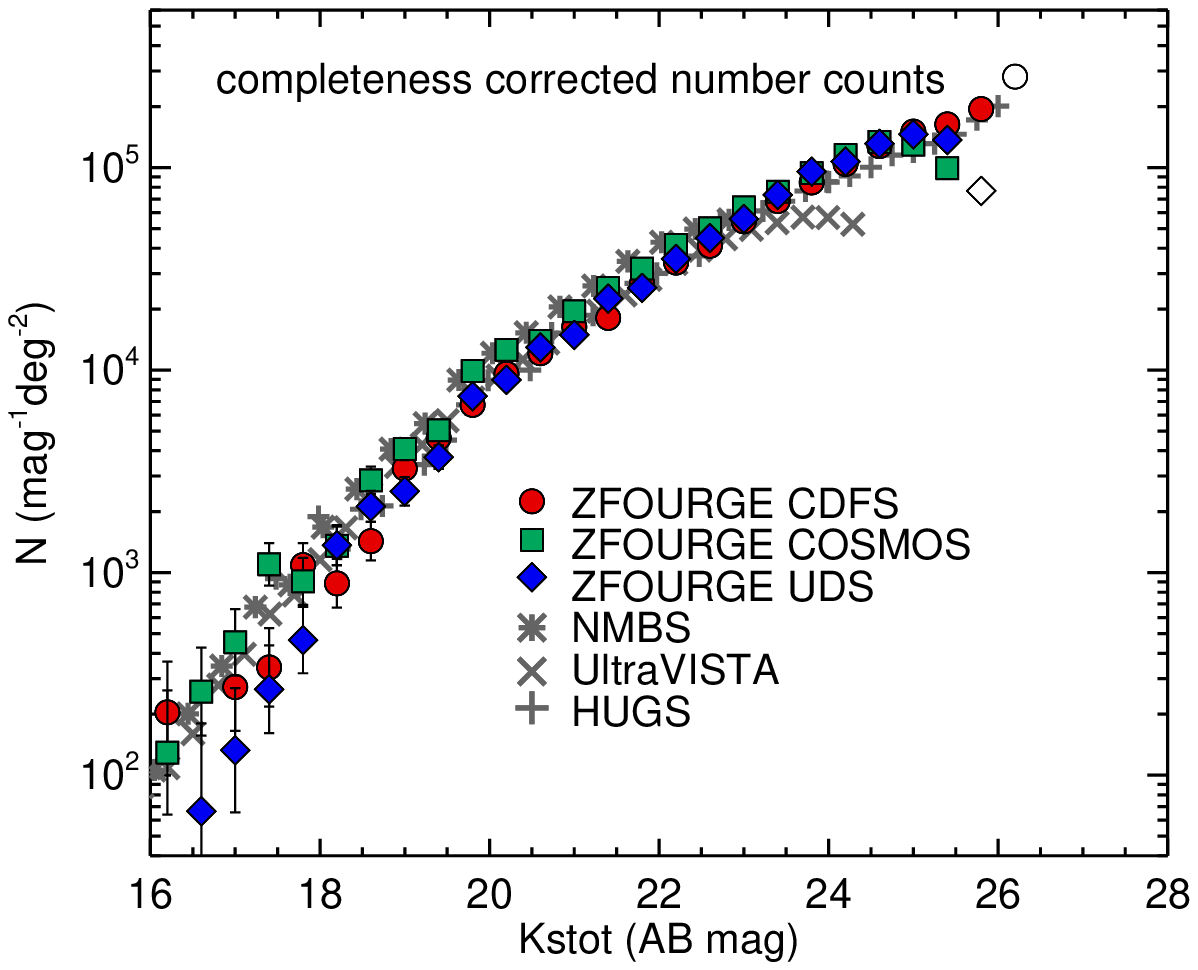}
\caption{Left: We test what fraction of sources is detected by a simulation, in which {mock} sources are inserted into the detection images. For a realistic approach, in which sources are allowed to overlap, we recover 80\% down to very deep magnitudes of $K_s=25.4-25.9$. Right: Completeness corrected number counts. We correct the observed counts in Figure \ref{fig:kcounts} using the completeness estimates in each field. Here we omit sources at magnitudes fainter than the 50\% completeness limits. Sources with $<80\%$ completeness are shown with open symbols. For comparison, the $K_s$-band {completeness corrected} number counts of similar galaxy surveys are indicated in grayscale.}
\label{fig:kcounts2}
\end{figure*}

\begin{table}
\caption{Completeness test results}
\begin{center}
\begin{tabular}{l c c c c}
\hline
\hline
& \multicolumn{2}{c}{with masking} &  \multicolumn{2}{c}{w/o masking}\\
& 80\% & 50\% & 80\% & 50\% \\
\hline
CDFS & 26.0 & 26.3 & 25.9 & 26.2\\
COSMOS & 25.5 & 25.6 & 25.4 & 25.6\\
UDS & 25.8 & 26.0 & 25.7 & 25.9\\
\hline
\end{tabular}
\end{center}
\label{tab:completeness}
\end{table}

We show the results of the two tests in the left panel of Figure \ref{fig:kcounts2}. Even if only stars are masked and sources are allowed to overlap (solid lines) we recover at least 80\% down to very deep $K_s$-band magnitudes of ${25.4-25.9}$ and 50\% down to $25.6-26.2$. These values correspond well with the turnover in $K_s$-band number counts in Figure \ref{fig:kcounts} and the stacked $K_s$-band image depths (Section \ref{sec:kstacks}). The 50\% and 80\% completeness limits of both tests are tabulated in Table \ref{tab:completeness}. {The slight elevation with a higher than 100\% completeness fraction at magnitudes $<24.5$ for the non-masking case is due to confusion with bright sources.}

We correct the number counts from Figure \ref{fig:kcounts} using the completeness estimates as function of observed magnitude from the more conservative test (obtained w/o masking, i.e., the solid curves) in each field and show these in the right panel of Figure \ref{fig:kcounts2}. We also include similar results from the NMBS \citep{Whitaker11}, UltraVISTA \citep{Muzzin13b} and HUGS \citep{Fontana14} surveys. NMBS and UltraVISTA have shallower depths, but much larger areas than ZFOURGE. Our number counts agree with these earlier results from the literature. They also show that ZFOURGE is one of the most sensitive surveys to date, comparable to HUGS and $1-2$ magnitudes deeper in $K_s$ than earlier groundbased surveys. Similar to NMBS, we find an excess of sources at brighter $K_s$-band magnitudes in COSMOS.

\section{Photometric redshifts}\label{sec:eazy}

\subsection{Template fitting}

Photometric redshifts were derived with EAZY	 \citep{Brammer08}, by fitting linear combinations of nine spectral templates to the observed SEDs. Of these, seven are the default templates described by \cite{Brammer08}, five of which are from a library of P\'EGASE stellar population synthesis models \citep{Fioc99}, one represents a young and dusty galaxy and another is that of an old, red galaxy \citep[see also][]{Whitaker11}. The final two templates represent an old and dusty galaxy and a strong emission line galaxy \citep{Erb10}. The code has the option to include a template error function, which we use, to account for systematic wavelength-dependent uncertainties in the templates. We also make use of a luminosity prior, based on the apparent magnitude calculated from the total $K_s$-band flux.

Offsets in the zeropoints may systematically affect the measured flux and therefore also the derived photometric redshifts. We correct for zeropoint offsets, by iteratively fitting EAZY templates to the full optical-near-IR observed SEDs. This procedure is described in detail by \cite{Whitaker11} and \cite{Skelton14}. Similar to \cite{Skelton14}, we use all sources in the fits, including those without a spectroscopic redshift available. We also use a two step process in which we first only vary the zeropoints of the $HST$-bands and then, keeping these fixed, we vary the zeropoints of the groundbased and $Spitzer$/IRAC data.

During this iterative fitting procedure, both the zeropoints and the templates were modified. These are separable corrections, as the templates are modified after shifting both the data and the best-fit SEDs to the rest-frame. Due to the wide range of galaxy redshifts and large number of filters in the catalogs, each part of the spectrum is sampled by a number of photometric bands. In small bins of rest-frame wavelength, we determined systematic offsets between the data and the templates and updated the templates. This allows the templates to reflect subtle features not initially included, such as the dust-absorption feature at $2175\mathrm{\AA}$. After adjusting the templates, zeropoint corrections are calculated in the observed frame. The process is repeated until zeropoint corrections in all bands except U or the IRAC bands become less than 1\% and this typically happens after three or four iterations. 

The zeropoint offsets are listed in Tables \ref{tab:anci0}, \ref{tab:anci1} and \ref{tab:anci2}. The zeropoints in these tables are the effective zeropoints, with galactic extinction and the zeropoint offsets incorporated. The offsets are typically of the order of 0.05 magnitude. The largest offsets occur for the COSMOS and UDS U-bands, which are known to have uncertain zeropoints \citep{Erben09,Whitaker11,Skelton14}. Template and zeropoint errors are hardest to separate from each other for the U- and IRAC $8\micron$ bands, as these lie at the blue and red ends of the spectra, without bracketing filters. 

The residuals between the best-fit templates and observed SEDs are excellent tracers of spatial variations in the zeropoint. We found small variations for all images. In particular, we were able to pinpoint small offsets between the different quadrants of the \fs\ images in UDS
. To alleviate the spatial effect, our final derivation for every filter includes two runs of the fitting process. After the first run we remove a 2 dimensional polynomial fit to the spatial residuals. This is directly incorporated into the catalogs, i.e., we apply a correction to all sources as a function of their x- and y-coordinates in the images and using the corresponding 2 dimensional offsets in each filter. Finally the fitting process is repeated in the way described above to obtain the final zeropoint offsets. 

The spatial variations in zeropoint of the VLT/VIMOS/$R-$band image are larger than in the other images and could not be described by a polynomial function. This image is very deep, so we do not wish to discard it. We therefore impose a minimum error on the flux of 5\%. {In Figure \ref{fig:xyresid} in Appendix \ref{sec:xyresid} we show the residual maps after subtracting the polynomial fits.}

We use the output parameter {\tt z\_peak}  from EAZY as indicator of the photometric redshift. {\tt z\_peak} is estimated by marginalizing over the redshift probability distribution function, $p(z)$. If $p(z)$ has more than one peak, {\tt z\_peak} only marginalizes over the peak with the largest integrated probability. 

\begin{table}
\caption{Explanation of EAZY photometric redshift catalog header}
\begin{threeparttable}
\begin{tabularx}{0.49\textwidth}{>{\hsize=0.15\hsize}X>{\hsize=0.85\hsize}X}
\hline
\hline
id&			ID number \\
z\_spec&			spectroscopic redshift (if no redshift available, z\_spec is set to -1) \\
z\_a&			photometric redshift derived without a K luminosity prior\\
z\_m1&			weighted redshift derived without a K luminosity prior\\
chi\_a& 			minimum $\chi^2$ derived without a K luminosity prior\\
z\_p& 			best-fit redshift after applying the prior\\
chi\_p&			minimum $\chi^2$ after applying the prior\\
z\_m2& 			weighted redshift after applying the prior\\
odds& 			parameter indicating presence of second $\chi^2$ minimum (1 if no minimum)\\
l68,u68&  		1 sigma confidence interval\\
l95,u95&  		2 sigma confidence interval\\
l99,u99&  		3 sigma confidence interval\\
nfilt& 			number of filters used in the fit\\
q\_z& 			quality parameter\\
z\_peak& 		default derived photometric redshift\\
peak\_prob& 		peak probability\\
z\_mc&			randomly drawn redshift value from redshift probability distribution\\
\hline
\end{tabularx}
\end{threeparttable}
\label{tab:eazyheader}  
\end{table}

We provide the full EAZY photometric redshift catalogs. See Table \ref{tab:eazyheader} for an explanation of the catalog header.

\subsection{Photometric redshift uncertainties determined by EAZY}\label{sec:pz}

\begin{figure*}
\includegraphics[width=\textwidth]{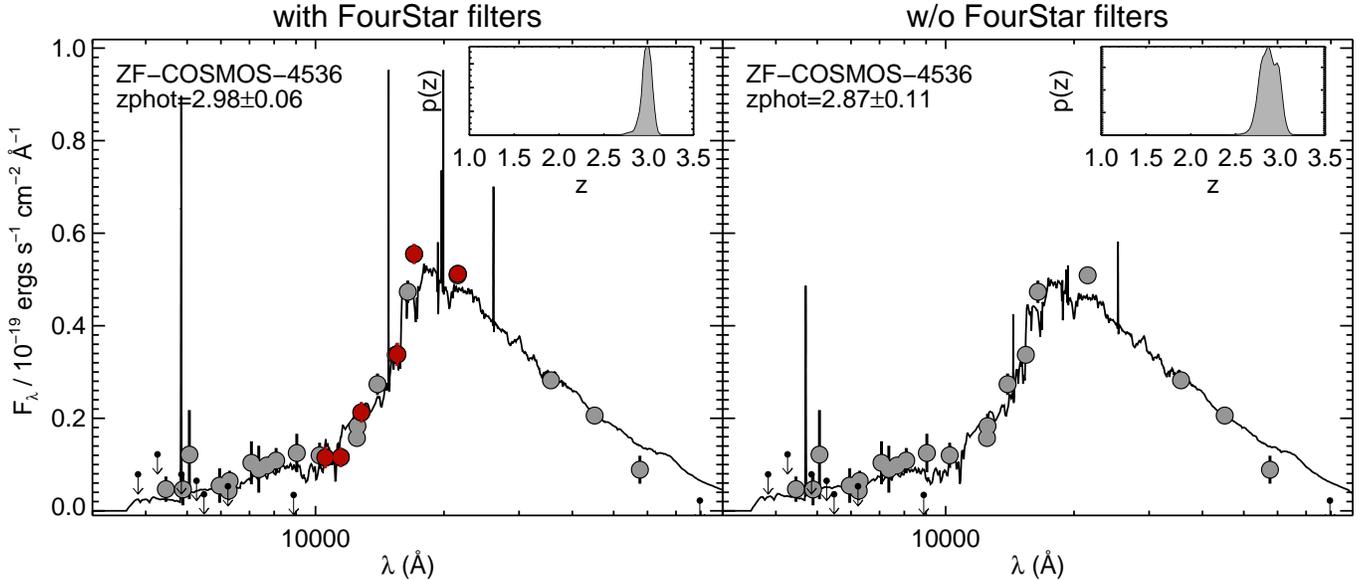}
\caption{An example galaxy at $z=2.1$ with a large \b4, traced by the \fs\ filters (indicated in red). Gray datapoints represent flux in ancillary filters, with downward pointing arrows representing upper limits. In the left panel we show the best-fit SED template derived using the \fs\ near-IR medium-bandwidth filters. In the right panel we show the best fit, without the \fs\ bands. The insets show the redshift probability functions corresponding to the fits. Including the \fs\ filters leads to a factor two better constraint on the photometric redshift.}
\label{fig:balmer_sed}
\end{figure*}

As a result of the use of near-IR medium-bandwidth filters, spectral features such as the \b4\ are better sampled for galaxies at $1.5<z<3.5$. In Figure \ref{fig:balmer_sed} we illustrate the ability of the \fs\ medium-bandwidth filters to constrain galaxy SEDs and redshift probability distibutions. We can determine a photometric redshift error due to the fitting process, using the $16th-84th$ percentiles from $p(z)$. For better constrained redshifts, $p(z)$ will be narrower and the error on $z_{phot}$ will be smaller. We show the SED of a galaxy at a redshift of $z=2.98\pm0.06$, with the uncertainty derived from the 68th percentile of the $p(z)$. This galaxy has a strong $4000\mathrm{\AA}$/Balmer feature, well sampled by the \fs\ medium-bandwidth filters. The photometric redshift derived without the use of medium-bandwidth filters in the near-IR, i.e., using only the available broadband groundbased $Y,\ J,\ H$ and $K/K_s$ or spacebased F125W, F140W and F160W filters, is $z=2.87\pm0.11$. The galaxy has a broader redshift probability distribution, $p(z)$, without the \fs\ filters, i.e., the redshift is less tightly constrained in the fit.

\begin{figure*}
\includegraphics[width=\textwidth]{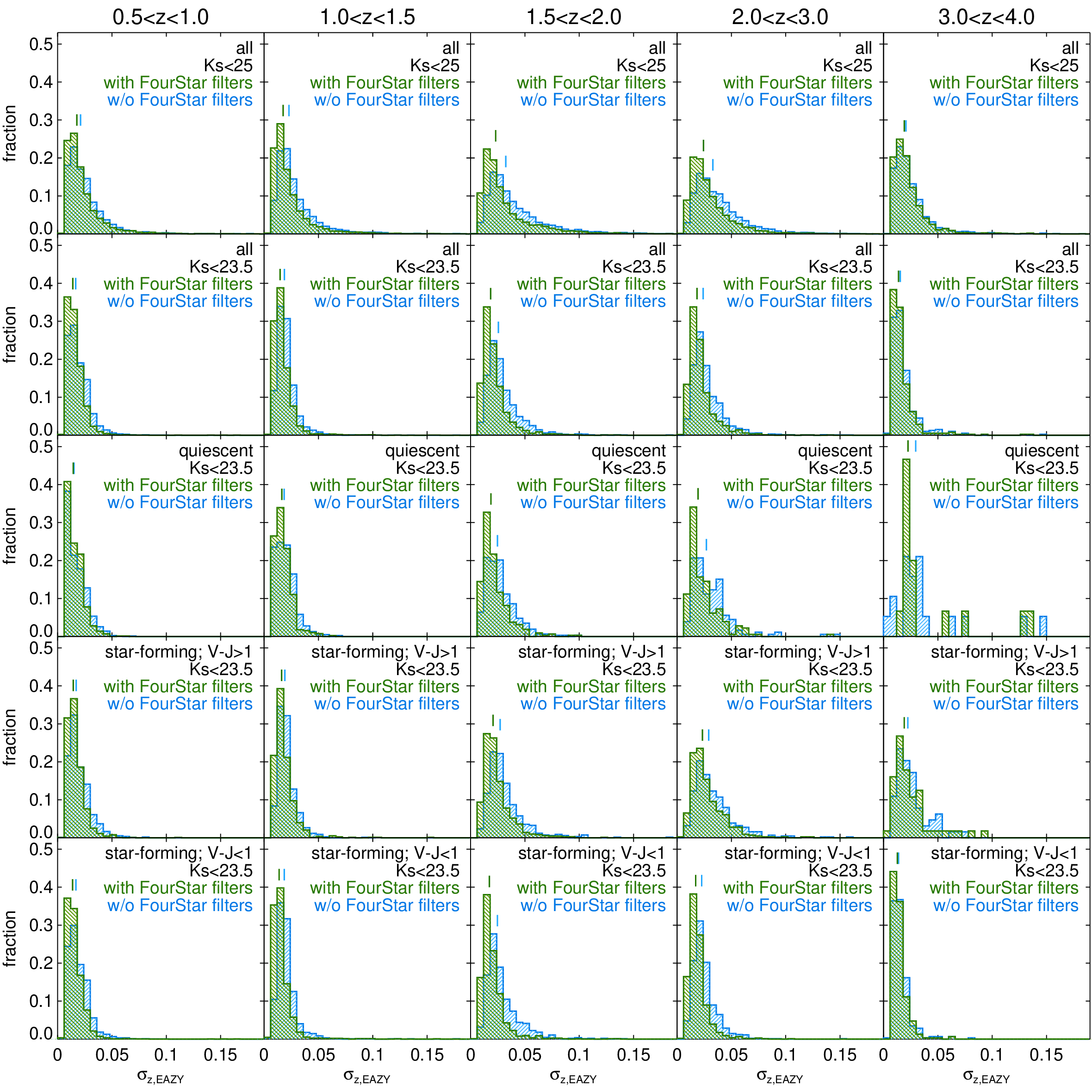}
\caption{Redshift error, $\sigma_{z,EAZY}$ histograms in redshift bins (from left to right), normalized to the total number of sources in each bin. In the first and second row we inspect general magnitude-limited samples, with $K_s<25$ AB and $K_s<23.5$. In the third to last rows of panels we show, respectively, the error histograms of quiescent galaxies, red star-forming galaxies with $V-J\geq 1$, and blue star-forming galaxies with $V-J<1$. The median $\sigma_{z,EAZY}$ is indicated just above the histograms in each panel, using the respective colors (green or {blue}) of the EAZY fits with and without near-IR medium bands. The photometric uncertaintes are systematically smaller if we include the \fs\ near-IR medium-bandwidth filters when fitting SED templates (green histograms).}
\label{fig:pz}
\end{figure*}

In Figure \ref{fig:pz} we show histograms of the errors, $\sigma_{z,EAZY}=p68(z)/(1+z)$, with $p68(z)$ the error from the 68th percentile of $p(z)$, in bins between $z=0.5$ and $z=4$. This is the redshift region where we expect the impact of the medium-bandwidth filters to be greatest. We show the histograms for a magnitude-limited sample, with $K_s<25$ in the top row, and with $K_s<23.5$ in the second row. We also show the histograms of different galaxy types, by splitting up the sample into quiescent and star-forming galaxies, using the UVJ technique \citep[e.g.,][]{Whitaker11}. The star-forming galaxies were additionally split into blue and red by their rest-frame $U-V$ and $V-J$ colors, which we explain further in Section \ref{sec:uvj}.

The histograms indicate that over a large range in redshift, the errors on the photometric redshifts are smaller if we include the \fs\ filters. This holds for all galaxy types. The effect is especially clear around $z=2$, and is noticable for higher redshifts as well. For example, at $1.5<z<2$, the median uncertainty is 40\% higher without the \fs\ filters, with $\sigma_{z,EAZY}=0.036$ compared to $\sigma_{z,EAZY}=0.025$. \citet{Whitaker11} find a similar trend with redshift, for the medium bands of NMBS. The peak of the histograms shifts towards higher $\sigma_{z,EAZY}$ with increasing redshift, up to $z=3$, except for blue star-forming galaxies (with blue $U-V$ and $V-J$ colors, see Section \ref{sec:uvj}), for which $\sigma_{z,EAZY}$ actually improves. A notable spectral feature for these glaxies is the Lyman Break at rest-frame $912\mathrm{\AA}$, which is moving through the optical medium-bandwidth filters at this redshift.

\subsection{Comparison with spectroscopic redshifts}
\begin{figure}
\includegraphics[width=0.49\textwidth]{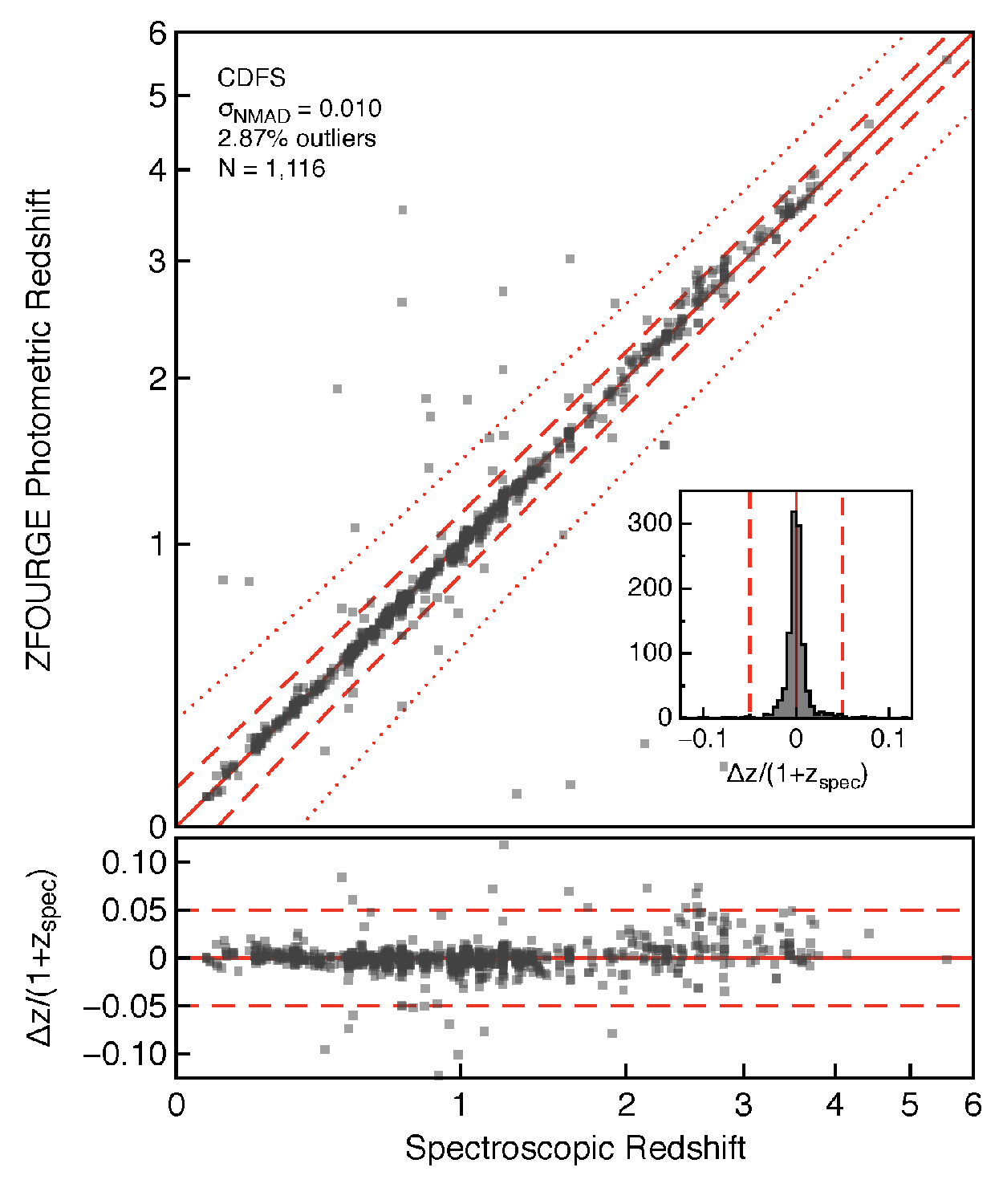}
\caption{Top: Photometric redshifts from ZFOURGE versus spectroscopic redshifts in CDFS. The NMAD scatter, the fraction of objects with $\Delta z/(1+z_{spec})>0.15$, and the number of galaxies with matches in both catalogs are shown in the upper left of the plot, while the histograms of $\Delta z/(1+z_{zspec})$ are shown as an inset in the bottom right of the plot. Bottom: the residual between the photometric and spectroscopic redshifts, divided by $1+{ z }_{ spec }$. The red solid, dashed and dotted lines indicate, respectively, $\Delta z/(1+z_{\mathrm{spec}})=0\pm 0,\ \pm0.05,\mathrm{and\ } \pm0.15$.}
\label{fig:zz0}
\end{figure}

\begin{figure}
\includegraphics[width=0.49\textwidth]{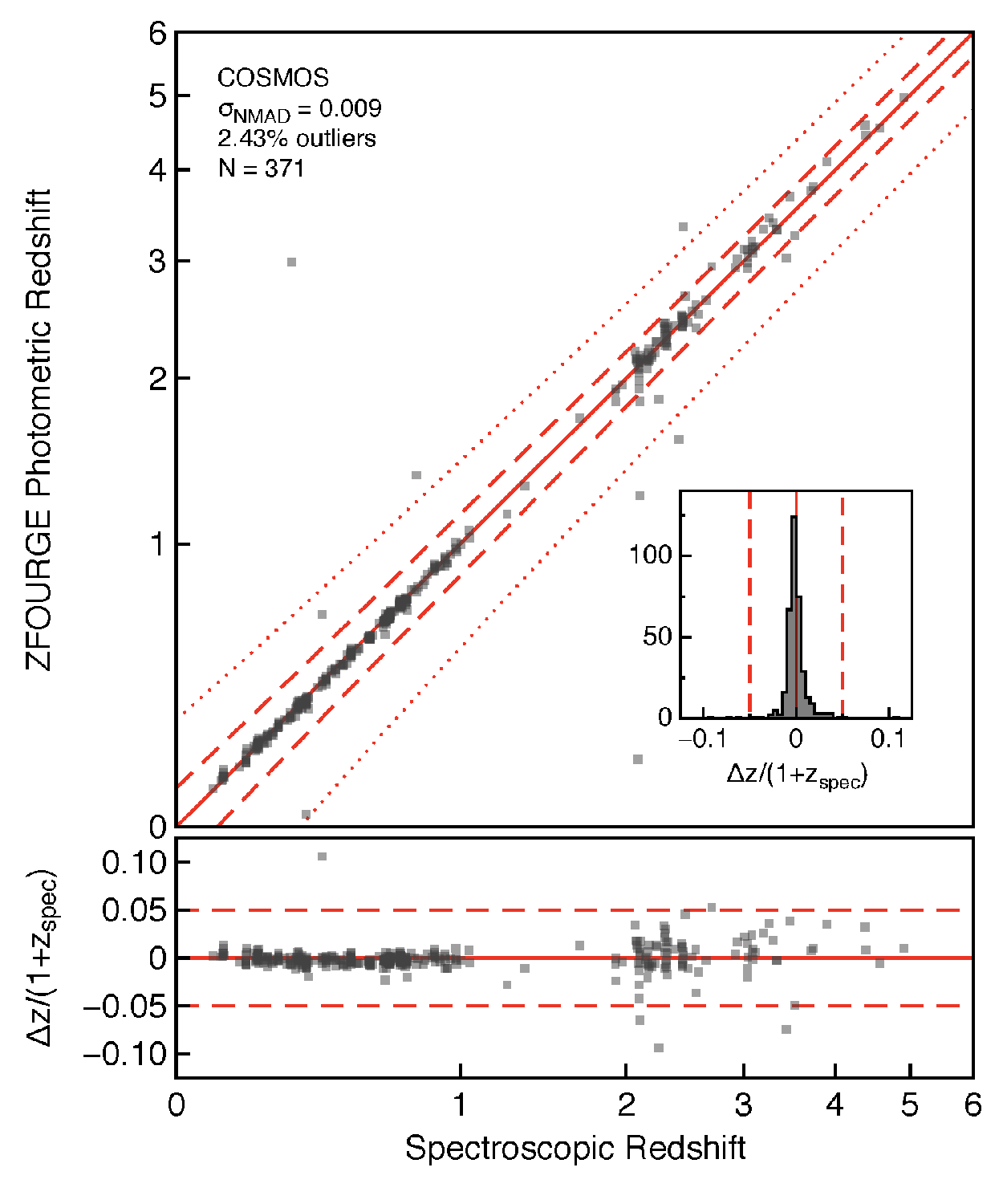}
\caption{Photometric versus spectroscopic redshifts for COSMOS (see caption of Figure \ref{fig:zz0}).}
\label{fig:zz1}
\end{figure}

\begin{figure}
\includegraphics[width=0.49\textwidth]{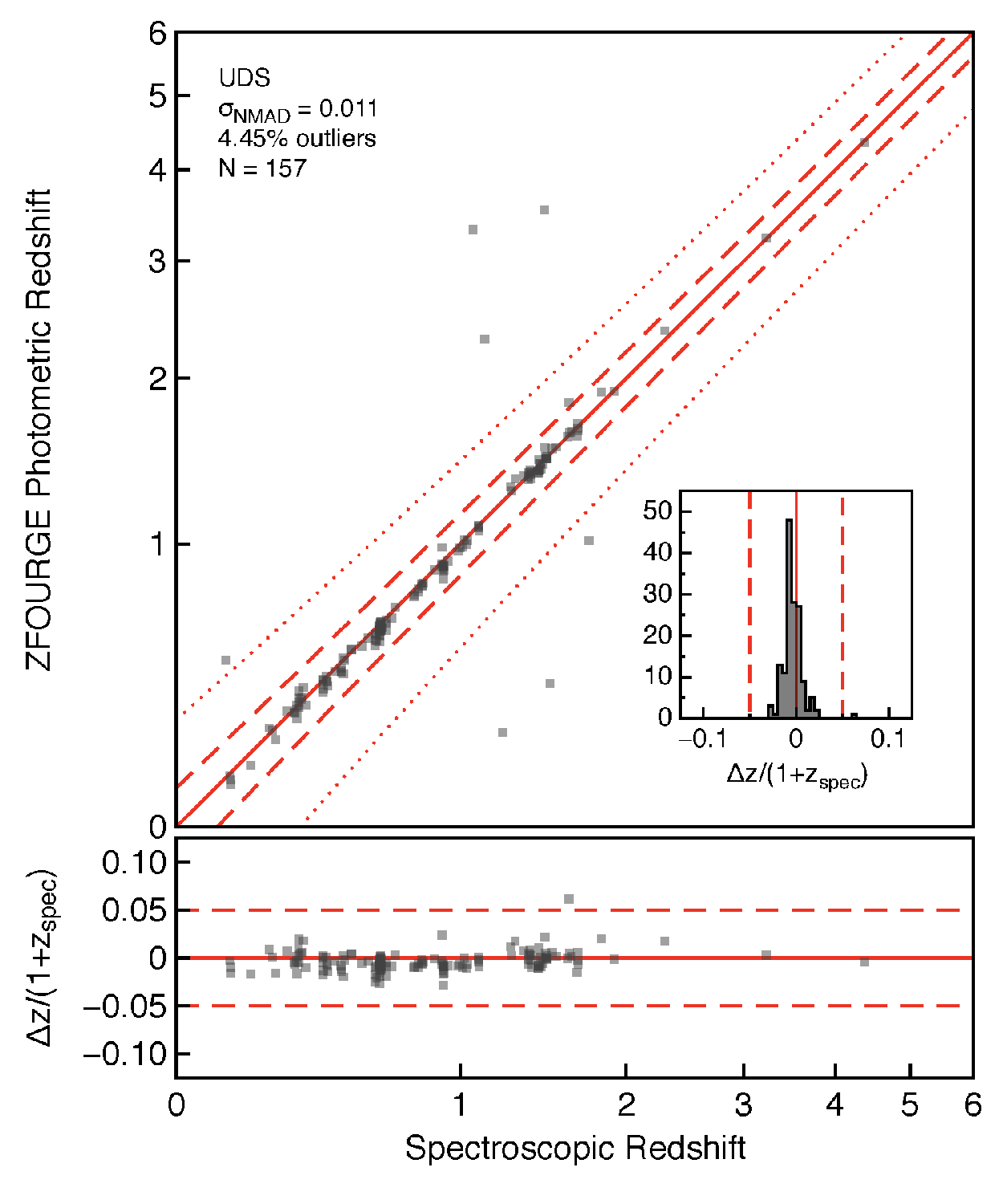}
\caption{Photometric versus spectroscopic redshifts for UDS (see caption of Figure \ref{fig:zz0}).}
\label{fig:zz2}
\end{figure}

A common comparison in the literature is to compare the photometric redshifts with spectroscopic redshifts. In Figures \ref{fig:zz0} to \ref{fig:zz2} we do this, using the compilation of publicly available spectroscopic redshifts in these fields provided by \citet{Skelton14}, with a matching radius of $1\arcsec$. We also included the first release from the MOSDEF survey \citep{Kriek15} and the VIMOS Ultra-Deep Survey \citep{Tasca16}. The overall correspondence is excellent, as indicated by the scatter in the difference between photometric and spectroscopic redshifts. We quantify the errors in the {photometric} redshifts, ${\sigma_{z}}$, using the normalized median absolute deviation (NMAD) of $\Delta z/(1+z)$, i.e., $1.48\times$ the median absolute deviation of  $\left| z_{phot} - z_{spec}\right| / (1+ z_{spec} )$. In CDFS ${\sigma_{z}}={0.010}$, in COSMOS ${\sigma_{z}}={0.009}$ and in UDS ${\sigma_{z}}={0.011}$. Only a small percentage are outliers, with $\Delta z/{ \left( 1+{ z } \right)  }>0.15$. In CDFS 2.9\% are outliers, in COSMOS {2.4}\% and in UDS {4.5}\%. At $z>1.5$ we find ${\sigma_{z}}={0.020}$, ${\sigma_{z}}={0.022}$ and ${\sigma_{z}}={0.013}$ in CDFS, COSMOS and UDS, respectively.

We have also compared with the unpublished redshifts of the ZFIRE survey \citep{Nanayakkara16}, and find an NMAD of ${\sigma_{z}}=\sim0.02$. Full results are shown in \citet{Nanayakkara16}.

\subsection{Redshift pair analysis}\label{sec:pairs}

The drawback of comparing to spectroscopic samples is that these are usually biased towards bright ($K_s<22$) star-forming galaxies, or unusual sources, such as AGN. Therefore these comparisons are not representative of the full photometric catalog and do not allow a careful study of how photometric redshift errors depend on galaxy properties. Here we present an alternative statistical analysis by looking at galaxy pairs. This method was first described and validated by \citet{Quadri10}. It does not rely on spectroscopic information and can be applied to the full catalogs, including faint sources. Therefore this technique provides us with a more representative photometric redshift uncertainty than possible by comparing to spectroscopic redshifts.

Due to clustering, close pairs of galaxies on the sky have a significant probability of being physically associated, and of lying at the same redshift. Other galaxy pairs will actually be chance projections along the line of sight, but this contamination by random pairs can be accounted for statistically, by randomizing the galaxy positions and repeating the analysis. Each true galaxy pair will give an independent estimate of the true redshift, and we can take the mean of the two values as our best estimate of the true redshift. The distribution of $\Delta z_{pairs}/ (1+z_{mean})$ of the pairs of galaxies can then be used to estimate the average photometric redshift uncertainties. It is a narrow distribution for robustly derived redshifts, or broader if the redshifts are very uncertain.

\begin{figure*}
\begin{center}
\includegraphics[width=0.49\textwidth]{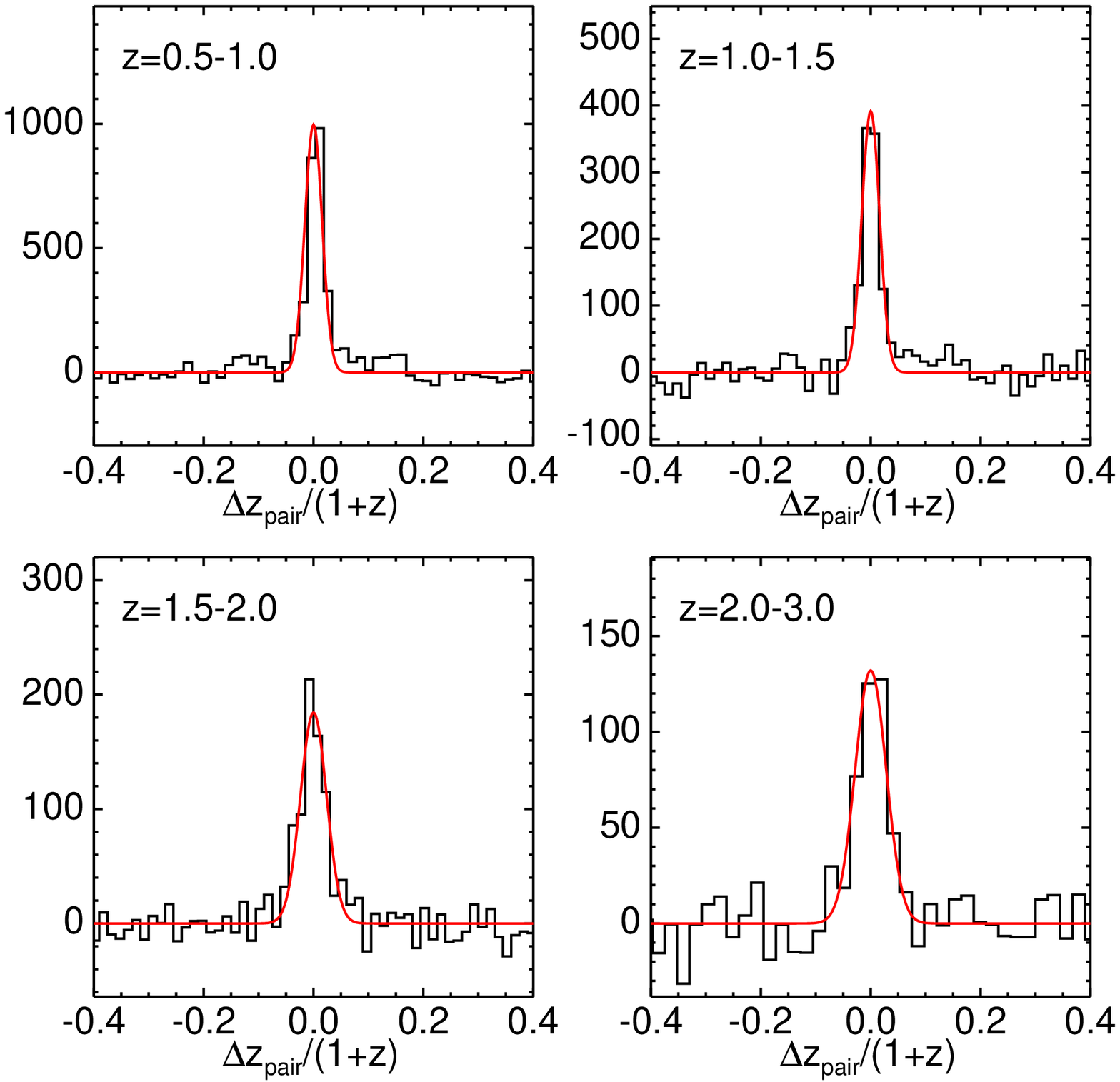}
\includegraphics[width=0.49\textwidth]{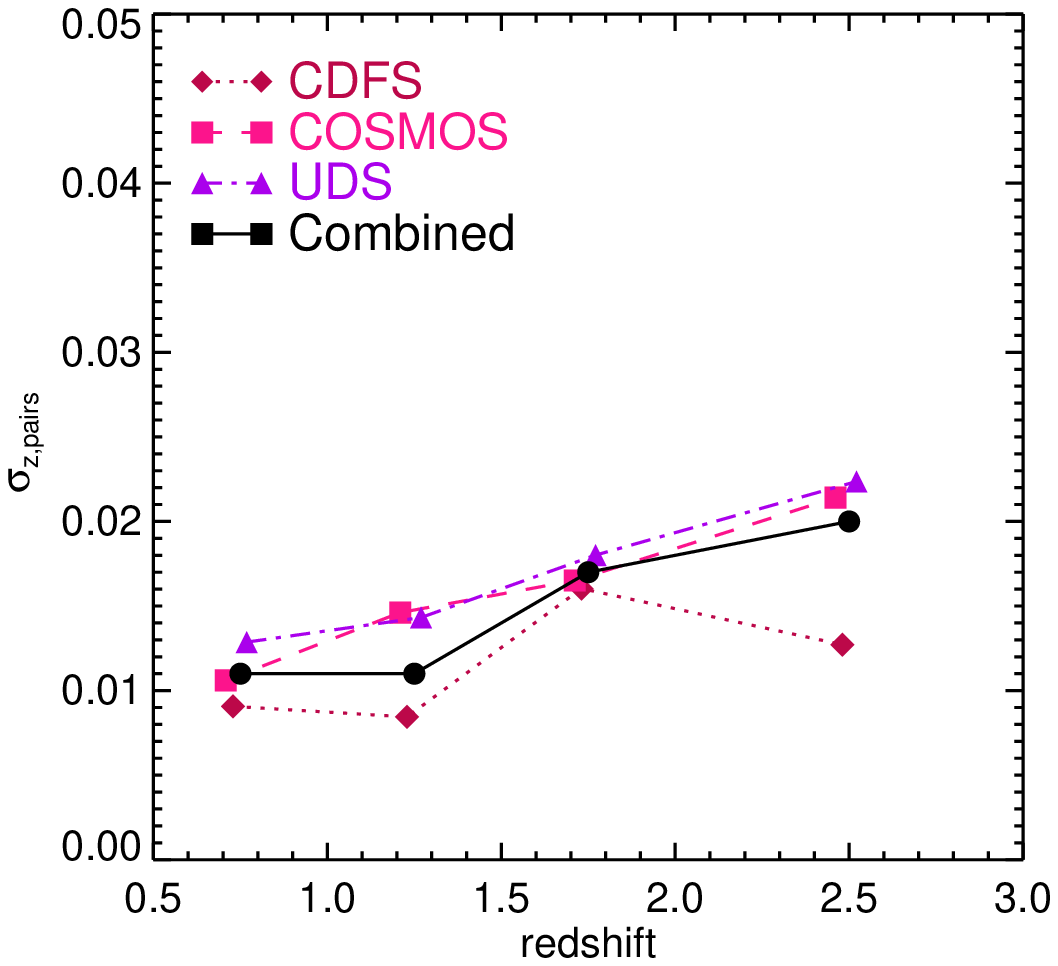}
\caption{{Analysis of photometric redshift accuracy using close pairs of galaxies.} Left: distribution of $\Delta z_{pairs}/ (1+z)$ for galaxies with $K_s<23.5$. We fit Gaussians to each histogram (red lines), from which we derive $\sigma_{z,pairs}$, the average uncertainty for individual galaxies, which is the standard deviation of the pair distribution divided by $\sqrt{2}(1+z_{mean})$. Right: $\sigma_{z,pairs}$ as determined in the left panel as a function of redshift (black solid line). We also show the results for the individual fields, as indicated in the legend. $\sigma_{z,pairs}$ increases with redshift, from $\sigma_{z,pairs}=0.01$ to $0.02$.}
\label{fig:zpairs1}
\end{center}
\end{figure*}

For illustration, we show the distributions of $\Delta z_{pairs}/ (1+z_{mean})$ in the left panel of Figure \ref{fig:zpairs1}, for pairs of galaxies with {\tt use=1} and total $K_s$-band magnitude $<23.5$, in four redshift bins. The pairs have angular separations between $2\farcs5$ and $15\ \arcsec$. To each distribution we fit a Gaussian and determined the standard deviation. As this is the standard deviation for the redshift differences, we divide by $\sqrt{2}$ to obtain the average redshift uncertainty for individual galaxies, $\sigma_{z,pairs}$, for a particular redshift bin, i.e., $\sigma_{z,pairs}$ is obtained from $\Delta z_{pairs}/(\sqrt{2} (1+z_{mean}))$. In the right panel we show $\sigma_{z,pairs}$ as a function of redshift. $\sigma_{z,pairs}$ increases with redshift, but in general is excellent: varying from 1\% to 2\% going from $z=0.5$ to $z=2.5$. Calculating $\sigma_{z,pairs}$ requires fairly large samples. This partly explains the scatter between results on individual ZFOURGE fields. Other reasons for differences between the fields are different image filter sets and image depths.

{An analysis of $\sigma_{z,pairs}$ can be affected by systematic errors in the photometric redshifts, leading to underestimates of the true redshift uncertainty. }
For example, because of systematic photometric errors, many sources could be fit with similiar, but wrong, redshifts. This is discussed in more detail by \citet{Quadri10}. {Furthermore, it is important to keep in mind that if all redshifts are systematically overestimated or underestimated, this will not be detected by this method.} We tested this scenario by inspecting pairs with at least one spectroscopic redshift available. We derived similar results, indicating that $\sigma_{z,pairs}$ for photometric pairs is not systematically affected. {The three fields in the survey also provide constraints, as we make use of different filtersets and systematics introduced between different filters will not be the same in each field. However, if systematics are introduced due to a particular choice of template, this is likely to go unnoticed.}

\begin{figure*}
\includegraphics[width=0.49\textwidth]{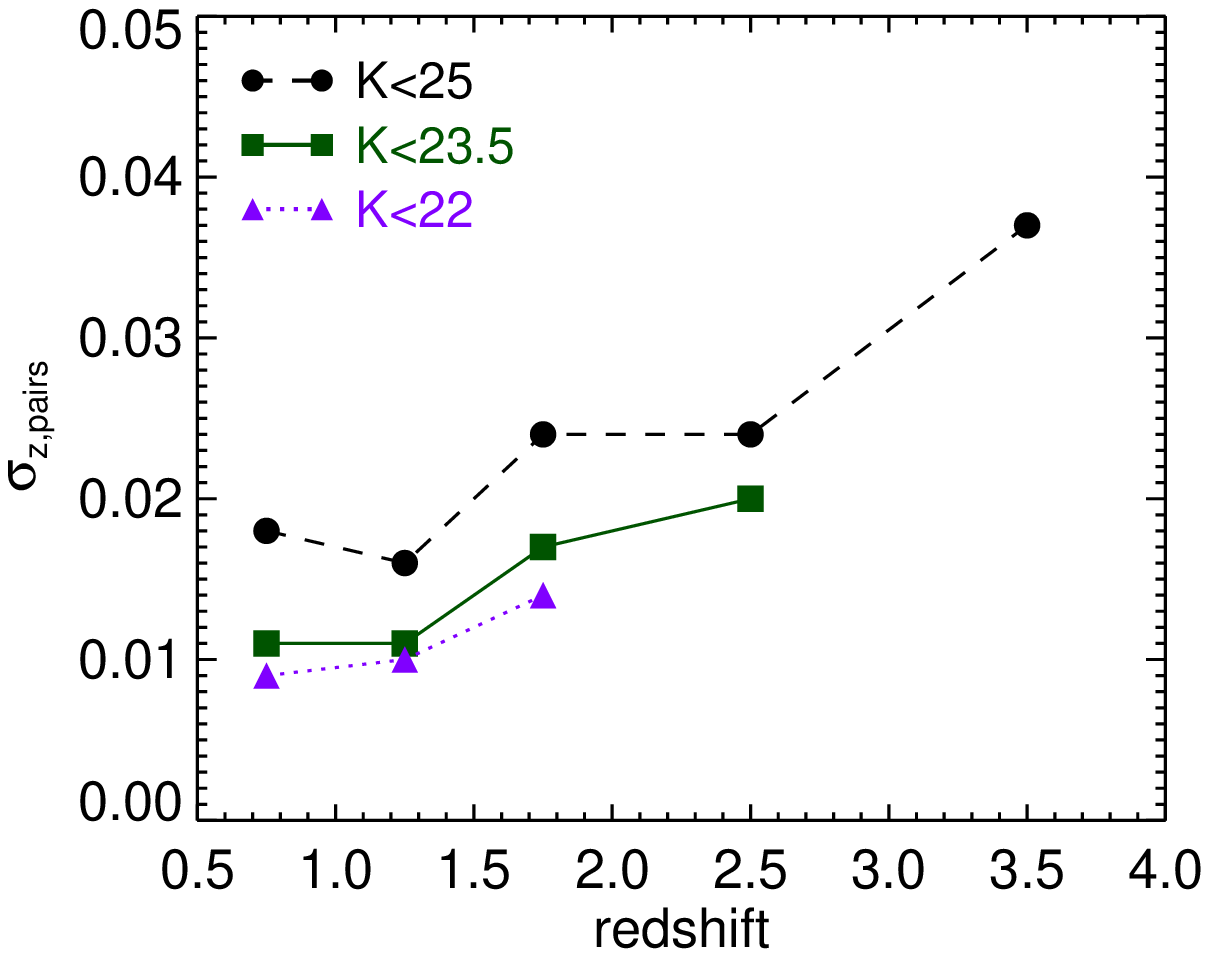}
\includegraphics[width=0.49\textwidth]{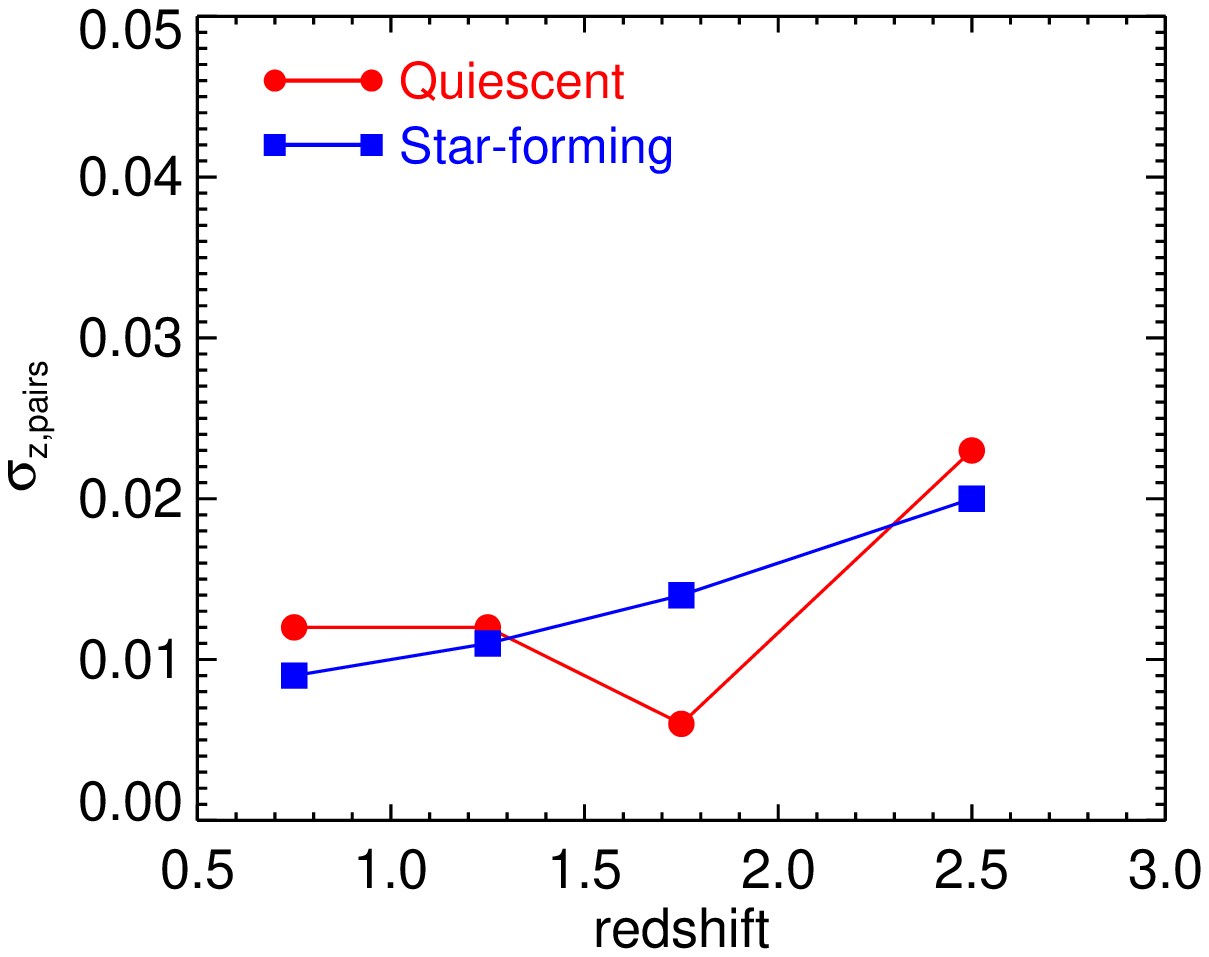}\\
\includegraphics[width=0.49\textwidth]{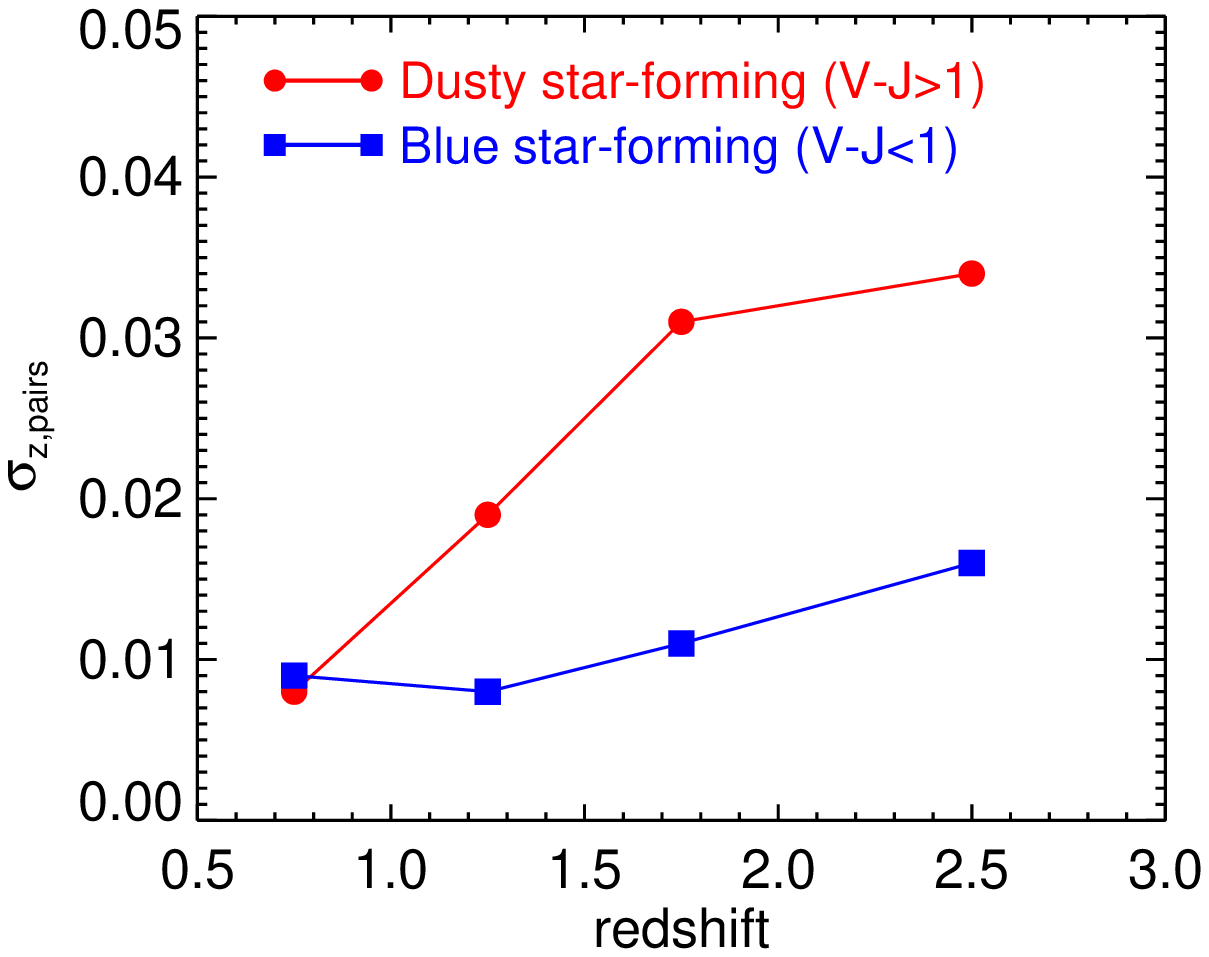}
\includegraphics[width=0.49\textwidth]{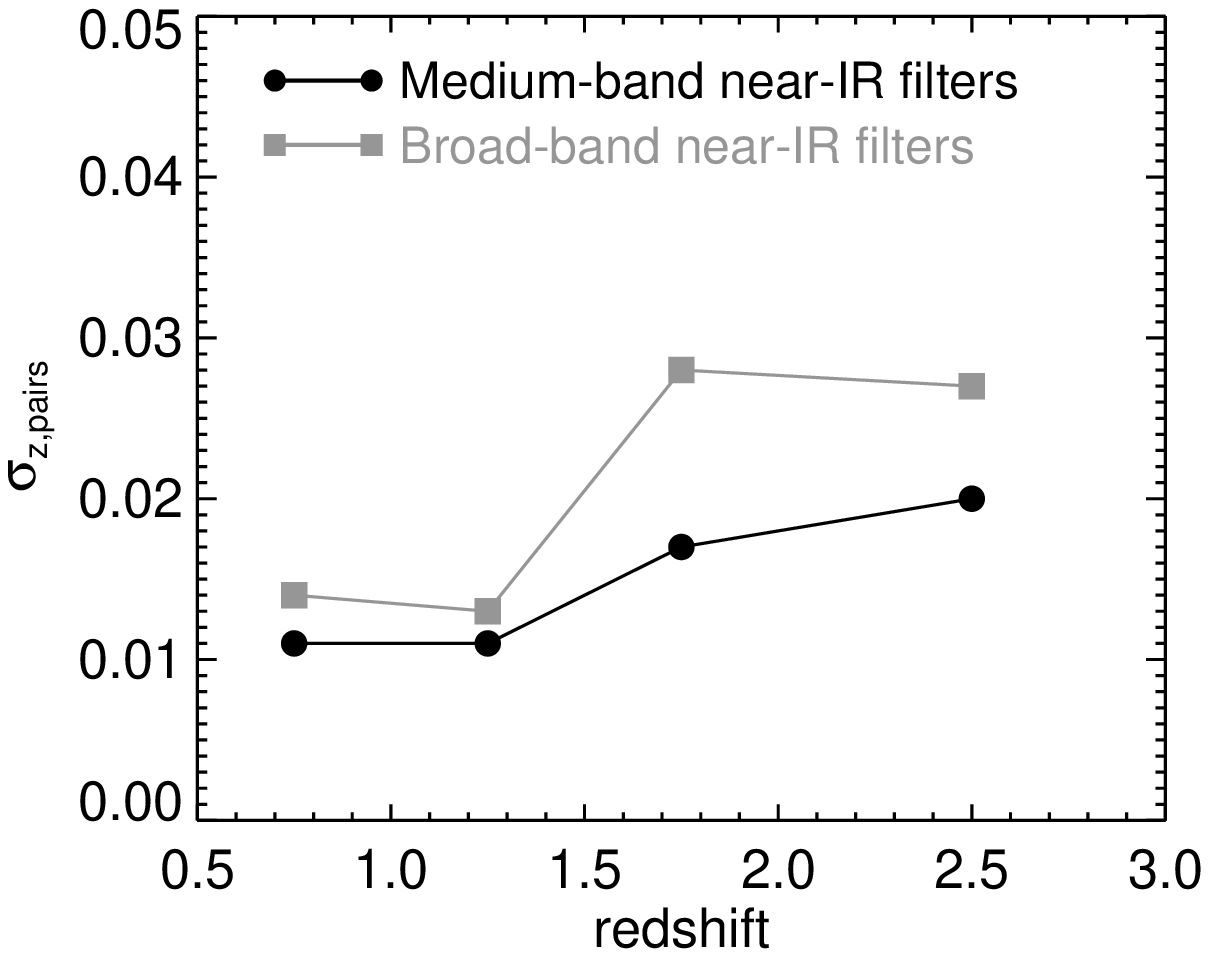}
\caption{$\sigma_{z,pairs}$ versus redshift, investigating trends with magnitude limits (top-left) or SED types (second and third panels), and investigating the effect of using the \fs\ filters (top-right). $\sigma_{z,pairs}$ tends to be smaller for brighter galaxies are considered, and for blue star-forming galaxies. $\sigma_{z,pairs}$ is clearly smaller if the near-IR medium band filters are used (compared to near-IR broadband), especially at $1.5 < z < 2.5$.}
\label{fig:zpairs2}
\end{figure*}

We expect $\sigma_{z,pairs}$ to be sensitive to various parameters, including the type of galaxy, the magnitude and redshift. As a first exploration we will here characterize how our photometric redshift uncertainty depends on these parameters. In the first panel of Figure \ref{fig:zpairs2} we show $\sigma_{z,pairs}$ versus $z$ for three different magnitude-limited samples, with $K_s<22$, $K_s<23.5$ (as above) and $K_s<25$. For the brightest galaxies, with $K_s < 22$, the uncertainty is very small, around 1\% up to $z=2$.  However, the uncertainty increases by roughly a factor towards fainter magnitudes up to $K\sim25$, which is near our completeness limit.

We have additonally investigated the dependence of $\sigma_{z,pairs}$ on galaxy type, using the same UVJ selected samples of quiescent, red star-forming and blue-starforming galaxies as in Section \ref{sec:pz}. The results are shown in the second and third panels of Figure \ref{fig:zpairs2}. Interestingly, the photometric redshifts of star-forming galaxies and quiescent galaxies are equally well constrained at most redshifts.  The exception occurs at intermediate redshift ($1.5 < z < 2$), where instead we find much smaller redshift uncertainties for quiescent galaxies. This is the redshift range where the \b4\ is moving through the $J_1$, $J_2$ and $J_3$ medium-bandwidth filters. In contrast, \citet{Quadri10} used shallower broadband photometry - with fewer optical filters - and found that quiescent galaxies have significantly better photometric redshifts at all redshifts. This emphasizes that the characteristics of photometric redshifts are dataset-dependent.

Comparing blue and red (dusty) star forming galaxies, we find that red galaxies have a factor $2-3$ worse $\sigma_{z,pairs}$, than do blue galaxies,{but these uncertainties are still small: $3-3.5\%$ at $z>2$.} The redshifts of these galaxies are {more} difficult to constrain, even with medium band photometry, as they have relatively featureless SEDs, and a degeneracy between redshift and the color of the reddest template allowed in the EAZY set \citep[e.g.,][]{Marchesini10,Spitler14}. Here we have split the sample at rest-frame $V-J=1$, but the effect will be stronger for dustier galaxies at redder $V-J$. This is a significant {issue} for star-forming galaxies with high mass or high SFRs, which often tend to be quite dusty. 

In the last panel of Figure \ref{fig:zpairs2} we compare $\sigma_{z,pairs}$ for the case where we have not {included} the near-IR medium band \fs\ filters in the EAZY fits {(but note that two of our three fields still include medium band filters in the optical)}. For the entire range considered here, the photometric redshifts are better derived if we do use the \fs\ medium bands. The effect is strongest at $z = 1.5 - 2.5$ where $\sigma_{z,pairs}$ using the \fs\ medium bands is $\sim50$\% smaller compared to $\sigma_{z,pairs}$ with the \fs\ filters removed. This confirms the efficacy of the {near-IR} medium bands at intermediate to high redshift.

\begin{figure*}
\includegraphics[width=0.49\textwidth]{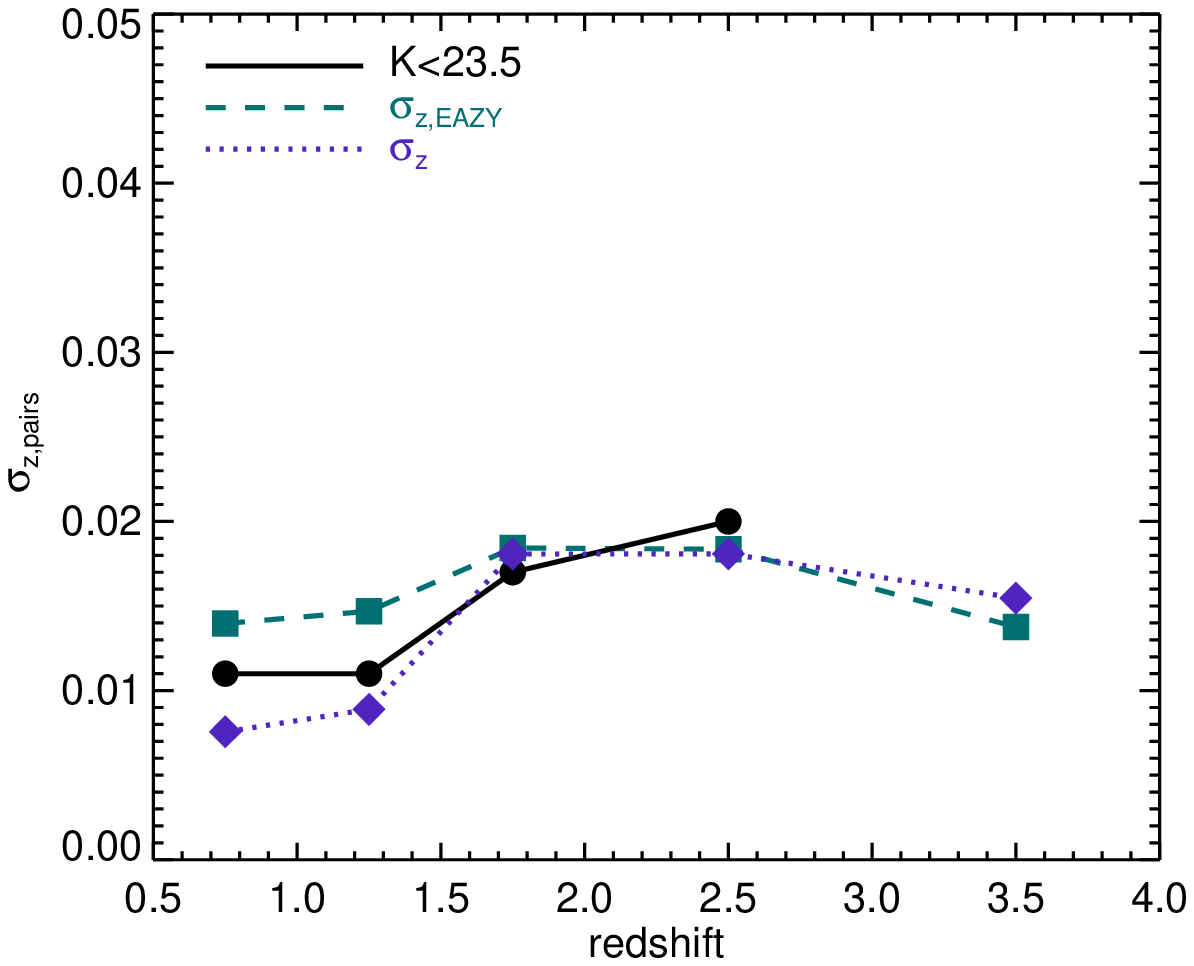}
\includegraphics[width=0.49\textwidth]{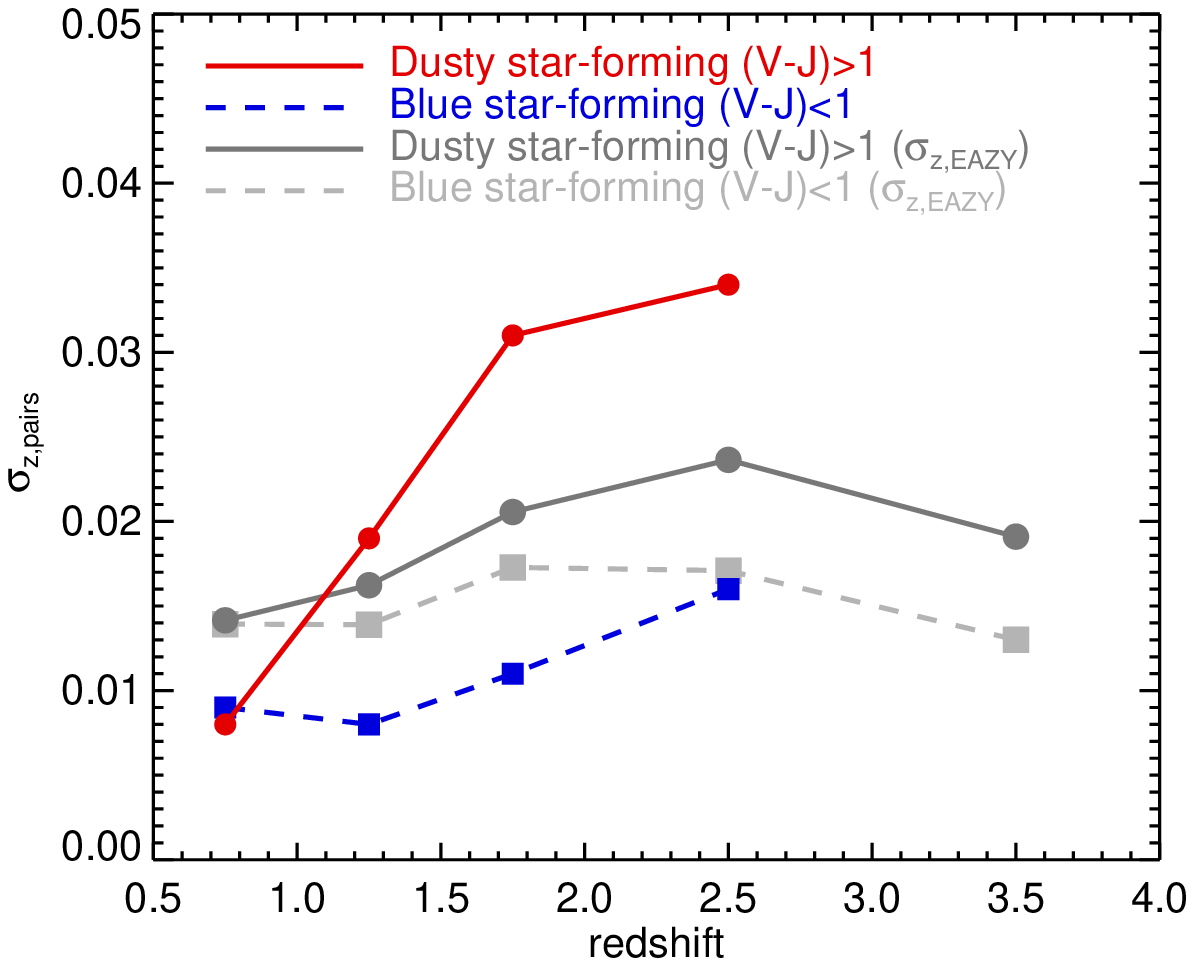}
\caption{Left: comparing the three different redshift quality tests, ${\sigma_{z}}$ from testing against spectroscopic samples, $\sigma_{z,pairs}$ from the redshift pair test, and  the photometric redshift uncertainty from the $p(z)$ from EAZY, $\sigma_{z,EAZY}$. These correspond well at all redshifts. Right: comparing $\sigma_{z,pairs}$ with $\sigma_{z,EAZY}$. $\sigma_{z,EAZY}$ tends to be underestimated for red star-forming galaxies.}
\label{fig:zmethod}
\end{figure*}

The pairs analysis also provides an interesting way to verify whether the redshift uncertainties that come from the EAZY template fits are reasonable. In the left panel of Figure \ref{fig:zmethod} we compare $\sigma_{z,pairs}$ to $\sigma_{z,EAZY}$, and find that they provide heartening agreement. This figure also shows that ${\sigma_{z}}$, the uncertainty estimated from comparing the photometric to the spectroscopic redshifts, provides a good estimate of the true uncertainties for the $K<23.5$ sample.

Although the redshift uncertainties estimated by EAZY appear quite reliable for the general population of galaxies, we find that they are {underestimated by a factor of 2} for dusty star-forming galaxies. In the right panel of Figure \ref{fig:zmethod} we compare $\sigma_{z,pairs}$ to $\sigma_{z,EAZY}$ from the EAZY fits. The pair redshifts of blue star-forming galaxies are better than we expect from the EAZY $p(z)$. However, for red and dusty star-forming galaxies the pair redshifts are 50\% worse than the EAZY $p(z)$. This effect increases with redshift and towards fainter magnitudes (not shown here). It  indicates that with current methods and state-of-the-art surveys, the degeneracy between rest-frame color and redshift for dusty galaxies cannot yet be accurately resolved, and we caution that photometric redshift uncertainties for faint dusty galaxies at $z>1.5$ are generally underestimated.

\subsection{Redshift distributions}

\begin{figure*}
\includegraphics[width=\textwidth]{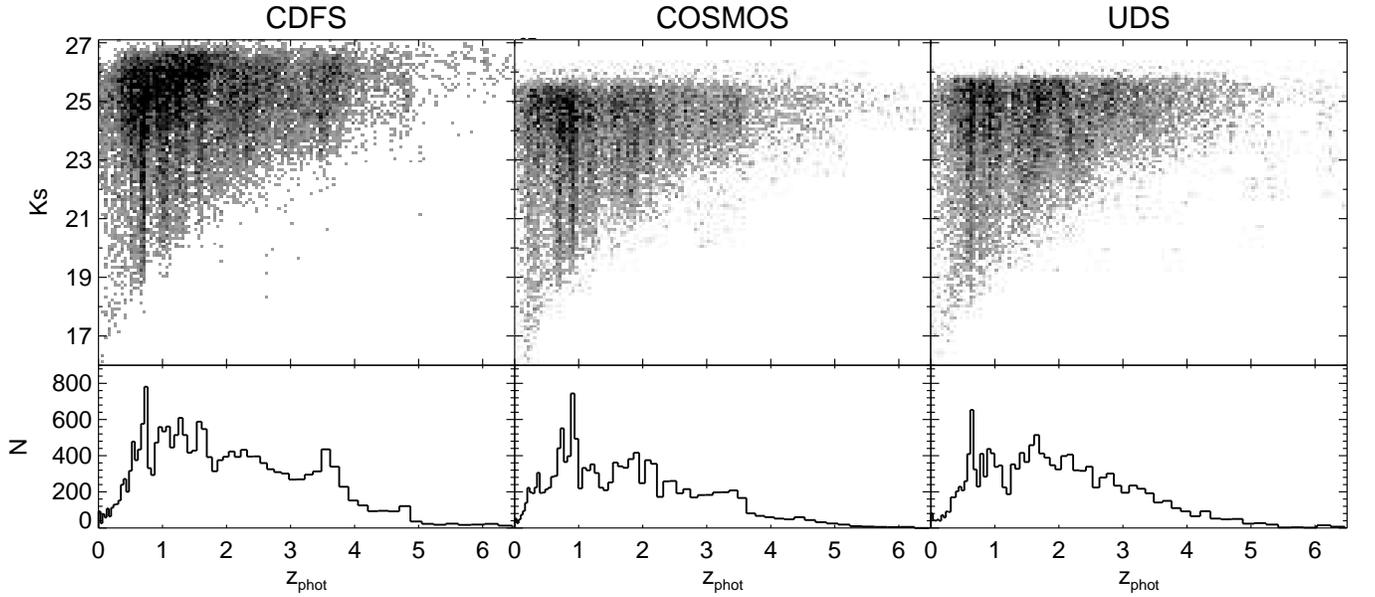}
\caption{Top panels: distribution of $K_s$-band magnitudes as function of redshift. The grayscale indicates the density in each point, with darker colors for higher densities. Bottom panels: photometric redshift ({\tt z\_peak}) distribution. Higher density peaks are clearly visible, for example the ZFOURGE identified cluster at $z=2.095$ in COSMOS \citep{Spitler12,Yuan14}.}
\label{fig:zk}
\end{figure*}

By improving the accuracy of the photometric redshifts, we can derive improved stellar masses and start identifying large scale structure. In Figure \ref{fig:zk} we plot the $K_s$-band magnitudes as function of {\tt z\_peak} (or $Z_{spec}$ where available). {The ZFOURGE redshift distributions for each field reveal density peaks} corresponding to known overdensities, e.g., at $z<1$ in COSMOS \citep[e.g.,][]{Kovac10,Knobel12}. {These include an overdensity at $z\sim2.1$, identified by \citet{Spitler12} using ZFOURGE photometric redshifts. This overdensity was spectroscopically confirmed at a redshift of $z=2.095$, with $\sigma_z/(1+z)=\sim2\%$ \citep{Yuan14}.}

\section{Stellar masses and star-formation rates}\label{sec:fast}

\begin{figure*}
\includegraphics[width=\textwidth]{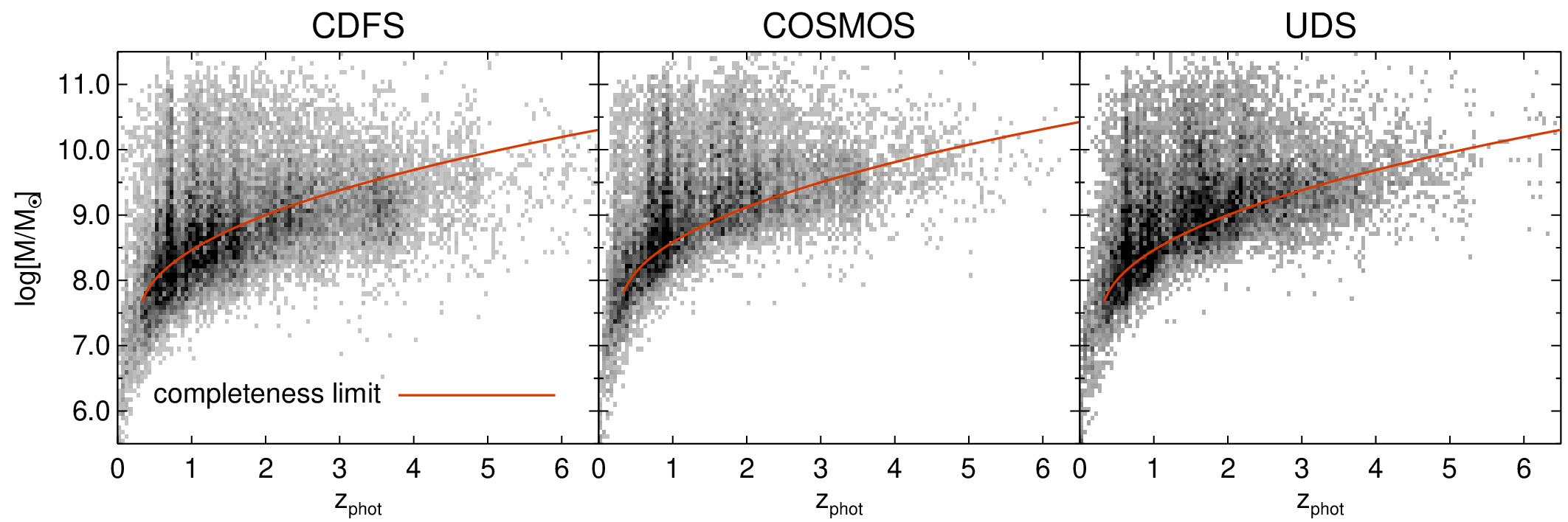}
\caption{Similar to the top panels of Figure \ref{fig:zk}, but with stellar mass instead of $K_s$-band magnitude. The solid red line indicates the 80\% mass completeness limit in each field.}
\label{fig:zm}
\end{figure*}

We estimated 80\% mass completeness limits using a method similar to \citet{Quadri12} \citep[see also][]{Marchesini09}. We selected galaxies within the range $K_s=24.0-24.7$ mag and scaling their fluxes to $K_s=25.0$ mag. Then we determined the 80th-percentile mass rank in narrow redshift bins. Galaxies above this value are the most massive objects that could plausibly fall below the $K_s$-band selection limit. A smooth function to these values is shown in Figure \ref{fig:zm}, as function of redshift. At $z=2$ we reach a completeness limit of $\sim 10^9$ \msol, and at $z=4$ we are complete above $\sim10^{9.5}$\msol. Beyond $z=4$ the completeness limit is extrapolated.

\begin{table}
\begin{center}
\caption{Explanation of the FAST stellar population catalog header}
\label{tab:fastheader}
\begin{threeparttable}
\begin{tabular}{l l}
\hline
\hline
id&			ID number \\
z&      		=z\_peak (or z\_spec if available) \\
ltau&     		log[tau/yr] \\
metal&      		metallicity (fixed to 0.020)\\
lage&        		log[age/yr]\\
Av&     		dust reddening\\
lmass&      		log[$M/M_{\Sun}$]\\
lsfr&     		log[SFR/($M_{\Sun}$/yr)]\\
lssfr&      		log[sSFR(/yr)]\\
la2t&      		log[age/$\tau$]\\
chi2&			minimum $\chi^2$\\
\hline
\end{tabular}
\end{threeparttable}
\end{center}
\end{table}

Stellar population properties (stellar mass, SFR, dust extinction, and age) were derived by fitting \cite{Bruzual03} models with FAST \citep{Kriek09}, assuming a \cite{Chabrier03} initial mass function, exponentially declining star formation histories with timescale $\tau$, solar metallicity and a dust law as described in \cite{Calzetti00}. For each source the redshift is fixed to the photometric redshift ({\tt z\_peak}) derived with EAZY, or the spectroscopic redshift if known. We limit dust extinction to $0\leq A_V\leq 4$, age to $7.5\leq log_{10}(\mathrm{age})\leq 10.1$ Gyr and $\tau$ to $7\leq \tau\leq 11$ Gyr. We provide the full FAST stellar population catalogs. See Table \ref{tab:fastheader} for an explanation of the catalog header.

\begin{table}
\caption{Explanation of the SFR catalog header}
\label{tab:sfrheader}
\begin{center}
\begin{threeparttable}
\begin{tabular}{l l}
\hline
\hline
id&			ID number \\
z &	phometric redshift (or spectroscopic redshift if available) \\
f\_24 & $Spitzer$/MIPS $24\micron$ flux (mJy) \\
e\_24 & $Spitzer$/MIPS $24\micron$ flux error (mJy) \\
f\_100\tnote{a} & Herschel/PACS $100\micron$ flux (mJy) \\
e\_100\tnote{a} & Herschel/PACS $100\micron$ flux error (mJy) \\
f\_160\tnote{a} & Herschel/PACS $160\micron$ flux (mJy) \\
e\_160\tnote{a} & Herschel/PACS $160\micron$ flux error (mJy) \\
L\_IR & total integrated IR luminosity \lsol \\
L\_UV & total UV luminosity \lsol \\
SFR & star formation rate (Equation \ref{eq:sfr}) \\
\hline
\end{tabular}
\begin{tablenotes}
\item[a] Herschel/PACS data only available in CDFS.
\end{tablenotes}
\end{threeparttable}
\end{center}    
\end{table}

SFRs, dust attenuations, ages and star formation histories of galaxies derived from SED fitting to UV, optical and near-IR photometry may be uncertain, especially if galaxies are highly dust-obscured. A different estimate of the SFRs can be obtained by inferring the total infrared luminosity (\lir\ $\equiv$ \lirtot) of galaxies and combining this with the luminosity emitted in the UV ($L_{UV}$ at rest-frame $2800\mathrm{\AA}$). $L_{UV}+L_{IR}$ provides an estimate of the total bolometric luminosity, which can be converted to SFR under the assumption that the galaxy is continuously forming stars \citep{Kennicutt98,Bell05}. In addition to the FAST catalogs, we provide catalogs with the net observed $L_{UV}+L_{IR}$ SFRs (see Table \ref{tab:sfrheader} for a description). 

We use the conversion from \citet{Bell05} to calculate SFRs from our data, scaled to a \citet{Chabrier03} IMF, 

\begin{equation}
SFR \: [M_{\odot} / \mathrm{yr}]   =   1.09 \times 10^{-10} \: ( \mathrm{L}_{\mathrm{IR}} +  2.2 \mathrm{L}_{\mathrm{UV}} )
\label{eq:sfr}
\end{equation}

\noindent
To derive $L_{IR}$ we use our extracted $24-160\micron$ photometry (Section \ref{sec:farir}), to which we fit a model spectral template to calculate the total luminosity. The model template is the averaged template from \citet{Wuyts08} (hereafter W08), generated by averaging the logarithm of the templates from the library of \citet{Dale02}. The motivation of this approach was to introduce a simple conversion of flux to luminosity, first proposed by W08 and later validated by \citet{Wuyts11} and \citet{Tomczak16}. So far public Herschel/PACS data only exists or the CDFS field. For the other fields we fitted the W08 template only to the $Spitzer$/MIPS/$24\micron$ photometry.

Total IR luminosities are then obtained by integrating these fits between $8-1000$\um\ in the \rf. \luv $= 1.5 \nu \mathrm{L}_{\nu, 2800}$ is the estimated rest-frame $1216-3000$\AA\ UV luminosity, that we derived with EAZY. Both \lir\ and \luv\ are in units of \lsol.
This conversion assumes that the total IR luminosity reflects the amount of obscured UV light from young stellar populations. Thus by adding its contribution to that of the unobscured UV luminosity (\luv) the net star-formation rate for galaxies can be measured.

\section{First validation of the UVJ diagram at $z=3$}\label{sec:uvj}

\begin{figure*}
\begin{center}
\includegraphics[width=\textwidth]{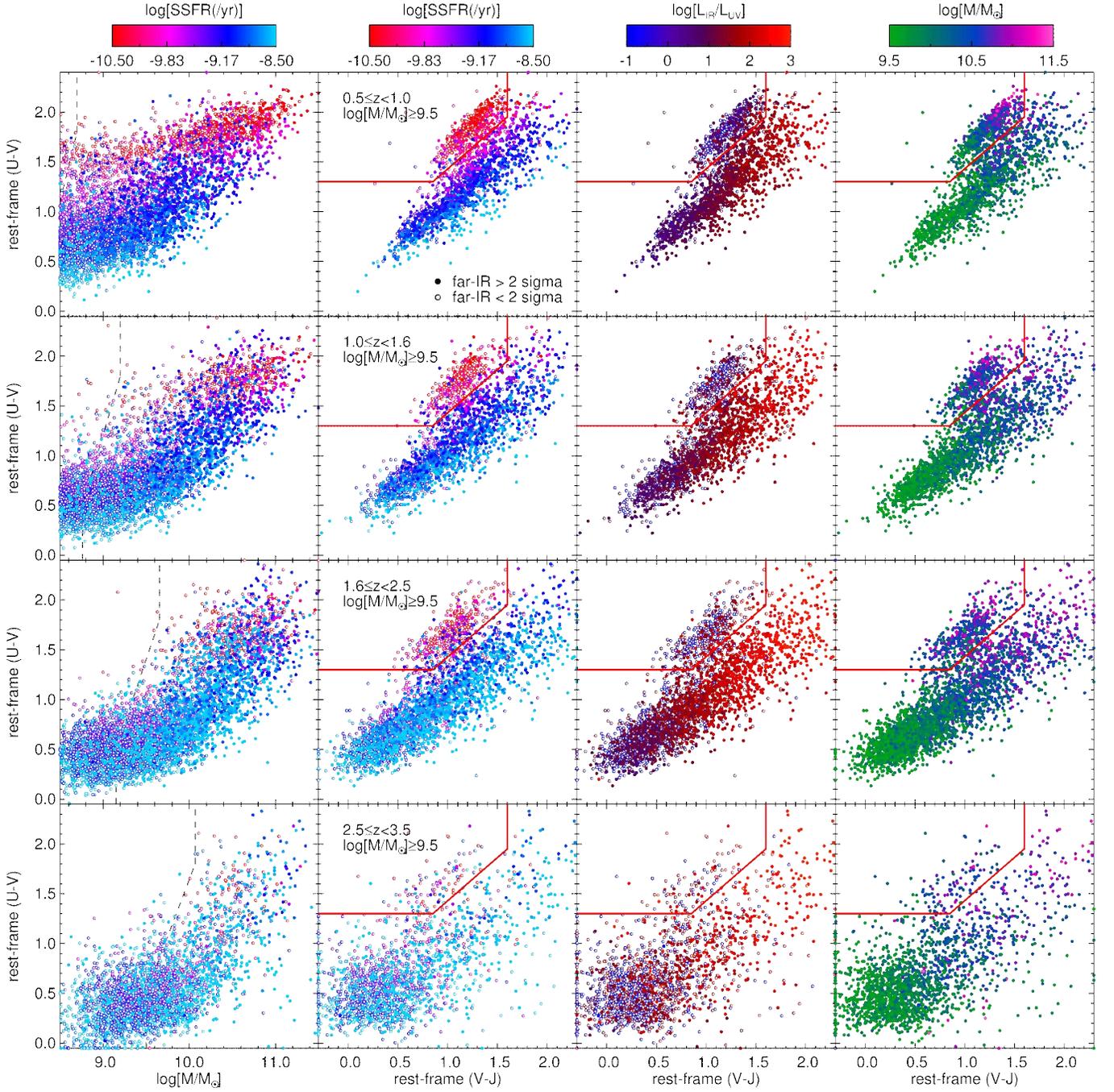}
\end{center}
\caption{Rest-frame $U-V$ versus stellar mass (left columns) or versus $V-J$ (second, third and fourth columns) for a mass complete sample of galaxies with {\tt use=1}, $SNR_{K_s}>10$ and stellar mass $M>10^{9.5}M_{\Sun}$, in four redshift bins (top to bottom). The vertical dashed lines indicate our stellar mass completeness limit. The red solid line in the UVJ diagrams separates quiescent (top left) from star forming (bottom left to top right) galaxies. Galaxies that are undetected in the far-IR at $<2\sigma$ are shown with open symbols. The sSFRs {decrease} towards blue $V-J$ and red $U-V$ colors, reflecting a gradient in age {(see also Figure \ref{fig:age})}. Galaxies with the lowest sSFRs are located in the quiescent region of the diagram, with red $U-V$ and blue $V-J$ colors. Galaxies span a large range in $log[L_{IR}/L_{UV}]$, ranging from $-1$ for the bluest UVJ star-forming galaxies to $3$ for the dustiest sources, and quiescent galaxies having low $log[L_{IR}/L_{UV}]$. A mass sequence is also visible, where massive galaxies tend to be redder.}
\label{fig:uvj}
\vspace{20pt}
\end{figure*}

{Using EAZY, we derived various rest-frame colors, for example in the Johnson/U and V-bands \citep{MaizAppelaniz06}, in the J-filter from the Two Micron All Sky Survey, and at $2800\mathrm{\AA}$ (using a tophat shaped transmission curve). Rest-frame colors were calculated by integrating the redshifted rest-frame filter bandpasses of the best-fit template for each individual source.  The process is described in more detail by \cite{Brammer11}, see also \cite{Whitaker11}. Rest-frame $2800\mathrm{\AA}$ luminosity can be used as a proxy of the UV luminosity of a galaxy, which in turn can be used to derive the unobscured SFR (see Section \ref{sec:fast}).} Rest-frame $U-V$ is historically used to distinguish between blue, late-type galaxies with active star formation and red early-type galaxies with low star formation. At low redshift these quiescent ``red and dead'' galaxies dominate at high mass and form a tight ``red sequence'' versus luminosity or stellar mass \citep[e.g.,][]{Tully82,Baldry04}. In Figure \ref{fig:uvj} (left column) we present the rest-frame $U-V$ versus stellar mass relation of our mass complete sample, color coded by specific star formation rate (sSFR$=$SFR/M*, with M* stellar mass).

\begin{figure*}
\begin{center}
\includegraphics[width=\textwidth]{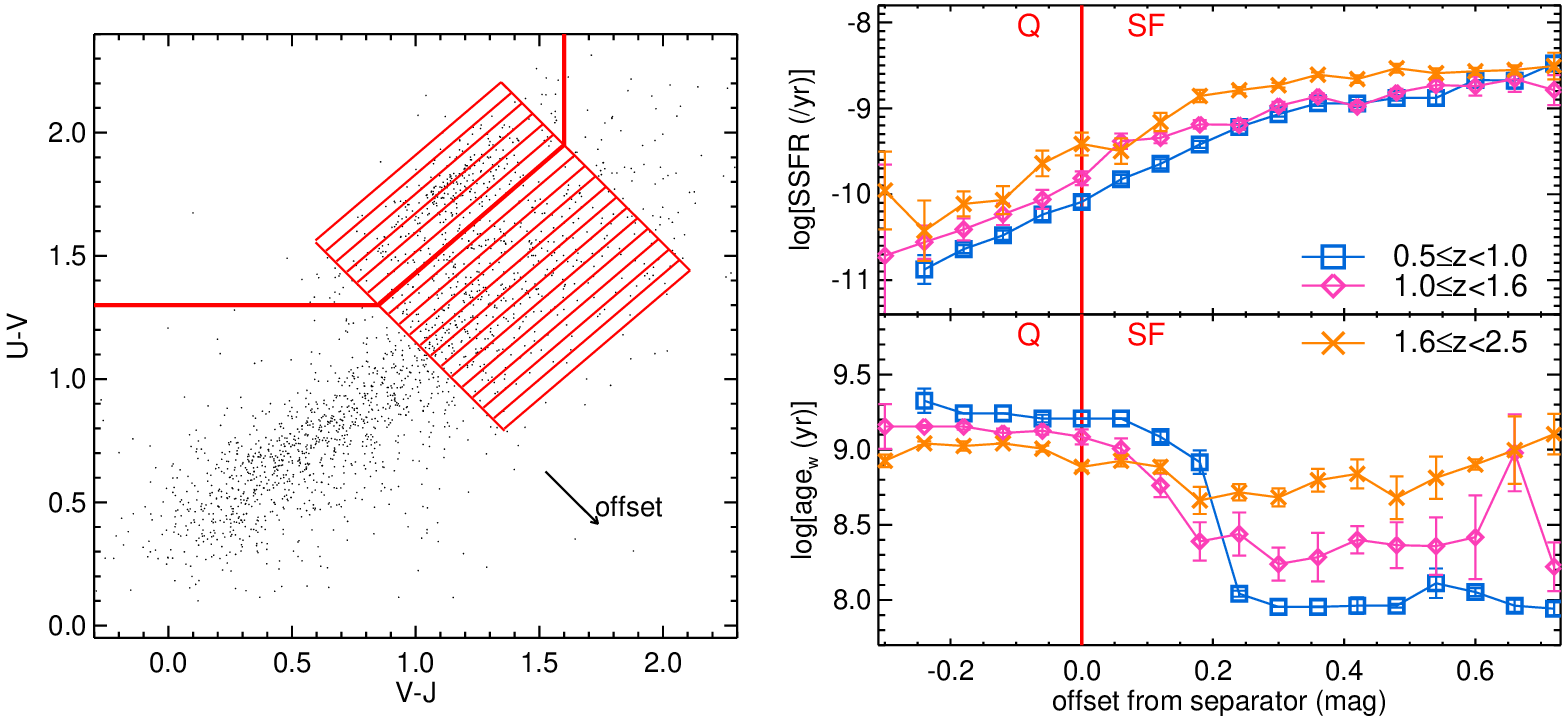}
\end{center}
\caption{{Left: Bins in UVJ space, set orthogonally to the red line that separates quiescent from star-forming galaxies. In these bins we measure the bootstrapped median SSFR and weighted age: age$_w$. The underlying datapoints are a randomly chosen subsample from Figure \ref{fig:uvj} to illustrate the general locus of galaxies in the diagram. Right: Gradient in SSFR (top) and age$_w$ (bottom) with increasing distance from the separator. Datapoints on the left of the red line represent quiescent galaxies (``Q'') and datapoints on the right star-forming galaxies (``SF''). SSFR increases towards bluer $U-V$ and $V-J$, while age$_w$ decreases.}}
\label{fig:age}
\end{figure*}

Clearly, at high redshift there are also large numbers of red massive galaxies with active star formation. These galaxies are red due to dust attenuation \citep[see e.g.,][]{Brammer09}, so a single $U-V$ color cannot be used anymore to separate quiescent and star forming galaxies.  We therefore use a two-color diagram, rest-frame $U-V$ versus rest-frame $V-J$ (hereafter UVJ), to efficiently separate quiescent from star forming galaxies \citep[see e.g.,][]{Labbe05,Wuyts07,Williams09}. In Figure \ref{fig:uvj} (second column from the left), we show the UVJ diagram for the same galaxies. Red star forming galaxies are now well separated from red quiescent galaxies, and a clear red sequence in UVJ is present up to $z = 3.5$. {At this high redshift a bimodality in color space was also found by \citet[e.g.][]{Whitaker11}. Here we aim to validate the UVJ selection criterion for such high redshifts by inspecting the sSFRs based on Herschel data.}

Several trends with UVJ color can be seen in the data. First, specific star formation rates show a gradient in color space, such that redder $U-V$ and bluer $V-J$ colors correspond to lower sSFR. This can be interpreted as a stellar ``age" gradient. We inspected the stellar ages derived from SED fitting for these galaxies {in bins towards redder $V-J$ and bluer $U-V$, shown in Figure \ref{fig:age}. Here the ages are weighted by the timescale $\tau$ from the exponentially declining star-formation histories in the models to better represent the age of the bulk of the stellar mass in the galaxy (see also \citet{NFS04}).} Interestingly, this gradient can even been seen amongst star-forming galaxies {with the highest sSFRS.} 

Secondly, star-forming galaxies span a large range in colors due to dust attenuation. This can be seen in the {third} column of Figure \ref{fig:uvj}, which shows the infrared excess IRX$ = \mathrm{log_{10}} L_{IR}/L_{UV}$, the ratio between dust absorbed emission over unattenuated UV emission from star formation. The IRX ranges from IRX$\sim-1$ for the bluest UVJ colors (colors typically found in dropout selected samples, \citep[e.g.,][]{Bouwens14} to IRX$\sim3$ for the dustiest sources with the reddest $U-V$ and $V-J$ colors. Quiescent galaxies are also characterized by very low $L_{IR}/L_{UV}$ ratios, despite their red $U-V$ colors and faint UV fluxes. SED fitting shows a similar trend with best-fit dust attenuation $A_V$. Galaxies at the tip of the red star-forming sequence also appear to have lower sSFRs, so the colors of the very reddest galaxies appear to be a combination of dust and old age \citep[see also][Bedregal, in prep]{Forrest16,Fumagalli16}. {However, we caution that these galaxies are also expected to have the least accurate photometric redshifts (Section \ref{sec:pairs}).}

Finally, there is a clear trend with stellar mass, such that massive galaxies tend to be redder. Up to the highest redshifts probed here ($z\sim3.5$) massive $M > 10^{10.5}$\msol\ galaxies are predominantly quiescent and/or dusty \citep[see e.g.,][]{Straatman14,Spitler14}. This is not a selection effect, as only galaxies are shown above our mass completeness limit. The $U-V$ vs stellar mass (left column) shows an absence of low mass quiescent galaxies at $z=2-3$, with massive galaxies quenching first, and lower mass quiescent galaxies rapidly building up from $z=2$ to $z=0.5$ \citep[see][]{Tomczak14}.

\begin{figure}
\includegraphics[width=0.49\textwidth]{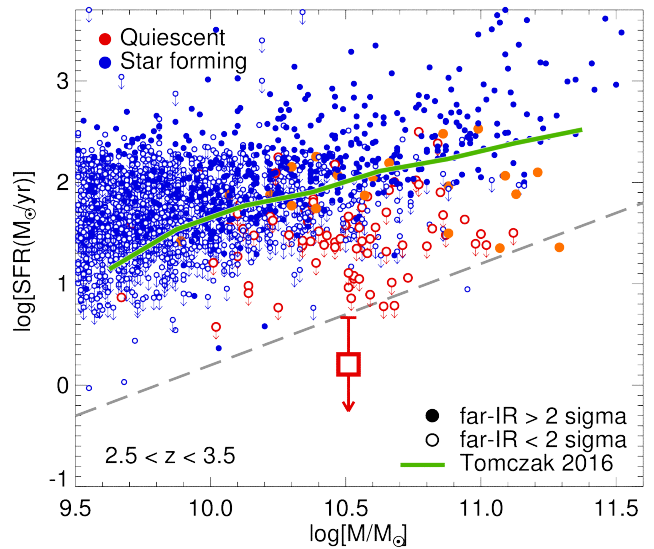}
\caption{SFR versus stellar mass for galaxies at $2.5<z<3.5$. We show UVJ selected star-forming galaxies in blue and UVJ selected quiescent galaxies in orange/red.  Galaxies with $<2\sigma$ measurements in the far-IR are shown with open symbols. Downward pointing arrows are $<1\sigma$ limits on SFR. The large square symbol represents the median $SFR_{IR}$ of 21 undetected quiescent galaxies in CDFS. The thick green line is the median SFR-stellar mass relation for star-forming galaxies from \citet{Tomczak16}, and the gray dashed line represents the criterion for quiescence at $z=3$, derived using sSFR$<(3 t_h)^{-1}$.}
\label{fig:msfr}
\end{figure}

The usage of the UVJ technique to identify quiescent galaxies has been validated to $z\lesssim 2.0$, by inspecting stacked WFC3 grism spectra \citep{Whitaker13} and stacked MIPS $24\micron$ data \citep{Fumagalli14}. However, verification at $z\gtrsim3$ has proved difficult. \citet{Straatman14} and \citet{Spitler14} identified quiescent galaxies in the UVJ diagram at even at higher redshifts, to $z\sim 4$, but at these extreme redshifts the IR observations are not deep enough, and the samples are too small, to rule out significant obscured star formation. Here, we place the strongest constraints yet on the SFRs of quiescent galaxies at $z\sim3$ by studying the infrared fluxes from ultradeep PACS Herschel data from GOODS-Herschel and PEP. Using the net UV+IR SFR based on the ultradeep Herschel data, we show the SFR versus stellar mass diagram in Figure \ref{fig:msfr}. The individual quiescent galaxies at $2.5 < z < 3.5$ are shown in red and orange. 80\% of the quiescent galaxies are not detected in the far-IR at $<2\sigma$ (open red symbols). 

To confirm a galaxy as quiescent based on sSFR, we adopt the definition of \citet{Damen09} which states that a galaxy is quiescent if their sSFR is $< (3 t_h)^{-1}$, with $t_h$ the Hubble time. At $z=3$ this limit corresponds to a sSFR$< 1.6 \times 10^{-10}\ \mathrm{yr^{-1}}$. The observed constraints on the sSFR of individual galaxies are not very strong, due to the limited  sensitivity in the IR at such high redshift. At $z = 3$ our best estimate of the SFR comes from the combination of ultradeep $Spitzer$/MIPS $24~\micron$ and Herschel/Pacs/$100$ and $160~\micron$ data in the CDFS (see Section \ref{sec:fast}). The $1\sigma$ limiting SFR for individual sources is SFR$_{IR} < 15$~\msun~$\mathrm{yr^{-1}}$. This means that for galaxies with stellar mass $5 \times 10^{10}$~\msun\ at $2.5 < z < 3.5$, we can constrain their individual sSFRs to be $< 3.0 \times 10^{-10}\ \mathrm{yr^{-1}}$. To place firmer constraints, we follow the procedure of \citet{Straatman14} and stack the $24-160\ \micron$ fluxes {of UVJ-selected quiescent galaxies}, limiting our sample to $M>10^{10}$~\msun, to gain more sensitivity in the far-IR. We calculate the median of the stack of 21 IR undetected ($< 2 \sigma$) galaxies in CDFS, finding a SFR$_{IR}= 1.6\pm3.1$~\msun$\ \mathrm{yr^{-1}}$, where the errors are derived by bootstrapping. At a mean stellar mass of $3.2 \times 10^{10}$~\msun, this translates into a sSFR$=0.5\pm1.0\times 10^{-10}\ \mathrm{yr^{-1}}$,{ easily meeting the \citet{Damen09} criterion for quiescence.}

This shows that UVJ selected quiescent galaxies at $z\sim3$ indeed harbour very low levels of star formation, a factor of $>15\times$ lower at 95\% confidence than the median of star-forming galaxies at the same redshift \citep{Tomczak16}. Quenching of star formation was thus very efficient, even at such early times. 

\section{Summary}\label{sec:sum}

In this paper we have presented the data products and public release of ZFOURGE, a near-IR galaxy survey {with the \fs\ instrument} aimed to increase our understanding of the evolution of galaxies at intermediate to high redshift. {We made use of} 5 medium-bandwidth near-IR filters ($J_1,J_2,J_3,H_s,H_l$), and ultra-deep $K_s$-band detection images, obtained by co-adding the \fs/$K_s$-band images with publicly available $K_s$-band data. The medium-bandwidth filters are particularly well suited to constrain the photometric redshifts of sources in the redshift range $1.5<z<3.5$. 

The $K_s$-band detection images have $5\sigma$ depths between 25.5 and 26.5 AB magnitude, and we detected $>70,000$ galaxies in the three extragalactic fields CDFS, COSMOS and UDS. {The {ZFOURGE \fs} observations were augmented with images over a large range in wavelength, $0.3-160\micron$, and fluxes were consistently derived using accurate PSF modelling and deblending methods. The ZFOURGE data release comprises source catalogs with photometry, photometric redshifts and stellar population properties, such as stellar mass derived with SED modelling, and SFRs derived from UV and IR luminosities.}

{Photometric redshifts were tested against} spectroscopic redshifts from the literature, resulting in ${\sigma_{z}}=0.010$ in CDFS, ${\sigma_{z}}=0.009$ in COSMOS and ${\sigma_{z}}=0.011$ in UDS. As spectroscopic samples of galaxies are often biased towards bright and blue galaxies, we performed another, independent test of the robustness of the photometric redshifts, by inspecting galaxy pairs. We found excellent results, with $\sigma_z=0.01-0.02$ for a $K<23.5$ magnitude limited sample, between $z=0.5$ an $z=2.5$, {and showed that} photometric redshifts are better constrained by 50\% if the near-IR \fs\ medium-bandwidth filters are included, compared to SED fitting with the \fs\ filters removed.

We investigated the efficacy of the UVJ diagram {to classify galaxies} beyond $z=2$ {and} illustrated how UVJ colors correlate with sSFR and infrared luminosity excess (dust attenuation). 
Using the UVJ diagram, we selected a sample of quiescent galaxies at $2.5<z<3.5$ and investigated their sSFR properties. We confirmed that these were indeed quiescent, with an average sSFR$=0.5\pm1.0\times 10^{-10}$ and $>15\times$ supressed SFRs, relative to the average stellar mass versus SFR relation of star forming galaxies, thereby for the first time validating the UVJ classification to $z=3.5$.

\section{Acknowledgements}
This paper, as well as the ZFOURGE data products, could not have come about without the wonderful help of many people. We would like to thank, first of all, the staff at Las Campanas Observatory: David Osip, Jorge Araya, Herman Olivares, Gabriel Prieto, Hugo Rivera, Geraldo Valladares, Jorge Bravo, Gabriel Martin and Mauricio Navarrete. We would also like to thank the Mitchell family for their continuing support. We further thank Lisa Kewley, Tiantian Yuan, Marijn Franx, Mariska Kriek, Adam Muzzin and Jesse van de Sande for useful discussions. We would like to thank the anonymous referee for constructive criticisms. This work was supported by the George P. and Cynthia Woods Mitchell Institute for Fundamental Physics and Astronomy, the National Science Foundation grant AST-1009707 and the NL-NWO Spinoza Grant. Australian access to the Magellan Telescopes was supported through the National Collaborative Research Infrastructure Strategy of the Australian Federal Government. GGK acknowledges the support of the  Australian Research Council Future Fellowship FT140100933. KEW gratefully acknowledges support by NASA through Hubble Fellowship grant \#HST-HF2-51368 awarded by the Space Telescope Science Institute, which is operated by the Association of Universities for Research in Astronomy, Inc., for NASA, under contract NAS 5-26555. KG and TN acknowledge support of Australian Research Council grants DP130101460 and DP130101667.

\bibliographystyle{apj}

\appendix

\section{PSF convolution}\label{apx:psf}

\begin{figure*}
\includegraphics[width=\textwidth]{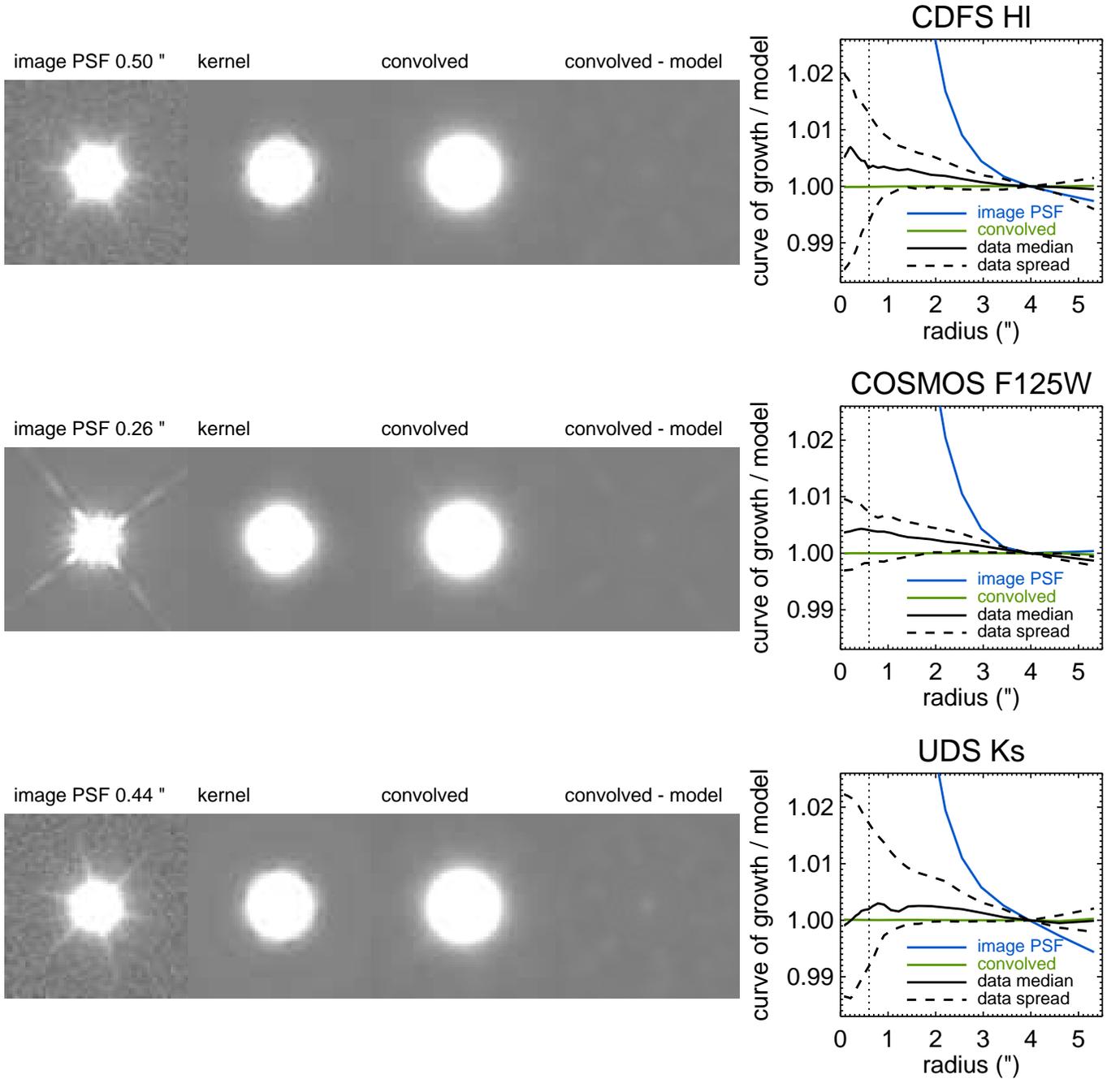}
\caption{Example PSF diagnostic plots. Here we show the groundbased \fs/$H_l$ and $K_s$ filters and the space based $HST$/WFC3/F125W filter, with various PSF widths. Postage stamps ($10\farcs65\times10\farcs65$) from left to right are: the median stacked PSF of the original science images, with their FWHM indicated at the top; a kernel derived using the deconvolution code developed by I. Labb\'e; the PSF from the leftmost panel, convolved using the corresponding kernel; the convolved PSF minus the Moffat model. The rightmost panels show curves of growth for a number of cases, divided by the model curve of growth and normalized at $r=4\arcsec$. A perfect comparison with the model means the ratio of curves of growth will be one at all radii. In blue we show the {ratio of the} curve of growth of the original PSF (leftmost postage stamp) {and the model PSF} and in green the same, but after convolving. The black lines represent the median and $1\sigma$ scatter of {curve of growth ratios for} individual point sources in the convolved images that were used for photometry.}
\label{fig:psf_selected}
\end{figure*}

In Section \ref{sec:conv} we have explained how images are convolved, such that their average point source profile matches a $FWHM=0\farcs9$ Moffat PSF, with the goal of obtaining consistent aperture photometry over all filters. Here we show example diagnostic plots (Figure \ref{fig:psf_selected}) of the kernel derivation. The four columns represent the original image PSF, the kernel used for convolution, the PSF after convolving with this kernel and the residual after subtracting the model Moffat profile. curves of growth representing the PSFs before and after convolution are shown in the rightmost panels. The green curve in particular shows the light profile of the convolved PSF divided by that of the model, which is close to unity. Whereas in Figure \ref{fig:gc} we show the curves of growth measured on the median PSF derived from the convolved images, we inspect here the median and $1\sigma$ scatter of the curves of growth of individual stars, finding similar residuals of $<2\%$ compared with the model PSFs.

\begin{figure*}
\includegraphics[width=\textwidth]{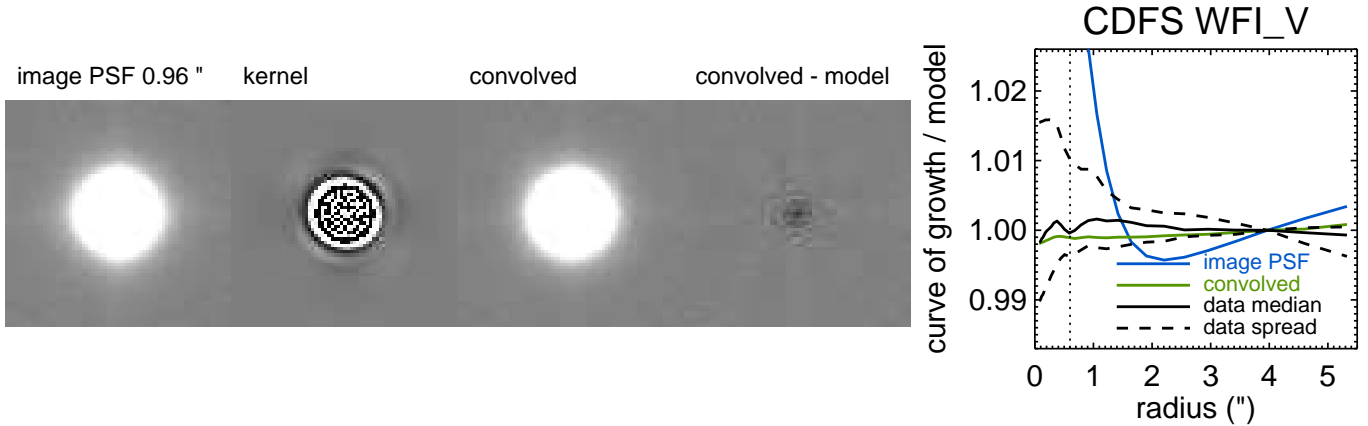}
\caption{Example of a case where the orginal PSF was worse than the target PSF. See Figure \ref{fig:psf_selected} for a description of the panels. Here the image was \textit{deconvolved}. A small residual is left after subtracting the model PSF from the deconvolved PSF, but the integrated flux at $r=0\farcs6$ corresponds well, to within 1\%.}
\label{fig:psf_decon}
\end{figure*}

Mathematically, we are able to produce PSF matched images in case of \textit{deconvolution}, when the image has a worse quality, i.e., a broader PSF, than the target $FWHM=0\farcs9$ Moffat profile. We have applied deconvolution in some cases with care, as we cannot make any assumptions on the true underlying light profile of galaxies. We find that the strongest residuals occur once the image PSF becomes much larger than the target PSF. After inspecting the residuals by eye and taking into account the aperture radius with a diameter of $1\farcs2$, we include images that have up to 15\% broader PSFs than the target PSF. We show an example in Figure \ref{fig:psf_decon}. While the deconvolved PSF shows some residual compared to the model, the curves of growth indicate that we capture the same amount of light within 1\% at $r=0\farcs6$, the aperture radius that we use to derive the catalog fluxes with. In total 11 of 92 UV to near-IR images were deconvolved.

\section{Comparison to the 3D-HST photometric catalogs}\label{sec:flux_comparisons}

\begin{figure*}
\includegraphics[width=0.9\textwidth]{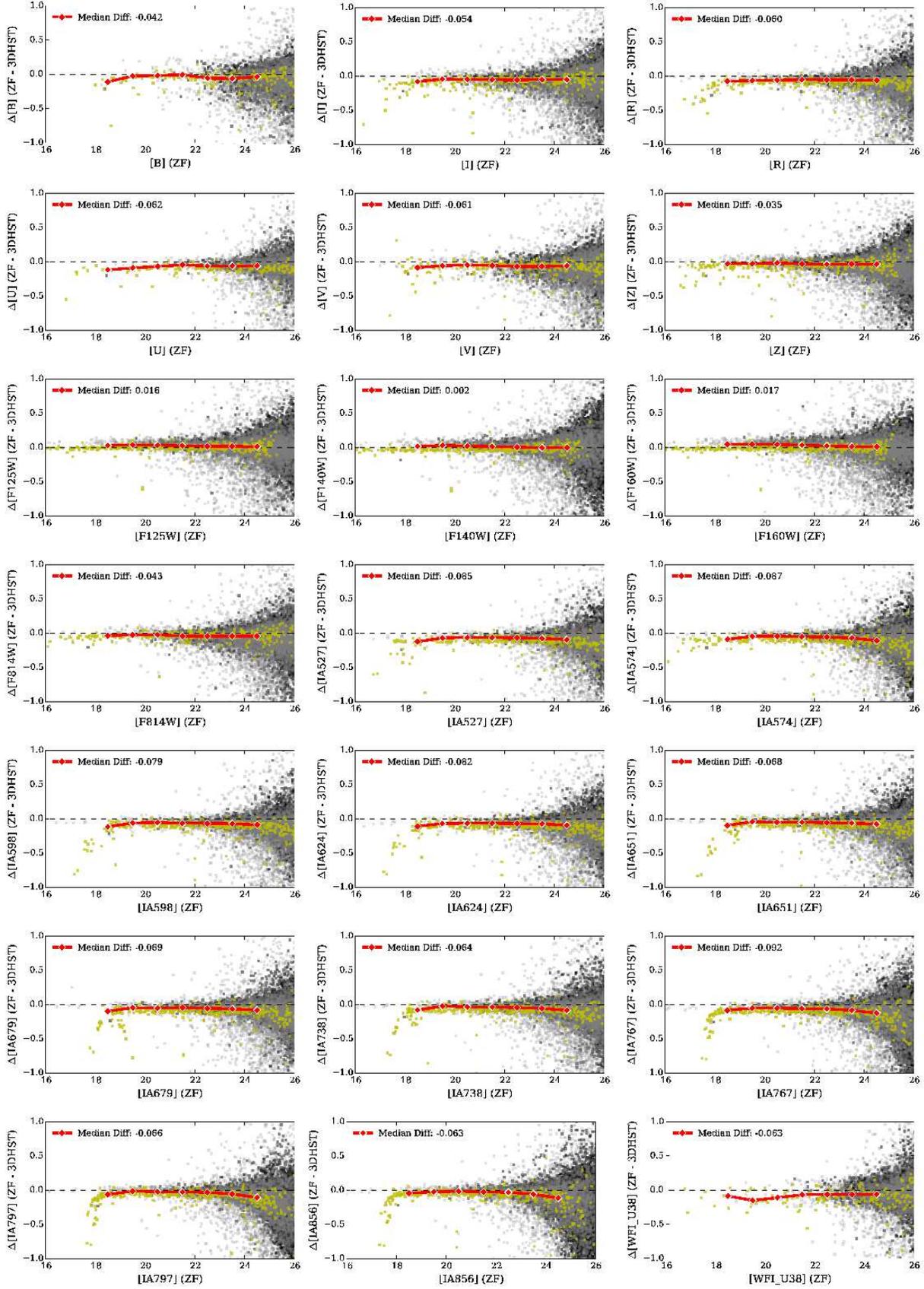}
\caption{The difference between ZFOURGE and 3D-HST {total} magnitudes plotted as a function of ZFOURGE magnitudes in CDFS for each band in common. Galaxies with {\tt use=1} are shown as black points, point sources with {\tt star=1} are shown as yellow points, and blended sources with SEflag$=2$ (and {\tt use=1}) are shown as grey points. The median magnitude difference for all galaxies is shown by the red solid line and large red diamond symbols in bins of 1 mag.}
\label{fig:deltamag_cdfs1}
\end{figure*}

\renewcommand{\thefigure}{\arabic{figure} (Cont.)}
\addtocounter{figure}{-1}

\begin{figure*}
\includegraphics[width=0.9\textwidth]{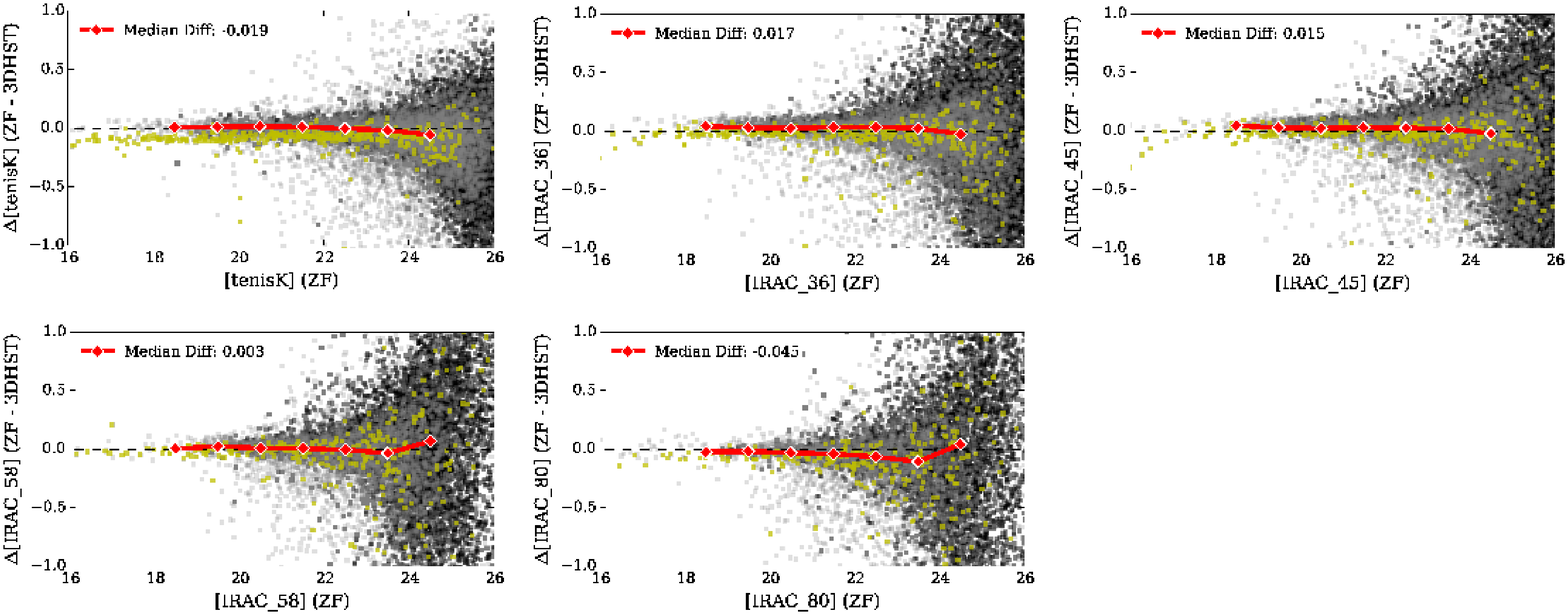}
\caption{}
\label{fig:deltamag_cdfs2}
\end{figure*}

\renewcommand{\thefigure}{\arabic{figure}}

\begin{figure*}
\includegraphics[width=0.9\textwidth]{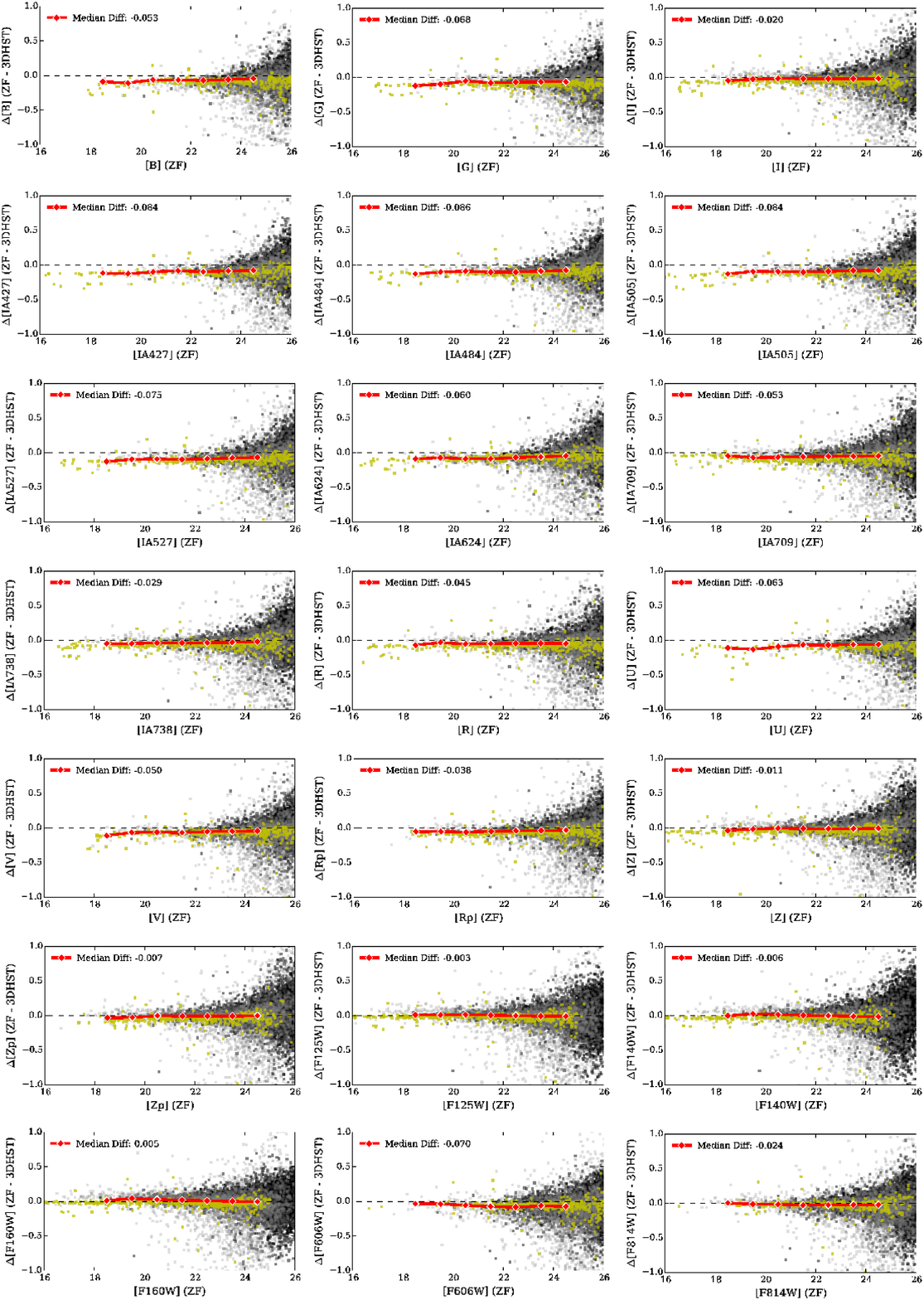}
\caption{The difference between ZFOURGE and 3D-HST {total} magnitudes plotted as a function of ZFOURGE magnitudes in COSMOS for each band in common. Symbols are the same as in Figure \ref{fig:deltamag_cdfs1}.}
\label{fig:deltamag_cosmos1}
\end{figure*}

\renewcommand{\thefigure}{\arabic{figure} (Cont.)}
\addtocounter{figure}{-1}

\begin{figure*}
\includegraphics[width=0.9\textwidth]{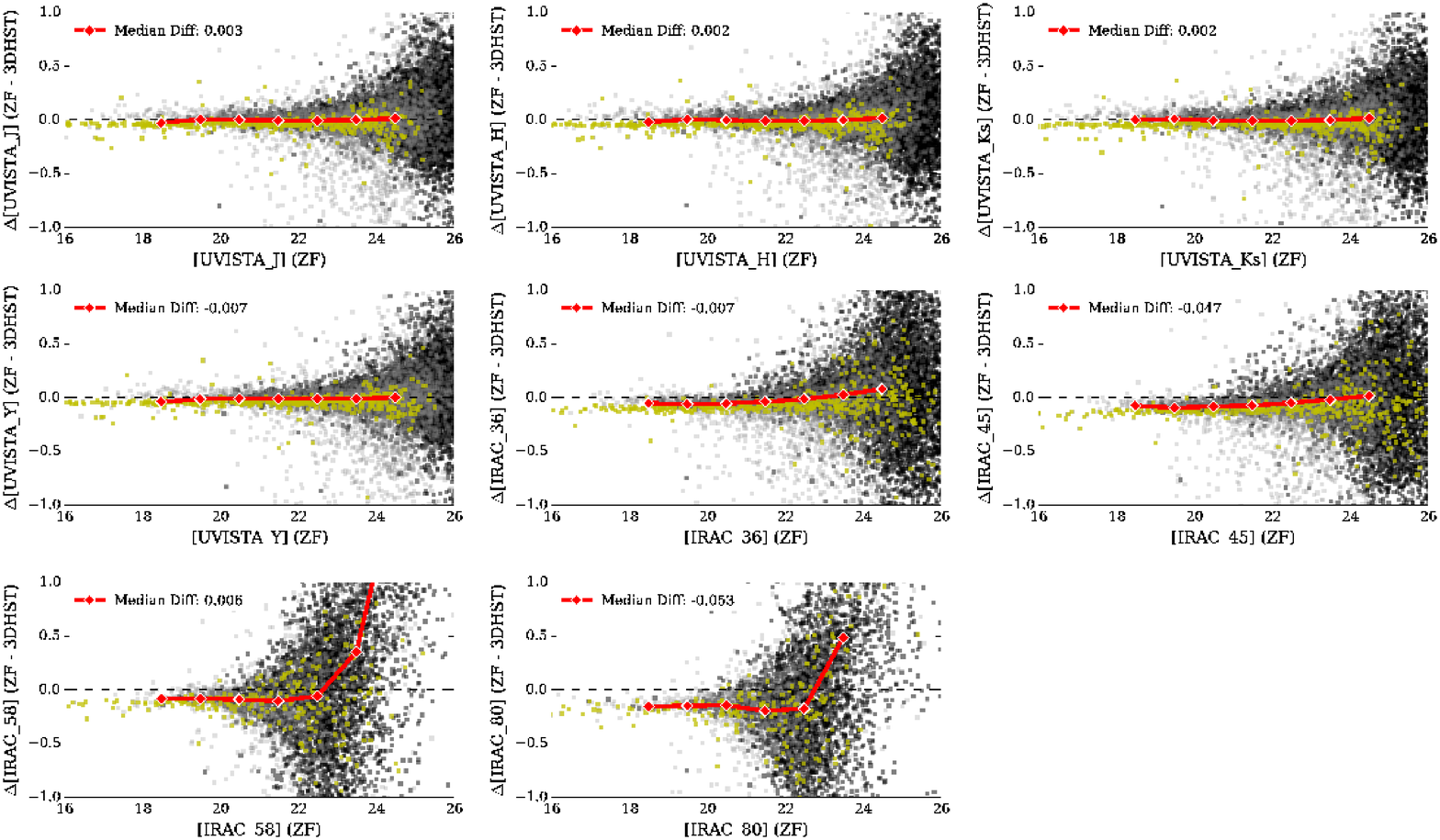}
\caption{}
\label{fig:deltamag_cosmos2}
\end{figure*}

\renewcommand{\thefigure}{\arabic{figure}}

\begin{figure*}
\includegraphics[width=0.9\textwidth]{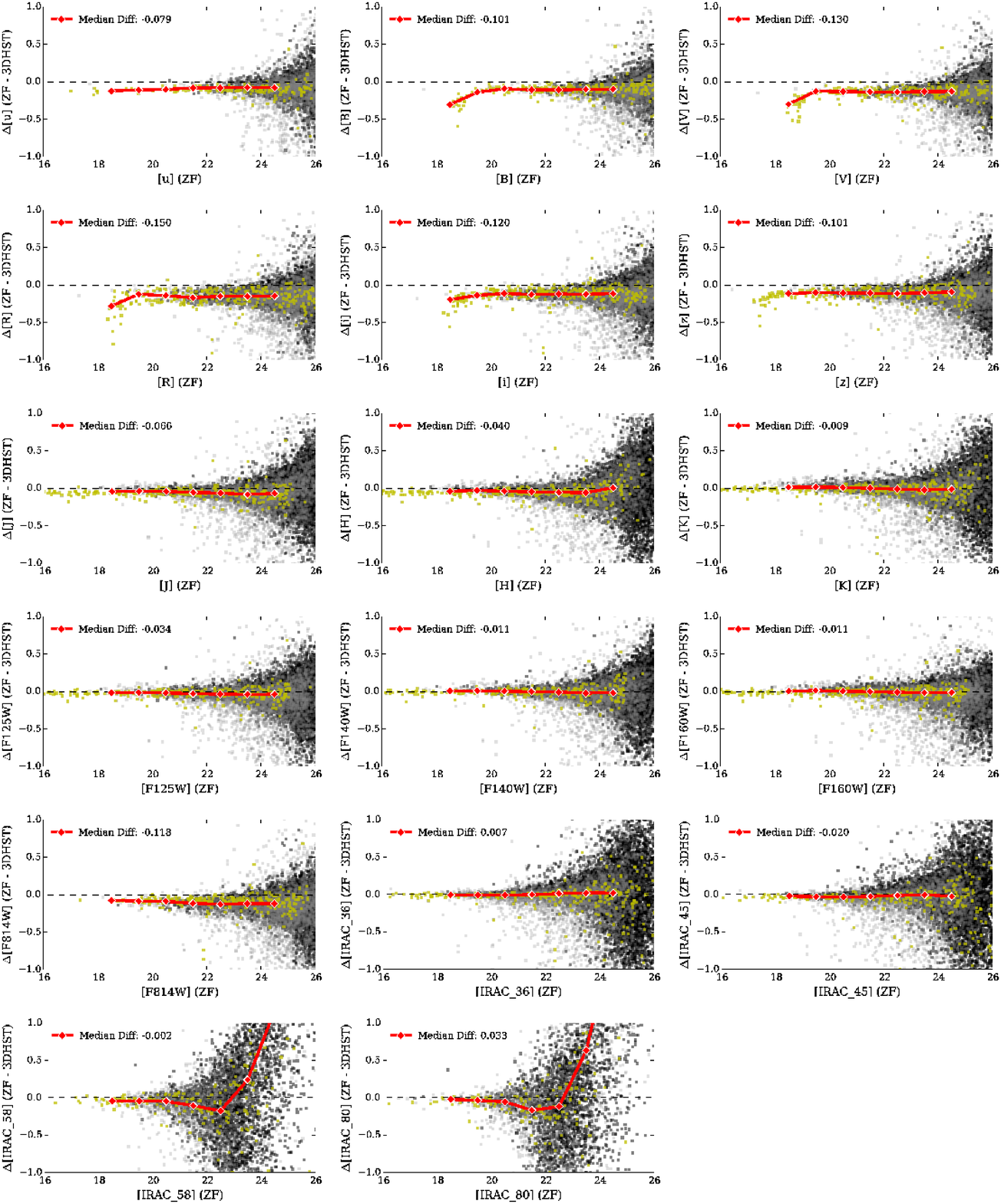}
\caption{The difference between ZFOURGE and 3D-HST {total} magnitudes plotted as a function of ZFOURGE magnitudes in UDS for each band in common. Symbols are the same as in Figure \ref{fig:deltamag_cdfs1}.}
\label{fig:deltamag_uds}
\end{figure*}

In this section we {compare} of the total magnitudes measured by ZFOURGE and those measured by 3D-HST, who make use of many of the same ancillary images. They also largely use the same methods to derive photometry. For each filter in common we calculate the difference in magnitude between crossmatched sources and show this versus total magnitude as per the ZFOURGE catalogs in Figures \ref{fig:deltamag_cdfs1} to \ref{fig:deltamag_uds}. For crossmatching we used a maximum angular separation of $1\arcsec$. We separately indicated sources that were flagged as possibly blended or contaminated by neighbours by SE ({\tt SEflags}$\geq2$). $\Delta$mag has somewhat more scatter for these sources at faint magnitudes, but overall the correspondence is quite good between the surveys, with $\Delta mag$ close to zero. The most notable exceptions are the $V,R,i$ and $z$-bands in UDS, which tend to be $\sim0.1$ magnitudes brighter in ZFOURGE.

\begin{figure*}
\begin{center}
\includegraphics[width=0.33\textwidth]{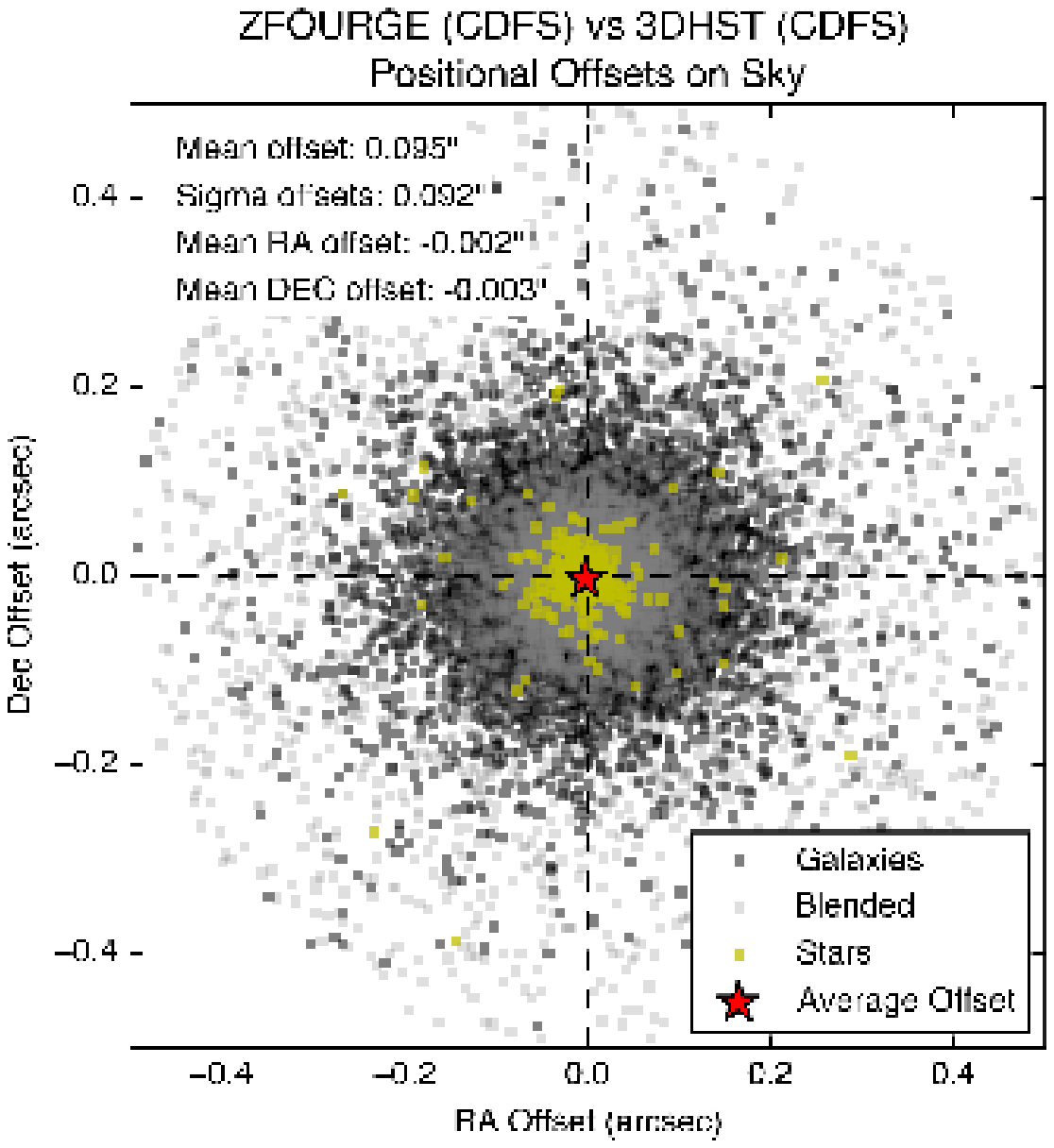}
\includegraphics[width=0.33\textwidth]{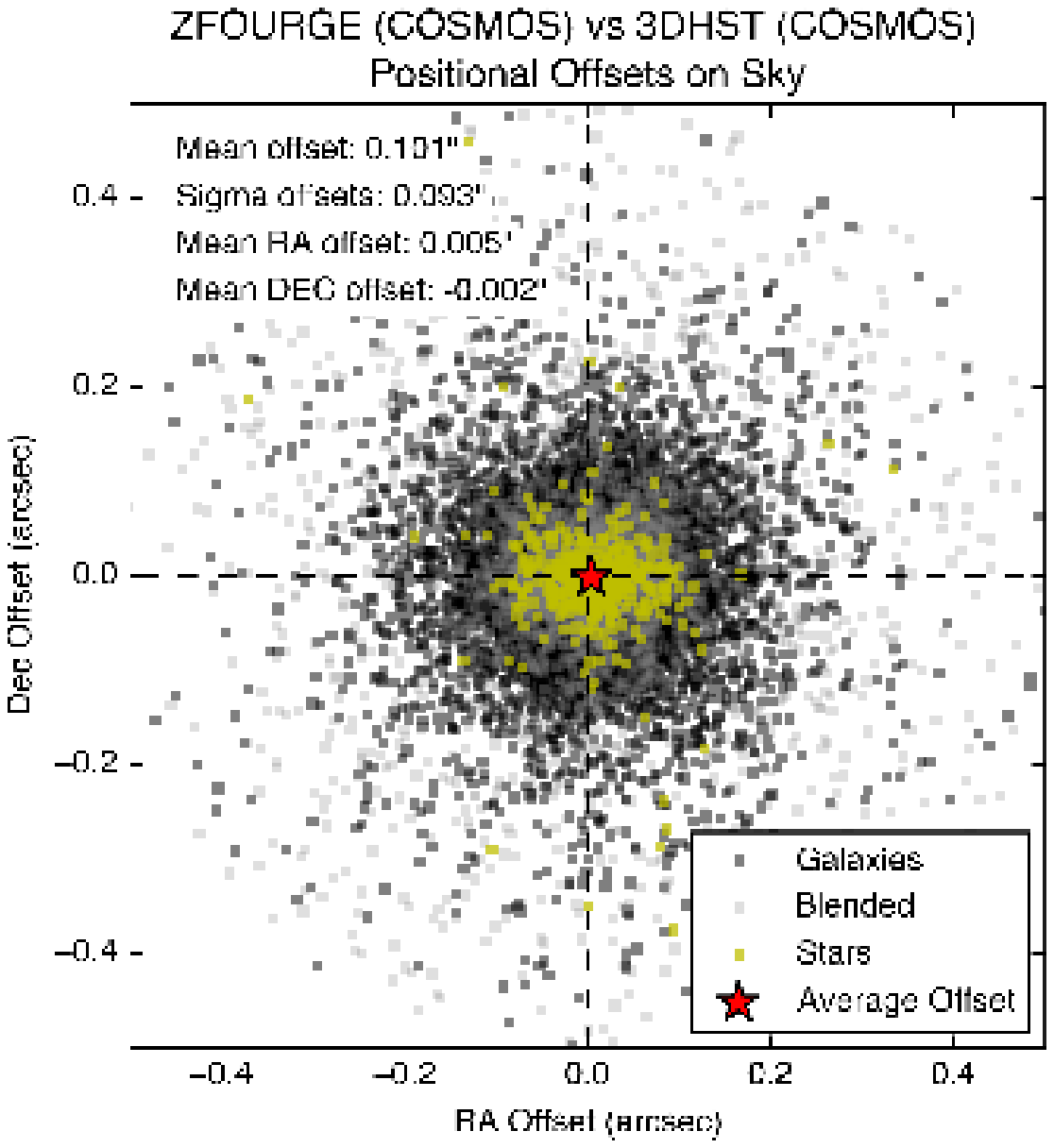}
\includegraphics[width=0.33\textwidth]{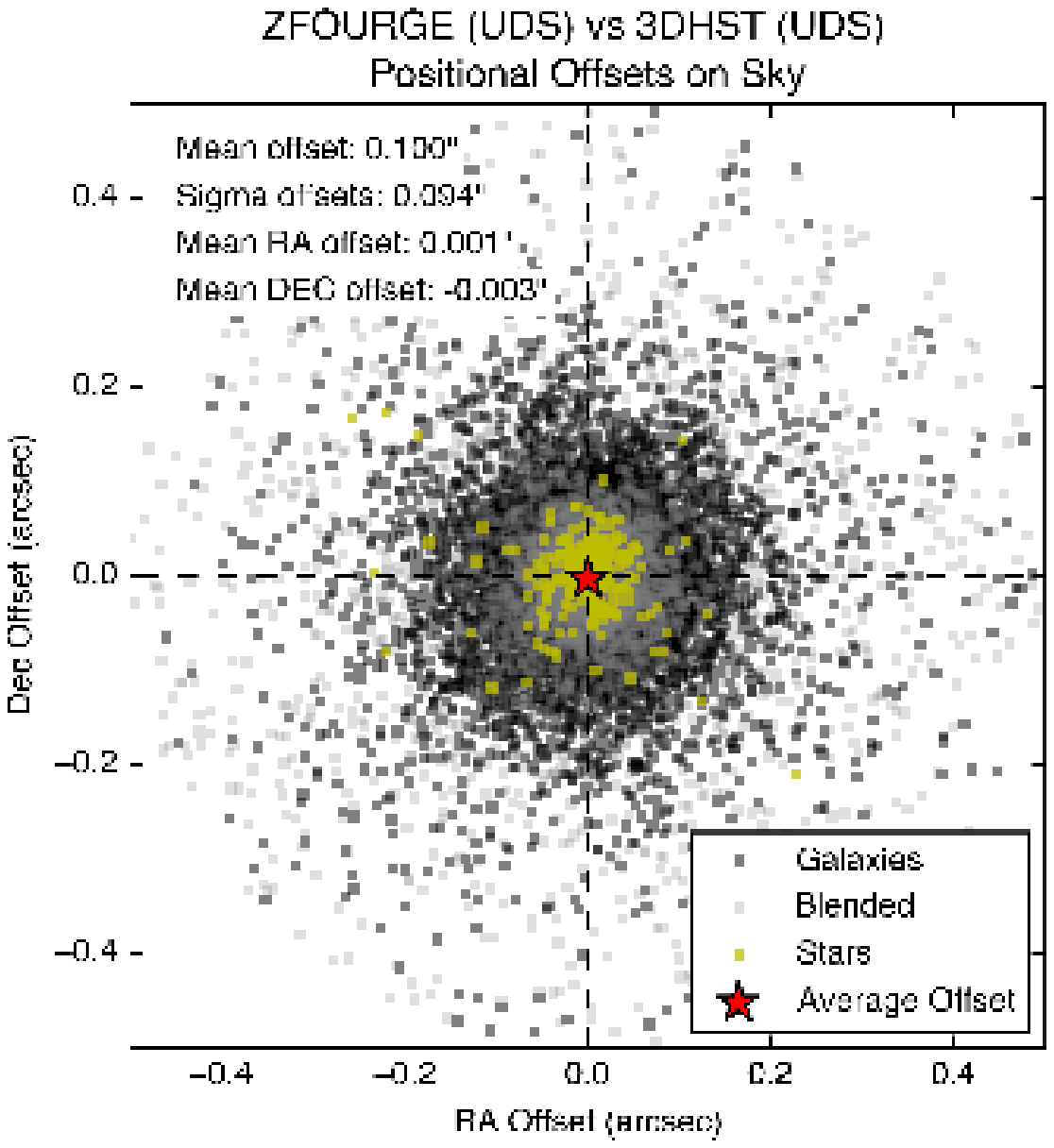}
\caption{Positional offsets between source locations in ZFOURGE and 3D-HST, using the same symbols as in Figures \ref{fig:deltamag_uds} to \ref{fig:deltamag_cosmos2}. The median offsets are indicated by red stars.}
\label{fig:astrm}
\end{center}
\end{figure*}

In Figure \ref{fig:astrm} we show the positional offsets between source locations in ZFOURGE and the 3D-HST survey. The median offsets are $\leq0\farcs005$ in RA or Dec in all fields, indicating the images are uniformly calibrated and can be reliably used for inter-survey comparisons.

\section{Spatial variation in the zeropoints}\label{sec:xyresid}

\begin{figure*}
\begin{center}
\includegraphics[width=\textwidth]{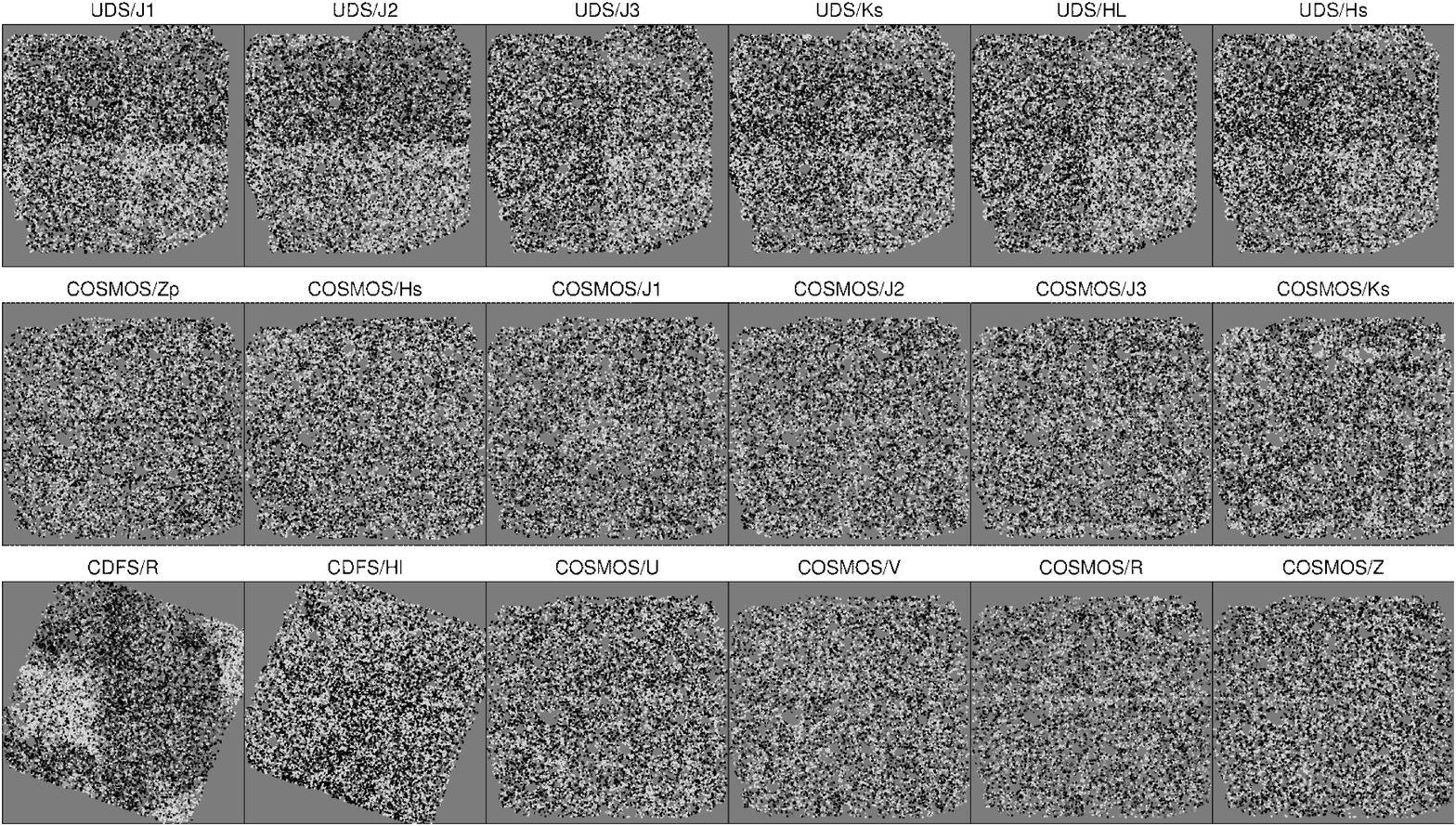}
\caption{Example spatial zeropoint residual maps of {after subtracting the 2-D polynomial fit}. The grayscale ranges from 0.95 to 1.05, i.e., a 5\% flux deviation. {CDFS/R had a complicated structure with large zeropoint variations, even after corrections. We resolved this by adding a 5\% systematic uncertainty as an error floor when fitting photometry.}}
\label{fig:xyresid}
\end{center}
\end{figure*}

In Figure \ref{fig:xyresid} we show example spatial zeropoint residual maps derived with EAZY, by comparing the best-fit templates to the observed galaxy SEDs. The residuals are of the order of $<5\ \%$. We fitted a two-dimensional polynomial to each offset map and used these to {derive} a correction to the flux of a specific filter, for all sources as a function of their x- and y-position. This was done for the full dataset. {With this method we were able to trace systematic offsets of $\pm \sim4\ \%$ between the four \fs\ detectors in UDS and correct these to $\pm \sim2\ \%$.} In one image strong, non-linear spatial effects stand out: this is in VIMOS/$R$ in CDFS. In this image the strong spatial varation could not be removed by this first order correction. Due to its large depth, we have kept the image in our sample, but applied a minimum error floor to the $V$-band flux of 5\% to take into account uncertainties on the zeropoint of the image.

\section{UVJ diagram field comparison}

\begin{figure*}
\begin{center}
\includegraphics[width=\textwidth]{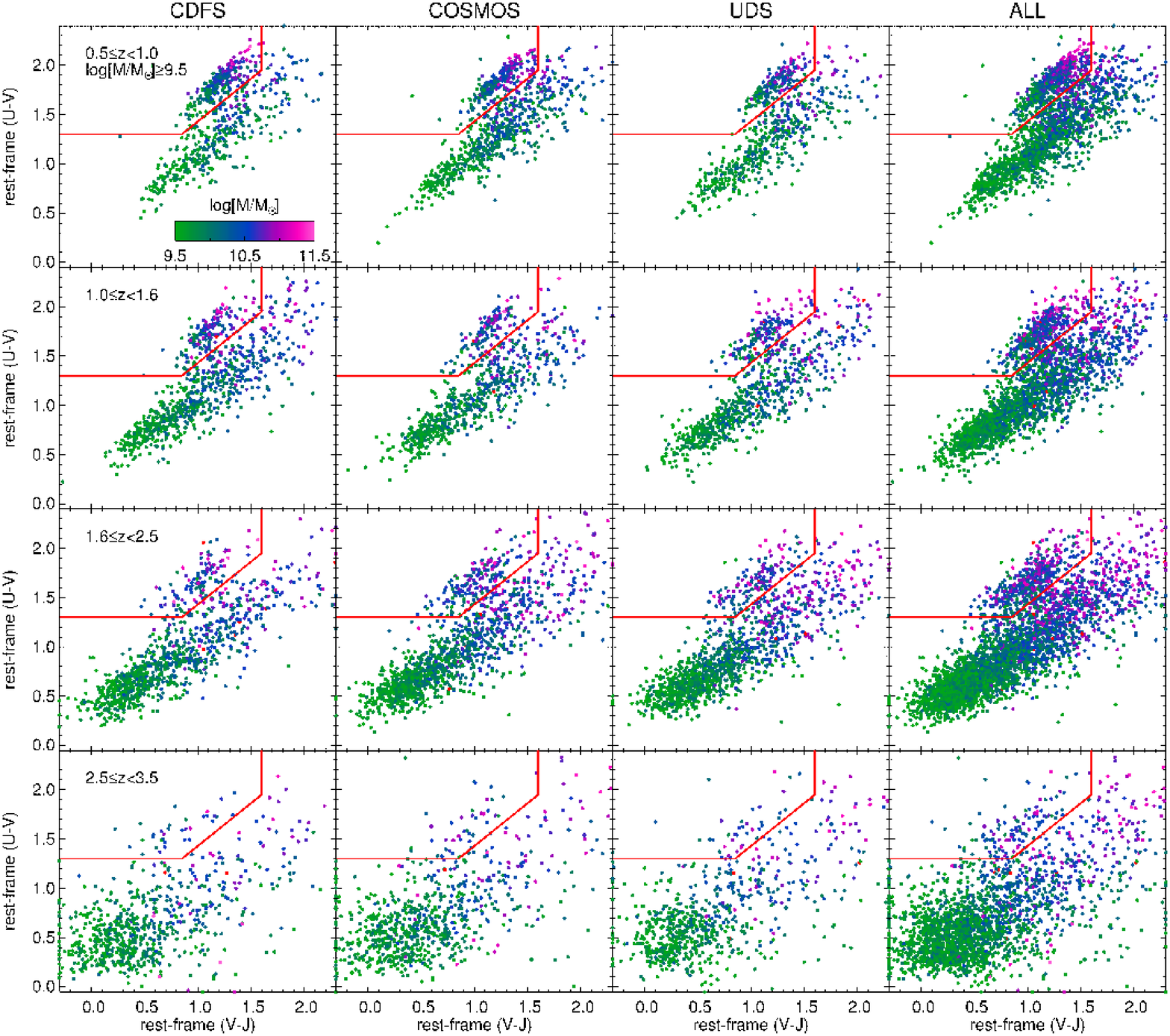}
\end{center}
\caption{Rest-frame $U-V$ versus $V-J$ diagrams of galaxies with {\tt use=1}, $SNR_{K_s}>10$ and stellar mass $M>10^{9.5}M_{\Sun}$. In the first three columns we show the three ZFOURGE fields. In the last column these are combined. From top to bottom we show bins of increasing redshift. The color scaling indicates stellar mass. The rest-frame $U-V$ and $V-J$ colors in each field show the same pattern, and the same location for the red sequence, indicating consistent photometry.}
\label{fig:uvjfd}
\end{figure*}

In this section we show the UVJ diagram color-coded by stellar mass (Figure \ref{fig:uvjfd}), as in the third column of Figure \ref{fig:uvj}. We show the same redshift bins, but split the diagrams into the three ZFOURGE fields. By comparing the rest-frame colors in the different fields we can look for inconsistencies, for example if the median locii of the datapoints are offset relative to each other. Here this is not the case, indicating consistent photometry.

\end{document}